\documentclass[a4paper,11pt]{article}
\pdfoutput=1 
\usepackage{array}
\usepackage{jheppub} 
\usepackage[T1]{fontenc} 
\usepackage{amsmath}   
\usepackage{amssymb}   
\usepackage{booktabs}  
\usepackage{cleveref}
\usepackage{makecell}
\usepackage{cellspace} 
\usepackage{slashed}
\newcommand{\rd}{\mathrm{d}}

\newcommand{\ri}{\mathrm{i}}

\title{\boldmath All Giant Graviton Two-Point Functions at Two-Loops}

\author[a]{Yu Wu\footnote{\emph{Note: The unusual ordering of authors instead of the standard alphabetical one in hep-th community is for students to get proper recognition of contribution under the current out-dated practice in China.}}}
\author[b,c]{Yunfeng Jiang}
\author[b]{Chang Liu}
\author[a,c,d]{Yang Zhang}

\affiliation[a]{Interdisciplinary Center for Theoretical Study, University of Science and Technology of China,\\ Hefei, Anhui 230026, China}

\affiliation[b]{School of physics \& Shing-Tung Yau Center, Southeast University,\\ Nanjing  211189, P. R. China}

\affiliation[c]{Peng Huanwu Center for Fundamental Theory, Hefei, Anhui 230026, China}
\affiliation[d]{Center for High Energy Physics, Peking University,
Beijing 100871, People’s Republic of China}

\emailAdd{wy626@mail.ustc.edu.cn}
\emailAdd{jinagyf2008@seu.edu.cn}
\emailAdd{chang\_liu\_nc3b@seu.edu.cn}
\emailAdd{yzhphy@ustc.edu.cn}

\abstract{We present a comprehensive two-loop computation of correlation functions involving two maximal giant gravitons and two arbitrary $R$-charge single-trace half-BPS operators in $\mathcal{N}=4$ Super-Yang-Mills theory. By combining the partially-contracted giant graviton (PCGG) method with the $\mathcal{N}=2$ harmonic superspace formalism, we achieve significant simplifications in perturbative calculations. The resulting correlation functions encode rich CFT data, from which we derive sum rules for the OPE coefficients. These sum rules are in perfect agreement with integrability predictions. Furthermore, at the integrand level, we find a hidden higher dimensional symmetry present at both one- and two-loop orders. This symmetry was discovered recently at strong coupling, which generalizes its counterpart in correlation functions of single-trace operators.}

\begin{document}
\begin{flushright}
{\tt USTC-ICTS/PCFT-25-35
}
\end{flushright}
\maketitle
\flushbottom

\section{Introduction}
\label{sec:intro}
Correlation functions are fundamental observables in quantum field theories. In a conformal field theory, they encode rich conformal data (scaling dimensions and OPE coefficients) and  through AdS/CFT duality, they are dual to scattering process in the bulk which offer useful insights into quantum gravity. In four-dimensional $\mathcal{N}=4$ super-Yang-Mills ($\mathcal{N}=4$ SYM) theory, the most extensively studied correlation functions are those of single-trace half-BPS operators, which are dual to gravitons and Kaluza-Klein (KK) modes in type-IIB supergravity. While two- and three-point functions of these operators are protected by supersymmetry and do not receive quantum corrections, their four-point functions are no longer protected and are highly non-trivial functions of the coupling constant. Computing these four-point functions is a key challenge in unraveling the full structure of $\mathcal{N}=4$ SYM theory. Over the past decades, significant progress has been made at both weak \cite{Eden:2011we,Eden:2012tu,Chicherin:2015edu,Chicherin:2018avq,Fleury:2019ydf,Coronado:2018ypq,Fleury:2016ykk,Bourjaily:2016evz,Bourjaily:2025iad} and strong coupling \cite{DHoker:1999kzh,Arutyunov:2000py,Rastelli:2016nze,Rastelli:2017udc,Arutyunov:2018neq,Arutyunov:2018neq,Alday:2019nin,Alday:2020dtb,Drummond:2019hel,Aprile:2019rep,Drummond:2022dxw,Caron-Huot:2018kta,Caron-Huot:2021usw,Huang:2024rxr,Huang:2024dxr,Wang:2025pjo,Fernandes:2025eqe} through cutting edge perturbative methods and bootstrap approach. Additionally, non-perturbative techniques such as integrability and supersymmetric localization have been developed to compute special classes of four-point functions \cite{Bargheer:2017nne,Bargheer:2018jvq,Fleury:2019ydf,Bargheer:2019kxb,Coronado:2018cxj,Bargheer:2019exp,Caron-Huot:2022sdy,Caron-Huot:2024tzr,Kostov:2019stn,Kostov:2019auq,Belitsky:2020qrm} and integrated correlators \cite{Binder:2019jwn,Chester:2020dja,Brown:2024tru,Brown:2025huy}, providing valuable finite coupling data.\par

Type-IIB supergravity/superstring theory in AdS$_5\times S^5$ hosts a diverse spectrum of objects beyond graviton and KK modes. In particular, due to the Myers effect \cite{Myers:1999ps,McGreevy:2000cw}, a graviton with large angular momentum in the presence of Ramond-Ramond flux will expand into an extended object called a 'giant graviton' \cite{McGreevy:2000cw,Balasubramanian:2001nh,Hashimoto:2000zp}. It is described by a D3 brane wrapping an $S^3\subset S^5$ in the internal space. The size of the D3-brane is bounded, a manifestation of the stringy exclusion principle, and the configuration with maximal size is termed a maximal giant graviton. In AdS/CFT correspondence, giant gravitons are dual to (sub-)determinant operators which are also half-BPS operators. Correlation functions involving these operators alongside the single-trace operators are of significant interest from both field theoretic and bulk perspectives. On the bulk side, they describe interactions between giant gravitons and KK modes, while on the field theory side, determinant operators represent baryonic states in large-$N$ gauge theory \cite{Berenstein:2002ke,Witten:1998xy,Witten:1979kh,Witten:1983tx}, serving as a valuable toy model for studying baryonic physics.\par

Correlation functions involving determinant operators have been much less studied, especially at quantum level compared to their single-trace counterparts. One of the main reasons is their computational complexity. With scaling dimensions of order $N_c$, the determinant operators are very heavy in the large-$N_c$ limit. Wick contractions among these operators is rather intricate and identifying the dominant contributions to the correlation functions in the large-$N_c$ limit is much more non-trivial  \cite{Corley:2001zk,deMelloKoch:2019dda,deMelloKoch:2004crq,Berenstein:2013md,Lin:2022wdr,Lin:2022gbu,Holguin:2022drf,Holguin:2023naq}. A large number of seemingly subleading non-planar contributions can add up together and lead to non-negligible contributions. As a result, exact computation of quantum corrections of the giant graviton correlation functions have only started recently. At weak coupling, one-loop results involving two maximal giant gravitons and two arbitrary single-trace half-BPS operators are obtained by the partially contracted giant graviton (PCGG) method and a direct Feynman diagram computation \cite{Jiang:2019xdz}. Progress at higher loops has been more limited: current two- and three-loop results are restricted to cases involving length-2 single-trace operators, obtained through by Lagrangian insertion and a bootstrap approach \cite{Jiang:2019xdz,Jiang:2023uut}. At strong coupling, correlation functions involving two maximal giant gravitons and two arbitrary KK modes have been computed very recently by superconformal bootstrap \cite{Chen:2025yxg}. At finite coupling, integrated correlation functions with two giants and two length-2 operators have been obtained by localization \cite{Brown:2024tru,Brown:2024yvt,Brown:2025huy}. In this work, we extend these results by presenting the complete two-loop computation of correlation functions involving two maximal giant gravitons and two single-trace operators of arbitrary length in the large-$N_c$ limit.\par

To achieve this, we developed a new method by merging the $\mathcal{N}=2$ harmonic superspace with the PCGG method, which we term the \emph{harmonic PCGG approach}. By exploiting supersymmetry and planarity, harmonic PCGG method reduces the number of required Feynman diagrams significantly. For example, in the previous one-loop computation \cite{Jiang:2019xdz}, one needs to compute 31 distinct Feynman diagrams while using the harmonic PCGG approach we only need to compute 1 diagram. At the two-loop level, as we will demonstrate, the complete result for two single-trace operators of arbitrary length can be obtained by computing only 26 diagrams.\par

Our one- and two-loop results reveal a remarkable structural pattern at the integrand level, which can be interpreted as a defect generalization of the hidden 10D conformal symmetry. This symmetry has a well-established counterpart in the four-point functions of single-trace operators, first identified at strong coupling \cite{Caron-Huot:2018kta} and later at weak coupling \cite{Caron-Huot:2021usw} in $\mathcal{N}=4$ SYM theory. Subsequent work has uncovered analogous structures in various other contexts \cite{Abl:2020dbx,Aprile:2020luw,Huang:2024dxr,Wang:2025pjo}. For giant graviton two-point functions, this symmetry was previously observed at strong coupling \cite{Chen:2025yxg}. Our results now demonstrate its persistence at weak coupling up to at least two-loop order. While the fundamental origin of this hidden conformal symmetry remains mysterious, it provides a powerful organizational principle: knowledge of length-2 operator results suffices to reconstruct correlation functions for operators of arbitrary lengths. This remarkable property offers significant computational advantages and suggests deeper underlying structure in the theory.\par

The rest of this paper is organized as follows. Section~\ref{sec:harmonicSpace} provides a concise introduction to the $\mathcal{N}=2$ harmonic superspace formalism. Building on this, Section~\ref{sec:hPCGG} presents a detailed development of the harmonic PCGG method. We then apply this method to compute correlation functions at one-loop (Section~\ref{sec:one}) and two-loop (Section~\ref{sec:two}) orders. The resulting structures reveal a hidden 10D conformal symmetry, which we analyze in Section~\ref{sec:hidden}. In Section~\ref{sec:integrbility}, we extract OPE data through conformal block decomposition and compare these results with integrability predictions, providing a consistency check of our approach. Finally, Section~\ref{sec:conclude} summarizes our findings and outlines future research directions. Additional technical details can be found in the appendices.

\section{Review of harmonic superspace}
\label{sec:harmonicSpace}
We review $\mathcal{N}=2$ harmonic superspace in this section, providing necessary background for later discussions as well as fixing notations and conventions. For a more detailed introduction to harmonic superspace and its applications, we refer to \cite{Eden:1998hh,Galperin1984,Eden:2000bk,Eden:2000mv,Eden:2010zz,Galperin:2001seg,Arutyunov:2002fh,Arutyunov:2003ad,Arutyunov:2003ae,Chicherin:2014esa} and references therein.

\subsection{Harmonic superfields}
A superfield is a function defined on a superspace, which extends Minkowski spacetime by the addition of Grassmann coordinates. In the standard $\mathcal{N}=2$ superspace, the superfield corresponding to a hypermultiplet $\Phi^i(x,\theta)$ depends on spacetime coordinate $x^{\mu}$ and two sets of Grassmann coordinates $\{\theta^{i\alpha },\bar{\theta}^{i\dot{\alpha}}\}$ $(i=1,2)$. Expanding the superfield with respect to the Grassmann coordinates yields coefficients that include both physical and auxiliary fields. To eliminate the auxiliary fields, one imposes the additional constraints
\begin{equation}
\label{cons1}
D_\alpha^{(i} \Phi^{j)}(x,\theta)=\bar{D}_{\dot{\alpha}}^{(i} \Phi^{j)}(x,\theta)=0,
\end{equation}
where $ D_\alpha^i$ and $\bar{D}_{\dot{\alpha}}^i$ are covariant spinor derivatives in the standard $\mathcal{N}=2$ superspace. However, these constraints put the physical fields on-shell at the same time. Therefore the standard $\mathcal{N}=2$ does not permit an off-shell formulation for supersymmetric quantum field theory.\par

\paragraph{Harmonic superspace} The resolution to this issue is to further enlarge the space by introducing harmonic variables $u_i^+$ (and their complex conjugates $(u_i^+)^* = u_i^-$) and defining harmonic superspace. The harmonic variables parametrize the coset space $S^2\sim SU(2)/U(1)$ and satisfy the following relations
\begin{equation}
    u_i^+ u^{-i}=\epsilon^{ij}u_i^+ u_j^-=1,\quad (u^{+i})^*=u_i^{-}=\epsilon_{ij}u^{-j},\quad \epsilon_{12}=-\epsilon^{12}=1.
\end{equation}
We may now consider superfields $\Phi^i(x,\theta,u)$ in harmonic superspace. The physical constraint \eqref{cons1} in the standard superspace is equivalent to imposing \emph{two} constraints: Grassmann analyticity and harmonic analyticity.
\paragraph{Grassmann analyticity} The Grassmann analyticity condition is given by
\begin{equation}
\label{cons2}
 D_\alpha^{+} \Phi(x,\theta,u)=\bar{D}_{\dot{\alpha}}^{+} \Phi(x,\theta,u)=0,  
\end{equation}
where the derivatives are defined as
\begin{equation}
 D_\alpha^{+}=u_i^{+}D_\alpha^i=\frac{\partial}{\partial \theta^{-\alpha}} , 
 \qquad \bar{D}_{\dot{\alpha}}^{+}=u_i^{+}\bar{D}_{\dot{\alpha}}^i=\frac{\partial}{\partial \bar{\theta}^{-\dot{\alpha}}} \,,
\end{equation}
with
\begin{equation}
\theta^{+\alpha}\equiv u_i^{+}\theta^{i\alpha},\qquad \bar{\theta}^{+\dot{\alpha}}\equiv u^{+}_i\bar{\theta}^{i\dot{\alpha}}\,.
\end{equation}
The Grassmann analyticity condition \eqref{cons2} can be solved by introducing the following coordinates
\begin{align}
\label{anaCoordinate}
x_{A}^{\alpha,\dot{\alpha}}=x^{\mu}\sigma_{\mu}^{\alpha,\dot{\alpha}}-4 \ri \theta^{(i\alpha}\bar{\theta}^{j)\dot{\alpha}}u_{i}^{+}u_{j}^{-}\,.
\end{align}
A superfield $q(x_A, \theta^{+}, \bar{\theta}^{+}, u)$ that depends only on $x_A^{\alpha\dot{\alpha}}$, $\theta^{+\alpha}$, $\bar{\theta}^{+\dot{\alpha}}$, and $u_i^{\pm}$ automatically satisfies the Grassmann analyticity condition. For simplicity, we will denote such superfields simply as $q(x_A)$.\par

Unlike \eqref{cons1}, the constraints \eqref{cons2} do not automatically place the physical fields of $q(x_A)$ on-shell. Thus, the path integrals of hypermultiplets remain well-defined in harmonic superspace. This allows us to use harmonic supergraphs to compute correlation functions of superfields, simplifying calculations compared to evaluating individual field components separately.

\paragraph{Harmonic analyticity} Expanding a harmonic superfield $q(x_A, \theta^+, \bar{\theta}^+, u)$ in terms of harmonic variables $u_i^{\pm}$ introduces an infinite tower of auxiliary fields. These fields can be eliminated by imposing the following condition
\begin{equation}
\label{H-anaylticity}
 D^{++}q(x_A,\theta^+,\bar{\theta}^+,u)=0\,,
\end{equation}
where $D^{++}$ is the harmonic derivative on the two-sphere $S^2$. Using the coordinate $x_A^{\alpha\dot{\alpha}}$ defined in \eqref{anaCoordinate}, $D^{++}$ can be written as
\begin{align}
\label{harmonic derivatives}
 D^{++}=u^{+i}\frac{\partial}{\partial u^{-i}}-4 \ri\,\theta^{+\alpha}\bar{\theta}^{+\dot{\alpha}}\frac{\partial}{\partial x_A^{\alpha,\dot{\alpha}}}. 
\end{align}
The condition \eqref{H-anaylticity} is called the harmonic analyticity (H-analyticity) condition. By imposing both the G-analyticity condition \eqref{cons2} and the H-analyticity condition \eqref{H-anaylticity}, we obtain the following expansion for the on-shell hypermultiplet
\begin{align}
q(x_A,\theta ^+,\bar{\theta}^+, u)=\phi^i(x_A) u_i^+ +\theta^{+\alpha}\psi_\alpha(x_A)+\bar{\theta}^+_{\dot{\alpha}}\bar{\kappa}^{\dot{\alpha}}(x_A)+4 \ri\,\theta^{+}\sigma^{\mu}\bar{\theta}^{+}\partial_{\mu}\phi^{i}(x_A)u^{-}_i.
\label{q}
\end{align}
The hypermultiplet consists of an $SU(2)$ doublet of scalars $\phi^{i}(x) (i=1,2)$, along with the Majorana spinors $\psi_{\alpha}$, $\bar{\kappa}^{\dot{\alpha}}$. All physical fields $\phi^i$, $\psi_{\alpha}$, $\bar{\kappa}^{\dot{\alpha}}$ satisfy  their respective free equations of motion $\Box\phi^{i}=\slashed{\partial}\psi=\slashed{\partial}\bar{\kappa}=0$.\par

The hypermultiplet $q(x_A)$ is complex and its conjugate $\tilde{q}(x_A)$ also satisfies Grassmann analyticity condition. After imposing harmonic analyticity condition, it can be expanded as
\begin{align}
    \tilde{q}\left(x_A, \theta^{+}, \bar{\theta}^{+}, u\right)=\bar{\phi}^i(x_A) u_i^{+}+\theta^{+\alpha} \kappa_\alpha(x_A)+\bar{\theta}_{\dot{\alpha}}^{+} \bar{\psi}^{\dot{\alpha}}(x_A)+4 i\bar{\theta}^{+}\bar{\sigma}^{\mu}\theta^{+}\partial_{\mu}\bar{\phi}^{i}(x_A)u^{-}_i.\label{tq}
\end{align}

\subsection{Action and Feynman rules}
\label{sec:FeyRule}
With harmonic superfields, we can construct supersymmetric invariant actions and define the path integral. The harmonic analyticity condition can be regarded as the equation of motion derived from the following action for hypermultiplets
\begin{equation}
 S_{\mathrm{HM}}=-2 \int \rd u\, \rd^4 x_A \rd^2 \theta^{+} \rd^2 \bar{\theta}^{+} \operatorname{Tr}\left(\tilde{q}D^{++} q\right).
\end{equation}
When coupled to the gauge sector, this action becomes
\begin{equation}
\label{guage}
S_{\mathrm{HM}/\mathrm{SYM}}=-2 \int \rd u\,\rd^4 x_A \rd^2 \theta^{+} \rd^2 \bar{\theta}^{+} \operatorname{Tr}(\tilde{q} D^{++} q+\ri g\, \tilde{q}V^{++} q),
\end{equation}
where $g$ is the gauge coulpling constant and $V^{++}$ is the real Grassmann analytic super gauge potential ( \emph{i.e.} $\tilde{V}^{++}=V^{++}$ and $V^{++}$ satisfies \eqref{cons2}).  In the Wess-Zumino gauge, the potential $V^{++}$ includes the gauge field $A_{\mu}$, the complex scalar $\phi$, the doublet of Majorana gluinos $\lambda_{\alpha}^i$, $\bar{\lambda}^{\dot{\alpha}i}$ and the real auxiliary field $Y^{ij}$:
\begin{align}
    V_{\textrm{WZ}}^{++}(x_A, \theta^{+}, &\bar{\theta}^{+}, u) = -2\ri \theta^+ \sigma^{\mu}\bar{\theta}^+A_{\mu}(x_A)-\ri\sqrt{2}(\theta^+)^2\bar{\phi}(x_A)+\ri\sqrt{2}(\bar{\theta}^+)^2\phi(x_A)\label{V}\\    &+4(\bar{\theta}^+)^2\theta^{+\alpha}\lambda_{\alpha}^i(x_A)u_i^{-}-4(\theta^+)^2\bar{\theta}^{+}_{\dot{\alpha}}\lambda^{\dot{\alpha} i}(x_A)u_i^{-}+3(\theta^+)^2(\bar{\theta}^{+})^2Y^{ij}(x_A)u^-_i u^-_j.\nonumber
\end{align}
The gauge sector action can be constructed from either the chiral superfield strength $W(x_L,\theta)$ or the antichiral superfied strength $\overline{W}(x_R,\bar{\theta})$
\begin{align}
\label{eq:N2action}
 S_{\mathcal{N}=2~\text{SYM}}=\frac{1}{2 g^2} \int \rd^4 x_L \rd^4 \theta \operatorname{Tr} W^2=\frac{1}{2 g^2} \int \rd^4 x_R \rd^4 \bar{\theta} \operatorname{Tr} \overline{W}^2.
\end{align}
where $x_{L},x_R$ are chiral and antichiral coordinates:
\begin{align}
x_{L}^{\alpha,\dot{\alpha}}=x^{\alpha,\dot{\alpha}}-2\ri\theta^{i\alpha}\bar{\theta}^{\dot{\alpha}}_i,\qquad x_{R}^{\alpha,\dot{\alpha}}=x^{\alpha,\dot{\alpha}}-2\ri\bar{\theta}^{\dot{\alpha}}_i\theta^{i\alpha}\,.
\end{align}
The chiral superfield strength $W(x_L, \theta)$ can be expressed in terms of superpotential $V^{++}$ as 
\begin{align}
    W=\frac{\ri}{4} u_i^{+} u_j^{+} \bar{D}_{\dot{\alpha}}^i \bar{D}^{j \dot{\alpha}} \sum_{r=1}^{\infty} \int \rd u_1 \ldots \rd u_r \frac{(-\ri g)^r V^{++}\left(u_1\right) \ldots V^{++}\left(u_r\right)}{\left(u^{+} u_1^{+}\right)\left(u_1^{+} u_2^{+}\right) \ldots\left(u_r^{+} u^{+}\right)},
\end{align}
where $u^{+}u_a^{+}\equiv u^{+i}\epsilon_{ij}u_a^{+j}$. The on-shell expansion of $W(x_L,\theta)$ is given by
\begin{align}
\label{W}
W(x_L,\theta)=\omega(x_L)+\theta_i^{\alpha}\lambda_{\alpha}^{i}(x_L)+\epsilon^{ij}\theta^{\alpha}_i\theta^{\beta}_j
F_{\alpha\beta}(x_L),
\end{align}
which includes a complex scalar $\omega$, an  $SU(2)$ doublet of chiral fermions $\lambda_{\alpha}^i$ and the self-dual gauge field strength $F_{\alpha\beta}$.

\paragraph{Feynman rules} We now present the Feynman rules necessary for one- and two-loop calculations. Consider the following correlation function in $\mathcal{N}=2$ harmonic superspace
\begin{equation}
\label{Ggenrical}    
G_n=\langle\mathcal{Q}_1\mathcal{Q}_2\cdots\mathcal{Q}_n\rangle=\int D q D V \mathcal{Q}_1\mathcal{Q}_2\cdots\mathcal{Q}_n e^{\ri(S_{\mathcal{N}=2 \text { SYM }}+S_{\mathrm{HM}/\mathrm{SYM}})},
\end{equation}
where the operators $\mathcal{Q}_i$ are constructed from the hypermultiplets $q(x_A)$ and $\tilde{q}(x_A)$, and the actions are defined in \eqref{guage} and \eqref{eq:N2action}. After rescaling the gauge potential superfield as
\begin{equation}
    V^{++}\rightarrow \frac{1}{g}V^{++},
\end{equation}
the action $S_{\mathrm{HM}/\mathrm{SYM}}$ becomes independent of $g$, while $S_{\mathcal{N}=2 \text { SYM }}$ is simply proportional to $g^{-2}$. Therefore, differentiating \eqref{Ggenrical} with respect to $g^2$ is equivalent to inserting a chiral Lagrangian density into the path integral
\begin{equation}
    g^2 \frac{\partial}{\partial g^2} G_{n}=\int 
                      \rd^4x_{n+1}\rd^4 \theta_{n+1} \langle  \mathcal{Q}_1\mathcal{Q}_2\cdots\mathcal{Q}_n\mathcal{L}_{n+1}\rangle,\quad \mathcal{L}_{n+1}=\frac{1}{2 g^2}\operatorname{Tr}W(x_{n+1}, u_{n+1})^2\,.
\label{diff}
\end{equation}
In this way, the $\ell$-loop correction of $n$-point correlation function can be converted to an integration of an $(n+\ell)$-point correlation function at tree-level.\par

\begin{figure}[h!]
\centering
\includegraphics[width=0.6\linewidth]{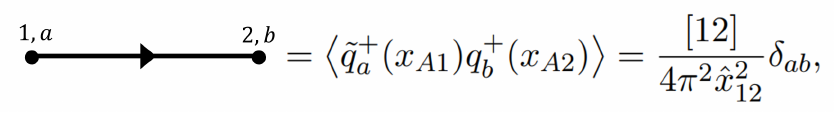}\label{HM}
\caption{Propagator of the hypermultiplet superfield.}
\label{fig:HMprop}
\end{figure}

\begin{figure}[h!]
\centering
 \includegraphics[width=0.65\linewidth]{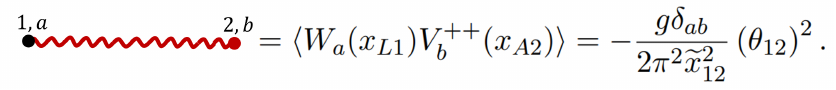}
 \caption{The gauge independent gluon propagator.}
 \label{WV}
\end{figure}

\begin{figure}[h!]
\centering
    \includegraphics[width=0.5\linewidth]{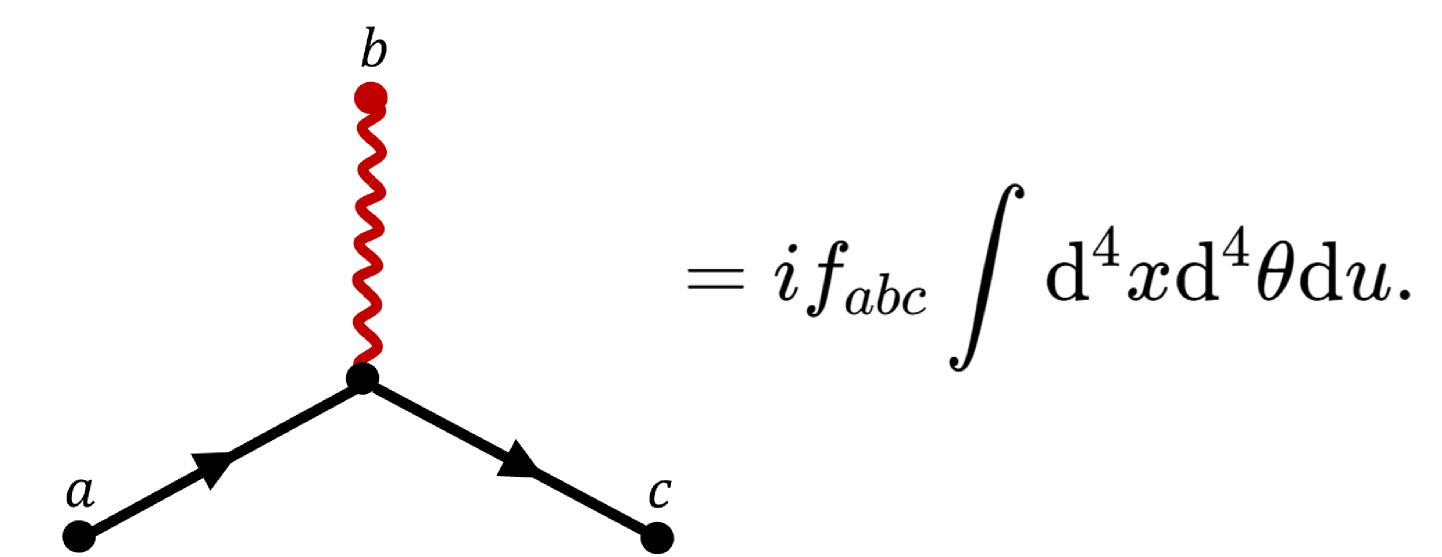}
   \caption{The vertex between a super gluon and matter fields.}
   \label{Vertex}
\end{figure}

\begin{figure}[h!]
\centering
    \includegraphics[width=0.7\linewidth]{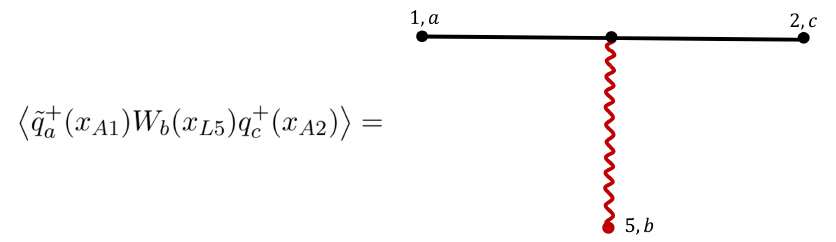}
    \caption{The diagram corresponding to the T-block $T_{125}$.}
    \label{Tblock}
\end{figure}

From the action $S_{\text{HM/SYM}}$, we derive the hypermultiplet propagator and the gauge-matter interaction vertex. The hypermultiplet propagator is shown in Figure~\ref{fig:HMprop},
where $a,b$ are color indices of the adjoint representation of the $SU(N)$  group and
\begin{align}
    \hat{x}_{12}^{\mu}\equiv x_{A1}^{\mu}- x_{A2}^{\mu}+\frac{2\ri}{[12]}\Big\{[1^{-}2]\theta_1^{+}\sigma^{\mu}\bar{\theta}_1^{+}+[2^{-}1]\theta_2^{+}\sigma^{\mu}\bar{\theta}_2^{+}+\theta_1^{+}\sigma^{\mu}\bar{\theta}_1^{+}+\theta_2^{+}\sigma^{\mu}\bar{\theta}_2^{+}
\Big\}\,.
\end{align}
We have introduced the following shorthand notations for the contraction of harmonic variables
\begin{align}
    [m~n]=-[n~m]=u_m^{+i}\epsilon_{ij}u_n^{+j},\quad [m^{-}~n]=u_m^{-i}\epsilon_{ij}u_n^{+j},\quad [m~n^{-}]=u_m^{+i}\epsilon_{ij}u_n^{-j}\,,\label{[ij]}
\end{align} 
which satisfy the following harmonic cyclic identity
\begin{align}
\label{eq:harmoniccI}
    [l~m][n~k]+[m~n][l~k]+[n~l][m~k]=0.
\end{align}
The gluon propagator depends on the gauge choice, but for our purposes, we only need the gauge independent propagator $ \langle W_a(x_{L1}) V_b^{++}(x_{A2})\rangle$, shown in Figure~\ref{WV}\,.
\\
The coordinate differences are defined as
\begin{align}  
\widetilde{x}_{12}^{\alpha\dot{\alpha}}\equiv x_{L1}^{\alpha\dot{\alpha}}-x_{A2}^{\alpha\dot{\alpha}}-4 \ri\theta^{- \alpha}_1\bar{\theta}^{+\dot{\alpha}}_2,\qquad \theta_{12}^{\alpha}\equiv\theta^{+\alpha}_1-\theta^{+\alpha}_2.
\end{align}
The interaction vertex between a gluon and two matter fields is given in Figure~\ref{Vertex}.
A fundamental building block for loop computations is the \textbf{T-block}, shown in Figure~\ref{Tblock}, which represents the tree-level three-point function of two G-analytic hypermultiplet superfields and one chiral superfield strength. Up to two-loop order, we need the expressions of the following T-blocks
\begin{align}
    T_{125}=\left\langle\tilde{q}_a(x_{A1}) W_b(x_{L5}) q_c(x_{A2})\right\rangle&=-\frac{2 \ri g^2 f_{a b c}}{(2 \pi)^4 x_{12}^2}\Big[ [21^{-}] \rho_1^2+[12^{-}] \rho_2^2-2\left(\rho_1 \rho_2\right)\Big]\label{Tblock1}\\
   T_{126}= \left\langle\tilde{q}_a(x_{A1}) W_b(x_{L6}) q_c(x_{A2})\right\rangle&=-\frac{2 \ri g^2 f_{a b c}}{(2 \pi)^4 x_{12}^2}\Big[ [21^{-}] \sigma_1^2+[12^{-}] \sigma_2^2-2\left(\sigma_1 \sigma_2\right)\Big]\label{Tblock2}
\end{align}
where $\rho_{r}$ and $\sigma_r$ are defined as
\begin{align}
\label{rhsi}
\rho_{r}=(\theta_{r}^{+}&-\theta_{5}^{i}u_{ri}^{+})x_{r5}^{-1},\quad
\sigma_{s}=(\theta_{s}^{+}-\theta_{6}^{i}u_{si}^{+})x_{s6}^{-1},\qquad r,s=1,\ldots,4\,.
\end{align}
Here, $x_{ij}^{-1}$ is short for
\begin{equation}
    (x_{ij}^{-1})^{\alpha \dot{\alpha}}=\frac{x_{ij}^{\alpha \dot{\alpha}}}{x_{ij}^2}\,.
\end{equation}

\section{Harmonic PCGG}
\label{sec:hPCGG}
In this section, we develop the harmonic PCGG method for computing loop corrections to giant graviton correlation functions.
\subsection{Set-up}
We study a class of four-point correlation functions in $\mathcal{N}=4$ SYM theory
\begin{align}
    G_{\{p,q\}}(x_1,\dots,x_4)=\langle \mathcal{D}(x_1)\mathcal{D}(x_2)\mathcal{O}_p(x_3)\mathcal{O}_q(x_4)\rangle\label{Gpp}
\end{align}
where 
\begin{align}
    \mathcal{D}(x_i)&\equiv\det(Y_i\cdot \phi(x_i)),\quad \mathcal{O}_p(x_j)\equiv \text{tr}((Y_j\cdot \phi)^p(x_j)) \quad i=1,2;\quad j=3,4.\nonumber
\end{align}
Here $Y_j\cdot\phi(x)\equiv\sum_{I=1}^6 Y^I\phi^I$ with $Y_{j}^{I}$ being a 6-dimensional null vector ($Y_j\cdot Y_j=0$). At weak coupling, $G_{\{p,q\}}$ admits a perturbative expansion
\begin{align}
    G_{\{p,q\}}=  G_{\{p,q\}}^{(0)}+  g^2 G_{\{p,q\}}^{(1)}+ g^4 G_{\{p,q\}}^{(2)}+ g^6 G_{\{p,q\}}^{(3)}+\dots,\label{expansion}
\end{align}
where $g^2=g^2_{\text{YM}} N_c/(16\pi^2)$ is the 'tHooft coupling constant. The goal of this work is to compute the two-loop correction $G_{\{p,q\}}^{(2)}$ for arbitrary lengths $p,q\ge 2$. We will first focus on the computation of $G_{\{p,p\}}^{(2)}$, the results for $G_{\{p,q\}}^{(2)}$ can be easily derived from $G_{\{p,p\}}^{(2)}$.

\paragraph{General structure} Superconformal symmetry constraints $G_{\{p,p\}}^{(k)}$ to the following form \cite{Eden:2000bk,Nirschl:2004pa}
\begin{align}
 G_{\{p,p\}}^{(k)}=\mathcal{R}_{\mathcal{N}=4}d_{12}^{N-p}\sum_{m+n+l=p-2}F^{(k)}_{[l,m,n]}(u,v)\mathcal{X}^l\mathcal{Y}^m\mathcal{Z}^n\,,\label{CN4}
\end{align}
where  $\mathcal{R}_{\mathcal{N}=4}$ is a universal factor
\begin{align}\label{R4}
\mathcal{R}_{\mathcal{N}=4}=&d_{12}^{2}d_{34}^{2}x_{12}^{2}x_{34}^{2}+d_{13}^{2}d_{24}^{2}x_{13}^{2}x_{24}^{2}+d_{14}^{2}d_{23}^{2}x_{14}^{2}x_{23}^{2}\\&+d_{12}d_{23}d_{34}d_{14}(x_{13}^2x_{24}^2-x_{12}^2x_{34}^2-x_{14}^2x_{23}^2)\nonumber\\&+d_{12}d_{13}d_{24}d_{34}(x_{14}^2x_{23}^2-x_{12}^2x_{34}^2-x_{13}^2x_{24}^2)\nonumber\\&+d_{13}d_{14}d_{23}d_{24}(x_{12}^2x_{34}^2-x_{14}^2x_{23}^2-x_{13}^2x_{24}^2)\nonumber.
\end{align}
The summation in \eqref{CN4} runs over all non-negative integers $m,n,l$ with the constraint $m+n+l=p-2$ and $\mathcal{X},\mathcal{Y},\mathcal{Z}$ are defined as
\begin{align}
\label{eq:calXYZ}
\mathcal{X}=d_{12}^2d_{34}^2,\qquad \mathcal{Y}=d_{13}^2d_{24}^2,\qquad \mathcal{Z}=d_{14}^2d_{23}^2,
\end{align}
with
\begin{align}
    d_{ij}\equiv\frac{Y_i\cdot Y_j}{(x_i-x_j)^2},\qquad x_{ij}^2\equiv(x_i-x_j)^2.
\end{align}
We will also denote $Y_i \cdot Y_j \equiv (y_i - y_j)^2 \equiv y_{ij}^2$. The coefficients $F_{[m,n,l]}^{(k)}$ are functions of the conformal cross ratios
\begin{align}
    u=z\bar{z}=\frac{x_{12}^2x_{34}^2}{x_{13}^2x_{24}^2},\qquad v=(1-z)(1-\bar{z})=\frac{x_{14}^2x_{23}^2}{x_{13}^2x_{24}^2}.\label{uv}
\end{align}
We also define the harmonic cross ratios for $y_i$
\begin{align}
    \sigma=\frac{y_{13}^2y_{24}^2}{y_{12}^2y_{34}^2},\qquad \tau=\frac{y_{14}^2y_{23}^2}{y_{12}^2y_{34}^2}.\label{sigmtau}
\end{align}

All dynamical information of the correlation function is encoded in the conformal invariant coefficient functions $F_{[l,m,n]}^{(k)}$, which will be the focus of our subsequent analysis.

\subsection{$\mathcal{N}=2$ reduction}
To apply $\mathcal{N}=2$ harmonic superspace techniques, we must rewrite quantities from $\mathcal{N}=4$ SYM to the $\mathcal{N}=2$ framework.  A vector multiplet in $\mathcal{N}=4$ SYM decomposes into a hypermultiplet (HM) and a gauge multiplet in $\mathcal{N}=2$ SYM. Correspondingly, the action splits as
\begin{equation}
S_{\mathcal{N}=4 ~\textrm{SYM}  }=S_{\textrm{HM}}+S_{\mathcal{N}=2~\textrm{SYM} }.
\end{equation}
The supermultiplet containing scalar fields $\phi^I$ $(I=1,\ldots,6)$ as the lowest components decomposes into the following four $\mathcal{N}=2$ supermultiplets
\begin{itemize}
\item A hypermultiplet $q=u_i^+\phi^i(x)+\ldots$
\item The complex conjugate $\tilde{q}=u_i^+\bar{\phi}_i(x)+\ldots$
\item Superfield strength $W=w(x)+\ldots$
\item Antichiral conjugate $\overline{W}=\bar{w}(x)+\ldots$\,.
\end{itemize}
The 6 scalar fields  $\phi^I$ in $\mathcal{N}=$ 4 SYM decompose into the $3+\bar{3}$ representation of $SU(3)$
\begin{equation}
    \phi^I\rightarrow \phi^A, \bar{\phi}_{A},\qquad A=1, 2, 3.
\end{equation}
The fields $\phi^A$ can be further projected from $SU(3)\rightarrow SU(2)\times U(1)$
\begin{equation}
    \phi^A\rightarrow\phi^i,\omega,\qquad i=1, 2,
\end{equation}
where $u_i^+\phi^i$ is the lowest component of hypermultiplet $q^+$ and $\omega$ is the lowest component of the superfield strength $W$. Similarly, the conjugate field $\bar{\phi}_{A}$ is decomposed into $\bar{\phi}_i$ and $\bar{\omega}$.\par

The propagators can be decomposed accordingly. First, we split the $SO(6)$ harmonic variable contractions into $SU(3)$ invariants
\begin{align}
\label{decompose1}
  & \langle \phi^I(x_1)\phi^J(x_2)\rangle=\frac{\delta_{IJ} Y_1^IY_2^J}{x_{12}^2}\\
   &\xrightarrow{SO(6)\rightarrow SU(3)}\langle\phi^A(x_1)\bar{\phi}^{B}(x_2)\rangle+\langle\bar{\phi}^A(x_1)\phi^{B}(x_2)\rangle=\frac{\left|1\bar{2}\right|+\left|\bar{1}2\right|}{x_{12}^2},\nonumber
\end{align}
where $\left|i\bar{j}\right|$, $\left|\bar{i}j\right|$ represents $\epsilon_{AB} i^A \overline{j^B}$, $\epsilon_{AB} \overline{i^A} j^B$, respectively. Here $i^{A}$ and $\bar{i}^A$ denote $SU(3)$ harmonic variables for $\phi^A$ and $\bar{\phi}^A$ respectively. Next, we project to $SU(2)$ by replacing $SU(3)$ contractions as $\left|i\bar{j}\right|\rightarrow[ij]$, $\left|\bar{i}j\right|\rightarrow-[ij]$ where $[ij]$ is defined in \eqref{[ij]} 
\begin{align}
\label{decompose2}
\langle&\phi^A(x_1)\bar{\phi}^B(x_2)\rangle=\frac{|1\bar{2}|}{x_{12}^2}\\
 &\xrightarrow{SU(3)\rightarrow SU(2)}\langle\phi^{i}(x_1)\phi^{j}(x_2)\rangle=\langle q^{+}(x_{A1})\tilde{q}^{+}(x_{A2})\rangle\Big|_{\theta_{1,2}=0}=\frac{[12]}{x_{A12}^2}\Big|_{\theta_{1,2}=0}.\nonumber
\end{align}
The single-trace operators $\mathcal{O}_p(x)$ and the determinant operators $\mathcal{D}(x)$ are constructed by the scalar field $Y\cdot\phi(x)$. In the $\mathcal{N}=2$ framework, we replace each scalar field by one of its $\mathcal{N}=2$ counterparts, this is called $\mathcal{N}=2$ projection. Since for each scalar field, we have several choices for the $\mathcal{N}=2$ projection, \emph{a priori} we need to take different projections in order to reconstruct the original $\mathcal{N}=4$ correlation function. However, as is shown in \cite{Arutyunov:2002fh}, there exists nice choices of the $\mathcal{N}=2$ projection which makes the reconstruction particularly simple. In the current case, the following projection turns out to be convenient
\begin{align}
\label{eq:N2reductionr}
G_{\{p,p\}}\to\mathcal{G}_{\{p,p\}}\equiv \langle \det \tilde{q}_1 \det q_2\, \text{tr} (\tilde{q}_3^{r} q_3^{p-r})\text{tr}(q_4^r \tilde{q}_4^{p-r})\rangle
\end{align}
In other words, we replace all scalar fields in the determinant operators $\mathcal{D}(x_1)$ and $\mathcal{D}(x_2)$ by $\tilde{q}_1\equiv\tilde{q}(x_{1A})$ and $q_2\equiv q(x_{2A})$ respectively. As for the single trace operators, we replace $r$ out of the $p$ scalar fields by $\tilde{q}_j\equiv \tilde{q}(x_{jA})$ while the rest $p-r$ scalar fields by $q_j\equiv q(x_{jA})$ ($j=3,4$). When $r=0$ or $p-r=0$, we call the projection a \emph{pure projection}. At the moment we do not fix the number $r$, in principle we need to compute the projected correlation functions for several different $r$ to recover the $\mathcal{N}=4$ SYM result. Nevertheless, as we will prove below, by using Grassmann analyticity and planarity, it turns out we only need to consider one fixed value of $r$ for each correlation function. For the coefficient $F_{[l,m,n]}$, we choose $r=m$ in \eqref{eq:N2reductionr} whose expansion involve $F_{[l,m,n]}$ and other coefficients $F_{[a,b,c]}$ with $|b-c|>|m-n|$. This allows us to recover $F_{[l,m,n]}$ recursively through a series of projections. Moreover, the projection \eqref{eq:N2reductionr} simplifies the contraction between the two determinant operators, leading to compact formulae for partially contracting the giant gravitons. After computing the correlation functions involving harmonic superfields, we can set all Grassmann variables $\theta_{1,2,3,4}$ to zero to extract correlation functions of the pure scalar operators.\par

The $\mathcal{N}=2$ correlation function $\mathcal{G}_{\{p,p\}}$ admits a perturbative expansion
\begin{align}
\mathcal{G}_{\{p,p\}}=\mathcal{G}_{\{p,p\}}^{(0)}+g^2\,\mathcal{G}_{\{p,p\}}^{(1)}+g^4\,\mathcal{G}_{\{p,p\}}^{(2)}+\cdots
\end{align}
Similar to the $\mathcal{N}=4$ case, the $\mathcal{N}=2$ superconformal symmetry constraints the loop corrections to the form \cite{Eden:2000bk,Eden:2000qp,Dolan:2000ut}
\begin{equation}
\label{eq:CN2}
    \mathcal{G}_{\{p,p\}}^{(k)}\Big|_{\theta_{1,2,3,4}=0}= \mathcal{R}_{\mathcal{N}=2}\Big(\frac{[12]}{x_{12}^2}\Big)^{N-p}\sum_{l+m+n=p-2}  X^l Y^m Z^n f_{[l,m,n]}^{(k)},
\end{equation}
where $\mathcal{R}_{\mathcal{N}=2}$ is the $\mathcal{N}=2$ universal factor given by
   \begin{align}     \mathcal{R}_{\mathcal{N}=2}=u\frac{[12]^2[34]^2}{x_{12}^4x_{24}^4}+\frac{[13]^2[42]^2}{x_{13}^4x_{24}^4}+\frac{[12][34][13][42]}{x_{12}^2x_{34}^2x_{13}^2x_{24}^2}(v-u-1).\label{RN2}
 \end{align}
The variables $X,Y$ and $Z$ are defined by
\begin{align}
    X=\frac{[12][34]}{x_{12}^2x_{34}^2},\qquad  Y=\frac{[13][24]}{x_{13}^2x_{24}^2},\qquad
    Z=\frac{[14][23]}{x_{14}^2x_{23}^2}\,,
\end{align}
which are the $\mathcal{N}=2$ counterparts of $\mathcal{X},\mathcal{Y}$ and $\mathcal{Z}$ defined in \eqref{eq:calXYZ}. The coefficients $F_{[m,n,l]}^{(k)}$ are linear combinations of the coefficients $f^{(k)}_{[m,n,l]}$. To establish their relations, we first expand the $\mathcal{N}=4$ propagators $d_{ij}^2$ using \eqref{decompose1} and \eqref{decompose2}, and then match the coefficients of harmonic factors $[ij]$ by comparing with \eqref{eq:CN2}. Explicit examples are provided in Appendix~\ref{app:reduce}. In summary, the $\mathcal{N}=2$ reduction translates the computation of $\mathcal{N}=4$ SYM correlators into determining $f_{[l,m,n]}^{(k)}$ within the $\mathcal{N}=2$ harmonic superspace framework.

\subsection{Lagrangian insertion}
Now we turn to the calculation of $\mathcal{G}_{\{p,p\}}^{(2)}$ in the $\mathcal{N}=2$ harmonic superspace. We use the Lagrangian insertion method to compute loop corrections. At two-loop order, we have
\begin{align}
\mathcal{G}_{\{p,p\}}^{(2)}= -\frac{1}{2 g^4}\int \rd^4x_{5}\rd^4 \theta_{5}\rd^4x_{6}\rd^4 \theta_{6} \langle \widetilde{\mathcal{D}}\mathcal{D}\mathcal{O}_p\mathcal{O}_p \mathcal{L}_5\mathcal{L}_6\rangle\,,
\end{align}
where the integrand is a six-point function at Born level. $\mathcal{N}=2$ superconformal symmetry fixes the integrand to be of the form \cite{Arutyunov:2002fh, Chicherin:2015edu}
\begin{align}
\label{factorized}
\langle \widetilde{\mathcal{D}}\mathcal{D}\mathcal{O}_p\mathcal{O}_p\mathcal{L}_5\mathcal{L}_6\rangle=\Theta_{5,6}\times \mathcal{A}(x_A, \theta^{+}, \bar{\theta}^{+}, u),
\end{align}
where $\Theta_{5,6}$ is the nilpotent six-point superconformal invariant \cite{Galperin:2001seg}. It carries a harmonic $U(1)$ charge of 2 at points 1 to 4 and an $R$-charge of 2 at points 5 and 6. The explicit form of $\Theta_{5,6}$ is determined by the superconformal symmetry and the Grassmann analyticity, given by \cite{Arutyunov:2003ad}:
\begin{align}
\Theta_{5,6}&=\quad\frac{\prod_{r=1}^4x_{r5}^2x_{r6}^2}{x_{56}^4}\frac{x_{12}^2x_{34}^2x_{13}^4x_{24}^4}{\mathcal{R}_{\mathcal{N}=2}}\\&\times\quad\left\{[12]^2[34]^2\tau_{14}\tau_{23}+[14]^2[23]^2\tau_{12}\tau_{34}+[12][23][34][41]\Big[\tau_{13}\tau_{24}-\tau_{12}\tau_{34}-\tau_{14}\tau_{23}\Big]\right\}\nonumber
\end{align}
where
\begin{align}
&\tau_{rs}=4(\rho_{r}\rho_{s})(\sigma_{r}\sigma_{s})+\rho_{r}^{2}\sigma_{s}^{2}+\rho_{s}^{2}\sigma_{r}^{2},
\end{align}
and we recall that $\rho_r$ and $\sigma_s$ have been defined in \eqref{rhsi}.

\paragraph{Superconformal frame} To obtain the correlation functions of the lowest components in the supermultiplet, we need to take $\theta_r=\bar{\theta}_r=0$ $(r=1,\ldots,4)$ in $\mathcal{G}_{\{p,p\}}^{(2)}$. In this case, $\Theta_{5,6}$ is simplified and given by
\begin{align}
\label{notation}
\Theta_{5,6}|_{\theta_{1,2,3,4}=0}=\theta_5^4\theta_6^4\frac{x_{12}^2x_{34}^2x_{13}^4x_{24}^4}{x_{56}^4}\mathcal{R}_{\mathcal{N}=2}.
\end{align}
Drawing on the experience from \cite{Arutyunov:2003ad,Arutyunov:2003ae, Chicherin:2015edu}, sometimes it is more convenient to perform the computations with  $\theta_{5,6}=\bar{\theta}_{5,6}=0$. In this case, $\Theta_{5,6}$ is even simpler
\begin{align}
\Theta|_{\theta_{5,6}=0}=\prod_{r=1}^4(\theta_r^+)^2.
\end{align}
In explicit computations, whether we take $\theta_r=0$, ($r=1,2,3,4$) or $\theta_{5,6}=0$ in the intermediate steps is a matter of choice and we call such a choice a superconformal frame. Due to the knowledge of the structure of $\Theta_{5,6}$, we can switch from one frame to the other straightforwardly.

\subsection{Partially contracted giant graviton}
The integrand \eqref{factorized} can be computed by the partially contracted giant graviton method. The idea is simple: we first perform a partial Wick contraction between the giant gravitons, leaving open legs to be contracted with single-trace operators and the Lagrangian density. After $\mathcal{N}=2$ reduction, the two giant gravitons become
\begin{align}
    &\det{(\tilde{q}_1)}=\frac{1}{N_c!}\epsilon_{\tilde{a}_1\cdots \tilde{a}_{N_c}}\epsilon^{\tilde{b}_1\cdots \tilde{b}_{N_c}}(\tilde{q}_1)^{\tilde{a}_1}{}_{\tilde{b}_1}\cdots(\tilde{q}_1)^{\tilde{a}_{N_c}}{}_{\tilde{b}_{N_c}},\\
    &\det{(q_2)}=\frac{1}{N_c!}\epsilon_{a_1\cdots a_{N_c}}\epsilon^{b_1\cdots b_{N_c}}({q_2})^{a_1}{}_{b_1}\cdots({q_2})^{a_{N_c}}{}_{b_{N_c}}.\nonumber
\end{align}
Performing partial contractions between the determinants, we obtain
\begin{align}
   \det(\tilde{q}_1)\det(q_2)\Big|_{\text{partial contractions}}=\sum_{\ell=0}^{N_c} \Big(\frac{[12]^2}{8\pi^2 x_{12}^2}\Big)^{N_c-\ell}\mathcal{G}_\ell(x_1,x_2),\label{DD}
\end{align}
where $\mathcal{G}_{\ell}(x_1, x_2)$ is the partially contracted giant graviton with $\ell$ pairs of un-contracted fields, given by
\begin{align}
\mathcal{G}_{\ell}(x_{1},x_{2})\equiv\frac{1}{(N_c!)^{2}}\left(\begin{array}{c}N_c\\N_c-\ell\end{array}\right)^{2}\epsilon_{\textcolor{blue}{\tilde{a}_{1}}\cdots\textcolor{blue}{\tilde{a}_{N_c-\ell}}\tilde{c}_{1}\cdots\tilde{c}_{\ell}}\epsilon^{\textcolor{blue}{\tilde{b}_{1}}\cdots\textcolor{blue}{\tilde{b}_{N_c-\ell}}\tilde{d}_{1}\cdots\tilde{d}_{\ell}}\epsilon_{\textcolor{blue}{a_{1}}\cdots \textcolor{blue}{a_{N_c-\ell}}c_{1}\cdots c_{\ell}}\epsilon^{\textcolor{blue}{b_{1}}\cdots\textcolor{blue}{b_{N_c-\ell}}d_{1}\cdots d_{\ell}}\label{PCGG}\\
\times\frac{\Big\langle(\tilde{q}_{1})^{\textcolor{blue}{\tilde{a}_{1}}}_{\textcolor{blue}{~~\tilde{b}_{1}}}\cdots(\tilde{q}_{1})^{\textcolor{blue}{\tilde{a}_{N_c-\ell}}}_{\textcolor{blue}{~~\tilde{b}_{N_c-\ell}}} (q_{2})^{\textcolor{blue}{\tilde{a}_{1}}}_{\textcolor{blue}{~~\tilde{b}_{1}}}\cdots(q_{2})^{\textcolor{blue}{\tilde{a}_{N_c-\ell}}}_{\textcolor{blue}{~~\tilde{b}_{N_c-\ell}}}\Big\rangle_{0}}{([12]/8\pi^{2}x_{12}^2))^{N_c-\ell}}(\tilde{q}_{1})^{\tilde{c}_{1}}_{~~\tilde{d}_{1}}\cdots(\tilde{q}_{1})^{\tilde{c}_{\ell}}_{~~\tilde{d}_{\ell}}(q_{2})^{c_{1}}_{~~d_{1}}\cdots(q_{2})^{c_{\ell}}_{~~d_{\ell}}.\nonumber
\end{align}
To highlight the patterns of index contractions, we colored the indices in (\ref{PCGG}). The prefactor in \eqref{DD} originates from the free propagators. 
\begin{align}
    \langle(\tilde{q}_1)^{a}_{~b}(q_2)^{c}_{~d}\rangle=\frac{[12]}{8\pi^2x_{12}^2}\delta^{a}_{d}\delta^{c}_{b}
\end{align}
By using the identities of the Levi-Civita tensor (see for example \cite{Berenstein:2003ah}), we can evaluate \eqref{PCGG} explicitly, yielding
\begin{align}
\label{PCGG1}
\mathcal{G}_\ell(x_1,x_2)=(N_c-\ell)!(-1)^{\ell} \sum_{\substack{k_1, \ldots, k_{\ell} \\ \sum_s s k_s=\ell}} \prod_{m=1}^{\ell} \frac{\left(-\operatorname{tr}\left[(\tilde{q}_1 q_2)^m\right]\right)^{k_m}}{m^{k_m} k_{m}!}.
\end{align}
Using the PCGG, the integrand \eqref{factorized} can be written as
\begin{align}
\langle \widetilde{\mathcal{D}}\mathcal{D}\mathcal{O}_p\mathcal{O}_p\mathcal{L}_5\mathcal{L}_6\rangle_0=
\sum_{\ell=0}^{N_c} \Big(\frac{ [12]^2}{8\pi^2 x_{12}^2}\Big)^{N_c-\ell}
\langle \mathcal{G}_{\ell}\mathcal{O}_p\mathcal{O}_p\mathcal{L}_5\mathcal{L}_6\rangle
\end{align}

\paragraph{Planarity} By large-$N_c$ counting, the dominant contribution in the large-$N_c$ limit comes from the single-trace part of the PCGG \eqref{PCGG1}, which can be viewed as a single-trace but non-local operator. The calculation of the integrand \eqref{factorized} can be done in three steps
\begin{enumerate}
\item List all Feynman diagrams that correspond to the planar Wick contractions between the PCGG and two single-trace operators, these diagrams will be called \emph{skeleton diagrams}.
\item Decorate each skeleton diagram in the previous step by inserting two Lagrangian densities while maintaining planarity, which correspond to quantum corrections of the skeleton diagrams.
\item Sum over all possible decorations from the previous step.
\end{enumerate}
The above procedure results in many Feynman diagrams to compute \emph{a priori}. However, as we will prove, planarity and harmonic analyticity imply that a lot of such diagrams actually do not contribute and many $f_{[l,m,n]}$ are vanishing.

\subsection{Harmonic analyticity}
In this subsection, we will prove that the coefficients $f_{[l,m,n]}^{(2)}$ and $F^{(2)}_{[l,m,n]}$ are vanishing for $|m-n|>1$. The proof is based on planarity and harmonic identification. Harmonic identification means we identify harmonic variables at two different points. Since $[jk]=-[kj]$, it follows that $[jj]=0$. Therefore, if we identify the harmonic variables $u_j^{\pm}$ and $u_k^{\pm}$, the resulting contraction of the harmonic variables is vanishing.  \par

Another important fact that we need is that the coefficients $f_{[l,m,n]}$ are independent of the harmonic variables $u_k^{\pm}$. This is a consequence of harmonic analyticity, which means that the correlator does not depend on $u_k^-$. This property comes from the fact that each operator of the correlator is annihilated by the harmonic derivative $u^+ \frac{\partial}{\partial u^-}$. We can choose the frame $\theta_5=\theta_6=0$, in which all harmonic charges are encoded in the factor $\prod_{i=1}^{4}\theta_i^{+2}$ and $X^lY^mZ^n$. This indicates that the coefficients $f_{[l,m,n]}$ have vanishing harmonic charge. If $f_{[l,m,n]}$ depend on harmonic variables, it must depend on both $u^+$ and $u^-$ to have zero total charge, but the dependence on $u^-$ violates the harmonic analyticity. Therefore $f_{[l,m,n]}$ are independent of $u_k^{\pm}$.\par
To prove $F^{(2)}_{[l,m,n]}=0$ for $|m-n|>1$, we start with the following lemma.
\paragraph{Lemma 1} The coefficients $f_{[0,m,n]}^{(2)}=0$ for $|m-n|>1$.\par
\noindent\emph{Proof}  The coefficient $f_{[0,m,n]}^{(2)}$ corresponds to the term $f_{[0,m,n]}^{(2)}Y^mZ^n$ in the correlation function. Since this term is independent of $X$, the harmonic identification $u_3^{\pm}=u_4^{\pm}$ does not affect $f_{[0,m,n]}^{(2)}$, as it is independent of the harmonic variables. \par

On the other hand, the identification $u_3^{\pm}=u_4^{\pm}$ forces any free propagator between $\tilde{q}_3$ and $q_4$ (and $q_3$ and $\tilde{q}_4$) to vanish, as these propagators are proportional to the factor $[34]=-[43]$. Consequently, only diagrams without free propagators connecting $\mathcal{O}_p(x_3)$ and $\mathcal{O}_p(x_4)$ can contribute to $f_{[0,m,n]}^{(2)}$. This restriction leaves three possible types of skeleton diagrams 
\begin{enumerate}
\item \textbf{No propagator} between $\mathcal{O}_p(x_3)$ and $\mathcal{O}_p(x_4)$; 
\item \textbf{One propagator} between $\mathcal{O}_p(x_3)$ and $\mathcal{O}_p(x_4)$. Attaching one of the Lagrangian densities to this propagator converts it into a T-block, which remains non-vanishing under the harmonic identification. 
\item \textbf{Two propagators} between $\mathcal{O}_p(x_3)$ and $\mathcal{O}_p(x_4)$. Attaching one Lagrangian density to each propagator converts both free propagators into T-blocks, preserving their contribution.\par
\end{enumerate}
The power of $Y^mZ^n$ can be determined explicitly for each type of skeleton diagrams. By planarity\footnote{This means we only consider the leading contribution of PCGG in the large $N$ limit, which is a single-trace operator of the form $\text{tr}(\tilde{q}_1q_2\cdots\tilde{q}_1q_2)$.}, it is not hard to see that the contributing skeleton diagrams all satisfy $|m-n|=0$ or $|m-n|=1$. Thus, all non-zero contributions to $f_{[0,m,n]}^{(2)}$ must obey $|m-n|\le 1$. It follows that $f_{[0,m,n]}^{(2)}=0$ for $|m-n|>1$.\par

The above result can be generalized to the following theorem.
\paragraph{Theorem 1} The coefficients $f_{[l,m,n]}^{(2)}=0$ for $|m-n|>1$.\\
\noindent\emph{Proof}  The coefficient $f_{[l,m,n]}^{(2)}$ corresponds to the term $f_{[l,m,n]}^{(2)}X^lY^mZ^n$ in the correlation function, with $l$ propagators between the operators $\mathcal{O}_p(x_3)$ and $\mathcal{O}_p(x_4)$. If we perform the harmonic identification $u_3^\pm=u_4^\pm$, the term vanishes because $X=0$. However, we can first divide it by the free propagator $\langle \tilde{q}_3q_4\rangle^l$ and then make the harmonic identification. This allows us to isolate the non-vanishing contributions for the coefficient $f_{[l,m,n]}^{(2)}$.
Following a similarly argument to the $f_{[0,m,n]}^{(2)}$ case, we conclude that the skeleton diagrams contributing to
\begin{align}
\lim_{u_3^\pm\to u_4^{\pm}}f_{[l,m,n]}^{(2)}\left(\frac{X}{\langle \tilde{q}_3q_4\rangle}\right)^lY^mZ^n
\end{align}
fall into three classes
\begin{enumerate}
\item \textbf{$l$ propagator} between $\mathcal{O}_p(x_3)$ and $\mathcal{O}_p(x_4)$; 
\item \textbf{$l+1$ propagator} between $\mathcal{O}_p(x_3)$ and $\mathcal{O}_p(x_4)$. Attaching one of the Lagrangian densities to one of the propagators (while respecting planarity) converts it into a T-block, which is non-vanishing after dividing by $\langle\tilde{q}_3q_4\rangle^l$ and the harmonic identification afterwards. 
\item \textbf{$l+2$ propagators} between $\mathcal{O}_p(x_3)$ and $\mathcal{O}_p(x_4)$. Attaching two Lagrangian densities to two propagators (while respecting planarity) converts them into T-blocks, preserving their contribution after dividing by $\langle\tilde{q}_3q_4\rangle^l$.\par
\end{enumerate}
Planarity enforces the constraint $|m-n|\le 1$ for all three types of diagrams. Therefore, if $|m-n|>1$, no planar diagram contributes and the corresponding $f_{[l,m,n]}^{(2)}=0$ in the large-$N$ limit.

\subsubsection{From $\mathcal{N}=2$ to $\mathcal{N}=4$}
Eventually we are interested in the coefficients $F_{[l,m,n]}^{(2)}$ in  $\mathcal{N}=4$ SYM, we will prove that these coefficients also satisfy the same constraint as their $\mathcal{N}=2$ counterparts $f_{[l,m,n]}^{(2)}$. We start with the following lemma.\\

\paragraph{Lemma 2} The coefficient $F_{[0,m,n]}^{(2)}=0$ if $|m-n|>1$. \\
\noindent\emph{Proof} For the correlation function $G_{\{p,p\}}^{(2)}$, the indices of $F^{(2)}_{[0,m,n]}$ are subject to the restriction $m+n=p-2$ with $m,n\ge 0$. For $p=2$, the only non-zero coefficient is $F_{[0,0,0]}^{(2)}$ (corresponding to $m=n=0$), so the lemma holds trivially. For $p=3$, we have $m+n=1$, yielding only two non-zero coefficients $F_{[0,1,0]}^{(2)}$ and $F_{[0,0,1]}^{(2)}$. Again, the lemma is satisfied since $|m-n|=1$. Henceforth, we assume $p\ge 4$. Without loss of generality, we restrict to $m\le n$, the proof for $m>n$ follows symmetrically.\par 

 For fixed $m$, we perform an $\mathcal{N}=2$ reduction \eqref{eq:N2reductionr} with $r=m$. At two-loop order, the expansion of this correlation function takes the form 
\begin{align}
\langle\det\tilde{q}_1\det q_2\text{tr}(\tilde{q}_3^{p-m}q_3^m)\text{tr}(q_4^{p-m}\tilde{q}_4^m)\rangle^{(2)}= \mathcal{R}_{\mathcal{N}=2}\Big(\frac{[12]}{x_{12}^2}\Big)^{N-p}
\left(Y^mZ^{p-m-2}f_{[0,m,p-m-2]}^{(2)}+\cdots\right)
\end{align}
where the ellipsis denotes lower order terms in $Y$. The reduction yields the relation
\begin{align}
\label{eq:recursionfFF}
f_{[0,m,p-m-2]}^{(2)}=a\,F_{[0,m,p-m-2]}^{(2)}+b\,F_{[0,m-1,p-m-1]}^{(2)}+c\,F_{[0,m-2,p-m]}^{(2)}
\end{align}
where $a,b,c$ depend on kinetic variables. We proceed the proof by induction on $m$.\par

As the starting point of the induction, we first prove the lemma for $m=0$ and $m=1$. Taking $m=0$, we have
\begin{align}
\label{eq:Ffp4}
F_{[0,0,p-2]}^{(2)}=f_{[0,0,p-2]}^{(2)}=0\,,\qquad p\ge 4
\end{align}
where the second equality follows from \textbf{Theorem 1}.\par

Now taking $m=1$, we have
\begin{align}
f^{(2)}_{[0,1,p-3]}=a\,F^{(2)}_{[0,1,p-3]}+b\,F^{(2)}_{[0,0,p-2]}=a\,F^{(2)}_{[0,1,p-3]}\,.
\end{align}
For $p=4,5$, the corresponding $F^{(2)}_{[0,1,p-3]}$ can be non-zero (consistent with $|m-n|\le 1$). For $p>5$, 
\begin{align}
F^{(2)}_{[0,1,p-3]}\propto f^{(2)}_{[0,1,p-3]}=0,
\end{align}
confirming the lemma for $m=1$.\par

Assume the lemma holds for $m-1$ and $m-2$ ($m\ge 3$), \emph{i.e.}
\begin{align}
&F^{(2)}_{[0,m-2,p-m]}=0\quad \text{if}\quad |p-2m+2|\ge 2\,,\\\nonumber
&F^{(2)}_{[0,m-1,p-m-1]}=0\quad \text{if}\quad |p-2m|\ge 2. 
\end{align}
We prove that $F^{(2)}_{[0,m,p-m-2]}=0$ when $|p-2m-2|\ge 2$. Focusing on $m<p-m-2$ (\emph{i.e.}, $m\le\tfrac{p}{2}-2$), the condition $|p-2m-2|\ge 2$ implies
\begin{align}
|p-2m|\ge4\quad \text{and}\quad |p-2m+2|\ge 6\,.
\end{align}
By the inductive hypothesis, $F^{(2)}_{[0,m-1,p-m-1]}$ and $F^{(2)}_{[0,m-2,p-m]}$ vanish. Substituting into the recursion relation \eqref{eq:recursionfFF}, we obtain
\begin{align}
F^{(2)}_{[0,m,p-m-2]}\propto f^{(2)}_{[0,m,p-m-2]}, 
\end{align}
where the last equality follows from \textbf{Lemma 1}. This completes the induction. Therefore we conclude that the lemma holds for all $m,n$ with $|m-n|>1$.


%

\paragraph{Theorem 2} The coefficients $F_{[l,m,n]}^{(2)}=0$ if $|m-n|>1$.\\
\noindent\emph{Proof}  For fixed $p$, the theorem is equivalent to the statement that $F^{(2)}_{[p-2-m-n,m,n]}=0$ for $|m-n|>1$. The strategy of the proof is similar to \textbf{Lemma 2}, but due to the mixing of harmonic channels in the $\mathcal{N}=2$ reduction, additional care is required. Without loss of generality, we assume $n-m>1$.\par

Consider the following projected correlation function
 \begin{align}
 \langle \det \tilde{q}_1 \det q_2 \text{tr} (\tilde{q}_3^{p-m} q_3^{m})\text{tr} (q_4^{p-m} \tilde{q}_4^{m})\rangle^{(2)}=\mathcal{R}_{\mathcal{N}=2}\sum_{a+b+c=p-2} &\Big(\frac{[12]}{x_{12}^2}\Big)^{N-a-b-c-2} X^a Y^b Z^c f_{[a,b,c]}^{(2)}.
 \label{eq:projectionEg}
 \end{align}
Using the harmonic cyclic identity \eqref{eq:harmoniccI}, $Y$ can be rewritten in terms of $X$ and $Z$, 
\begin{align}
Y=X\,\frac{x_{12}^2x_{34}^2}{x_{13}^2x_{24}^2}+Z\,\frac{x_{14}^2x_{23}^2}{x_{13}^2x_{24}^2}
\end{align}
Substituting this into \eqref{eq:projectionEg} yields
\begin{align}
\label{eq:specialProject}
  &\langle \det \tilde{q}_1 \det q_2 \text{tr} (\tilde{q}_3^{p-m} q_3^{m})\text{tr} (q_4^{p-m} \tilde{q}_4^{m})\rangle^{(2)}\\\nonumber
         =&\,\mathcal{R}_{\mathcal{N}=2}\sum_{a,b,c} \Big(\frac{[12]}{x_{12}^2}\Big)^{N-a-b-c-2}
     X^a \left(X\frac{x_{12}^2x_{34}^2}{x_{13}^2x_{24}^2}+Z\frac{x_{14}^2x_{23}^2}{x_{13}^2x_{24}^2}\right)^b Z^c f_{[a,b,c]}^{(2)}.
\end{align}
Focusing on the harmonic channel $X^{p-2-m-n}Z^{m+n+2}$, the contributing $f_{[a,b,c]}^{(2)}$ must satisfy
\begin{equation}
    \begin{aligned}
        m+n\leq b+c
    \end{aligned}
\end{equation}
where $b\leq m$ (due to the limit on the number of contractions between $q_2$ and $\tilde{q}_4$). Combining these inequalities gives 
\begin{equation}
    c-b\geq n-m> 1.
\end{equation}
By \textbf{Theorem 1}, $f_{[a,b,c]}^{(2)}=0$ for $c-b>1$.  On the other hand, the coefficients $F_{[a,b,c]}^{(2)}$ contribute to the channel $X^{p-2-m-n}Z^{m+n+2}$ after reduction if 
\begin{equation}
    b+c\geq m+n\,,
\end{equation}
which implies
\begin{equation}
    c-b\geq n-m.
\end{equation}
However, \textbf{Theorem 1} already enforces $f^{(2)}_{[a,b,c]}=0$ for $c-b>1$, so for $n-m>1$, the relevant $f^{(2)}_{[a,b,c]}$ vanish.\par

The vanishing of $F^{(2)}_{[p-2-m-n,m,n]}$ follows inductively. For $F^{(2)}_{[0,m,p-m-2]}$, the result holds by \textbf{Lemma 2}. Assuming $F_{[a,b,c]}^{(2)}=0$ for all $c-b>n-m$, the harmonic expansion reduces to
\begin{equation}
    F_{[p-2-m-n,m,n]}^{(2)}+\sum_{c-b>n-m}F_{[a,b,c]}^{(2)}=\sum_{c-b\geq n-m}f_{[a,b,c]}=0.
\end{equation}
By the induction hypothesis
\begin{equation}
    \sum_{c-b>n-m}F_{[a,b,c]}^{(2)}=0,
\end{equation}
and thus
\begin{equation}
    F_{[p-2-m-n,m,n]}^{(2)}=0.
\end{equation}
By induction, $F_{[l,m,n]}^{(2)}=0$ for all $|m-n|>1$, completing the proof.

\subsection{Lightcone OPE relations} \label{light-cone}
Another simplification comes from the lightcone OPE relations, originally derived for correlation functions of single-trace operators. These relations exploits the OPE structure of two single-trace half-BPS operators and is applicable to any correlation functions involving at least two such operators. We refer to \cite{Chicherin:2015edu} for a detailed derivation. In the current context, the lightcone OPE relations imply
\begin{equation}
\label{eq:sumFF}
 \sum_{k=-n}^m\left.\left(x_{13}^2 x_{24}^2\right)^{k+n}\left(x_{14}^2 x_{23}^2\right)^{m-n-k}\left(F_{\left[l+1,m-k,n+k\right]}^{(2)}-F_{\left[l,m-k,n+k\right]}^{(2)}\right)\right|_{x_{34}^2=0}=0,
\end{equation}
where we have taken the lightcone limit $x_{34}^2\to 0$. By \textbf{Theorem 2}, $F_{[l,m,n]}^{(2)}$ vanishes when $|m-n|>1$, therefore the summation on the left hand side of \eqref{eq:sumFF} actually reduces to one term, simplifying the relation to
\begin{equation}
\label{lcr}
(F_{\left[a+1,b,c\right]}^{(2)}-F_{\left[a,b,c\right]}^{(2)})|_{x_{34}^2=0}=0,\qquad |b-c|\le 1\,.
\end{equation}
This recursive structure reduces the problem of of computing all $F_{[a,b,c]}^{(2)}$ in the lightcone limit to determining only $F_{[0,b,c]}^{(2)}$ with $|b-c|\le 1$.\par

\paragraph{The missing terms} Working in the lightcone limit excludes terms that vanish when $x_{34}^2=0$. These terms can be systematically computed. We will present their detailed evaluation in Sections~\ref{sec:one} and~\ref{sec:two}, where we derive the full one- and two-loop results.

\subsubsection{Harmonic identification}
In the preceding subsections, we demonstrated for each $p$, we only need to compute very few coefficients $F_{[0,b,c]}^{(2)}$. These coefficients can be evaluated using Feynman diagrams, which can be further simplified through harmonic identifications. Some diagrams vanish automatically upon identifications, while for the non-vanishing ones, the identifications simplify the expressions of the key building blocks, such as the T-blocks, making computations more tractable.\par

As an illustrative example, consider the case $p=4$. Using the Lagrangian insertion method, the integrand takes the form
\begin{align}
\label{eq:exampleExpand}
\langle\tilde{\mathcal{D}}\mathcal{D}\mathcal{O}_4\mathcal{O}_4\mathcal{L}_5\mathcal{L}_6\rangle_{\theta_{5,6}=0}=&\prod_{r=1}^{4}(\theta^{+}_r)^2\Big(\frac{[12]}{x_{12}^2}\Big)^{N-4}\Big\{\frac{[12]^2[34]^2 }{x_{12}^4x_{34}^4}A_{[2,0,0]}^{(2)}
+\frac{[12][34][13][24]}{x_{12}^2x_{34}^2x_{13}^2x_{24}^2} A_{[1,1,0]}^{(2)}\nonumber\\&+\frac{[12][34] [14][23]}{x_{12}^2x_{34}^2x_{14}^2x_{23}^2} A_{[1,0,1]}^{(2)}
    + \frac{[13][24][14][23]}{x_{13}^2x_{24}^2x_{14}^2x_{23}^2}A_{[0,1,1]}^{(2)}\Big\}.
\end{align}
where $A_{[m,n,l]}^{(2)}$ is the integrand of $f_{[m,n,l]}^{(2)}$, \emph{i.e.}
\begin{align}
    f_{[m,n,l]}^{(2)}=\int d^4x_5\int d^4\theta_5\int d^4x_6\int d^4\theta_6 A_{[m,n,l]}^{(2)}\Big|_{\theta_r=0},\quad r=(1, 
  \cdots, 4).
\end{align}
Here, we omit coefficients $A_{[l,m,n]}^{(2)}$ with $|m-n|>1$ in \eqref{eq:exampleExpand}, as they correspond to vanishing contributions.
For brevity, we also suppress the spacetime dependence $x_{ij}^2$ in the free propagators henceforth. The four relevant coefficients are $A_{[2,0,0]}^{(2)}$, $A_{[1,1,0]}^{(2)}$, $A_{[1,0,1]}^{(2)}$ and $A_{[0,1,1]}^{(2)}$. In the lightcone limit, $A_{[2,0,0]}$ reduces to $A_{[0,0,0]}$  which already showed up for $p=2$. Likewise, $A_{[1,1,0]}^{(2)}$ and $A_{[1,0,1]}^{(2)}$ reduce to $A_{[0,1,0]}^{(2)}$ and $A_{[0,0,1]}^{(2)}$, respectively, via lightcone OPE relations, and already appeared for $p=3$. Thus, for $p=4$, the only new coefficient is $A_{[0,1,1]}^{(2)}$.

To compute $A_{[0,1,1]}^{(2)}$,  we first make the harmonic identification $u_3^{\pm}=u_4^{\pm}$\footnote{The identification $u_j^\pm=u_k^{\pm}$ will be written as $u_j=u_k$ in what follows.}, yielding
\begin{align}
\langle\tilde{\mathcal{D}}\mathcal{D}\mathcal{O}_4\mathcal{O}_4\mathcal{L}_5\mathcal{L}_6 \rangle_{\theta_{5,6}=0}\Big|_{u_3\equiv u_4}=\prod_{r=1}^{4}(\theta^{+}_r)^2[12]^{N-4}[13]^2[23]^2A_{[0,1,1]}^{(2)}.
\end{align}
Next, we factor out $[12]^{N-4}$
\begin{align}
\frac{\langle\tilde{\mathcal{D}}\mathcal{D}\mathcal{O}_4\mathcal{O}_4\mathcal{L}_5\mathcal{L}_6 \rangle_{\theta_{5,6}=0}\Big|_{u_3\equiv u_4}}{[12]^{N-4}} =\prod_{r=1}^{4}(\theta^{+}_r)^2 [13]^2[23]^2A_{[0,1,1]}^{(2)}\,,
\end{align}
and then identify $u_1=u_2$, giving
\begin{align}
\frac{\langle\tilde{\mathcal{D}}\mathcal{D}\mathcal{O}_4\mathcal{O}_4\mathcal{L}_5\mathcal{L}_6 \rangle_{\theta_{5,6}=0} \Big|_{u_3\equiv u_4} }{[12]^{N-4}} \Big|_{u_1\equiv u_2}= \prod_{r=1}^{4}(\theta^{+}_r)^2 [13]^4 A_{[0,1,1]}^{(2)}.   
\end{align}
Finally, we factor out $[13]^4$ and impose $u_1=u_3$ :
\begin{align}\label{harid}
    \Bigg(\frac{1}{[13]^4}\frac{\langle\tilde{\mathcal{D}}\mathcal{D}\mathcal{O}_4\mathcal{O}_4\mathcal{L}_5\mathcal{L}_6 \rangle_{\theta_{5,6}=0}\Big|_{u_3\equiv u_4}}{[12]^{N-4}} \Big|_{u_1=u_2}\Bigg)\Bigg|_{u_1\equiv u_3}=\prod_{r=1}^{4}(\theta^{+}_r)^2A_{[0,1,1]}^{(2)}.
\end{align}
The harmonic identification \eqref{harid} implies the following requirements for the contributing Feynman diagrams: they do not have free propagators between $\mathcal{O}_4(x_3, u_3)$ and $\mathcal{O}_4(x_4, u_4)$ and contain at most four propagators between $\tilde{\mathcal{D}}(x_1,u_1)$ and $\mathcal{O}_4(x_3, u_3)$\footnote{Note that some $[14]$ will become $[13]$ upon the identification $u_4=u_3$.}. Such diagrams will be listed in Section~\ref{sec:two}.

\section{One-loop computations}
\label{sec:one}
In this section, we revisit the one-loop computation of $G_{\{p,p\}}^{(1)}$, originally performed in \cite{Jiang:2019xdz} using the PCGG method. Here, we employ the harmonic PCGG method, demonstrating its greater efficiency. Remarkably, after harmonic identification, essentially we only need to compute one Feynman diagram, compared to the 31 distinct Feynman diagrams involving various fields required in the usual PCGG method.\par

We consider the $\mathcal{N}=2$ reduced correlation function
\begin{align}
\mathcal{G}^{(1)}_{\{p,p\}}=\langle\det\tilde{q}_1\det q_2\text{tr}(\tilde{q}_3^{p-m}q_3^m)\text{tr}(q_4^{p-m}\tilde{q}_4^m)\rangle^{(1)}
\label{eq:oneloopProject}
\end{align}
where the superscript means that we compute that correlation function at one-loop order. The lightcone OPE relation implies a recursive structure among correlation functions involving single-trace operators of different lengths. As we will see, the recursive structure guarantees that it is sufficient to compute only two coefficients $F^{(1)}_{[0,\lfloor \frac{p-2}{2}\rfloor,\lfloor \frac{p-1}{2}\rfloor]}$ and $F^{(1)}_{[0,\lfloor \frac{p-1}{2}\rfloor,\lfloor \frac{p-2}{2}\rfloor]}$ for each length $p$. For this purpose, it is convenient to choose $m=\lfloor \frac{p-2}{2}\rfloor$ in \eqref{eq:oneloopProject}. Using the Lagrangian insertion approach, the integrand is given by
\begin{align}
\langle\tilde{\mathcal{D}}\mathcal{D}\mathcal{O}_p\mathcal{O}_p\mathcal{L}\rangle=\Theta_5
\left(\frac{[12]}{x_{12}^2}\right)^{N-p}\sum_{l+m+n=p-2}X^lY^mZ^n A_{[l,m,n]}^{(1)},
\end{align}
where $\Theta_5$ is the superconformal nilpotent invariant with the following property
 \begin{align} 
 \label{eq:Theta5F1}    
 \Theta|_{\theta_r^+=0}&=\theta_5^4\mathcal{R}_{\mathcal{N}=2}x_{12}^2x_{34}^2x_{13}^2x_{24}^2\,,\quad r=1,\cdots,4\,.
 \end{align}
In what follows, we perform the computations in the frame $\theta_r^+=0$ where $\Theta_5$ takes the form in \eqref{eq:Theta5F1}, note that this choice of frame is different from the two-loop case. As discussed in the previous section, the harmonic analyticity implies that the coefficients $A_{[l,m,n]}^{(1)}$ do not depend on harmonic variables, which allows us to apply harmonic identification. We will perform the explicit computation for the first few values of $p$, from which we can see a clear pattern and it is straightforward to generalize to generic $p$.
\textbf{Theorem 1} and \textbf{Theorem 2} proven in the previous section also hold at one-loop, thus we have $A^{(1)}_{[l,m,n]}=0$ for $|m-n|>1$.
\paragraph{$p=2$ case} For $p=2$, we only need to compute one coefficient $A^{(1)}_{[0,0,0]}$, \emph{i.e.} 
\begin{align}
\langle\tilde{\mathcal{D}}\mathcal{D}\mathcal{O}_2\mathcal{O}_2\mathcal{L}\rangle=\theta_5^4 \mathcal{R}_{\mathcal{N}=2}
\left(\frac{[12]}{x_{12}^2}\right)^{N-2}A_{[0,0,0]}^{(1)}\,.
\end{align}
{We first perform the identification $u_3=u_4$, yielding
\begin{equation}
 \langle\tilde{\mathcal{D}}\mathcal{D}\mathcal{O}_2\mathcal{O}_2\mathcal{L}\rangle|_{u_3=u_4}=  \theta_5^4[13]^2[23]^2 x_{12}^2x_{34}^2 \left(\frac{[12]}{x_{12}^2}\right)^{N-2}A_{[0,0,0]}^{(1)}.
\end{equation}
Then factorize out the factors $[12]^{N-2}$ and $[13]^4$ consecutively and perform the identifications $u_1=u_2$ and $u_1=u_3$, leading to
\begin{align}
\frac{1}{[13]^4}\left.\left(\frac{1}{[12]^{N-2}}\left.\left( \langle\tilde{\mathcal{D}}\mathcal{D}\mathcal{O}_2\mathcal{O}_2\mathcal{L}\rangle|_{u_3=u_4}\right)\right|_{u_1=u_2} \right)\right|_{u_1=u_3}=\theta_5^4 x_{12}^2x_{34}^2 A_{[0,0,0]}^{(1)}
\end{align}
After identifying $u_1=u_2$ and $u_3=u_4$, {the only non-vanishing diagrams are those containing $N-2$ propagators between 1 and 2 and no free propagators between 3 and 4 after the Lagrangian insertion.} There is only one such Feynman diagram, given in Figure~\ref{dd22one}. The corresponding Feynman rules simplify due to the harmonic identification, leading to
\begin{equation}
    (\rho_1-\rho_2)^2(\rho_3-\rho_4)^2|_{u_1=u_2,u_3=u_4}=[13]^2\frac{x_{12}^2x_{34}^2}{\prod_{r=1}^4 x_{r5}^2}.
\end{equation}
Therefore
\begin{equation}
    F_{[0,0,0]}^{(1)}=\frac{1}{x_{13}^2x_{24}^2}\int \rd^4 x_5 \frac{x_{13}^2x_{24}^2}{\prod_{r=1}^4 x_{r5}^2}=\frac{1}{x_{13}^2x_{24}^2}F^{(1)}(z,\bar{z})\,,
\end{equation}
where $ F^{(1)}(z,\bar{z}) $ is the one-loop conformal integral
\begin{align}
    F^{(1)}(z,\bar{z}) &= \frac{x_{13}^2 x_{24}^2}{\pi^2} \int \frac{\mathrm{d}^4x_5}{x_{15}^2 x_{25}^2 x_{35}^2 x_{45}^2} \nonumber \\
    &= \frac{1}{z - \bar{z}} \left( 2\mathrm{Li}_2(z) - 2\mathrm{Li}_2(\bar{z}) + \ln(z\bar{z}) \ln\frac{1 - z}{1 - \bar{z}} \right).
\end{align}
\begin{figure}[h!]
    \centering
    \includegraphics[width=0.25\linewidth]{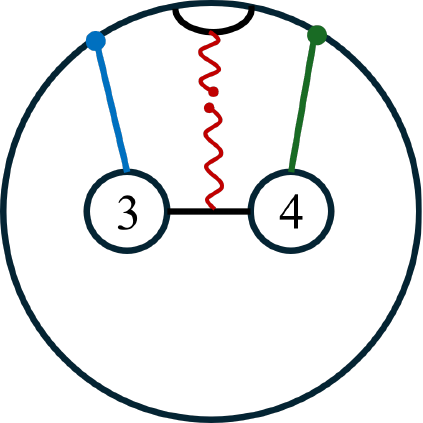}
    \caption{The only non-vanishing planar one-loop diagram for $\langle\tilde{\mathcal{D}}\mathcal{D}\mathcal{O}_2\mathcal{O}_2\mathcal{L}\rangle$ after harmonic identification.}
    \label{dd22one}
\end{figure}
}
\paragraph{$p=3$ case} For $p=3$, in $\mathcal{N}=4$ SYM, we have 
\begin{equation}
\langle\mathcal{D}\mathcal{D}\mathcal{O}_3\mathcal{O}_3\rangle=\mathcal{R}_{\mathcal{N}=4}
d_{12}^{N-3}(\mathcal{X}\, F_{[1,0,0]}^{(1)}+\mathcal{Y}\,F^{(1)}_{[0,1,0]}+\mathcal{Z}\,F^{(1)}_{[0,0,1]}). 
\label{eq:DD33N4}
\end{equation}
We choose $m=0$ in \eqref{eq:oneloopProject}, which is a pure projection. After $\mathcal{N}=2$ reduction, the one-loop integrand reads
\begin{align}
\langle\tilde{\mathcal{D}}\mathcal{D}\mathcal{O}_3\mathcal{O}_3\mathcal{L}\rangle=\theta_5^4 \mathcal{R}_{\mathcal{N}=2}
\left(\frac{[12]}{x_{12}^2}\right)^{N-3}(X\, A_{[1,0,0]}^{(1)}+Z\,A^{(1)}_{[0,0,1]}),
\label{eq:DD33N2}
\end{align}
where $A_{[l,m,n]}^{(1)}$ corresponds to $F_{[l,m,n]}^{(1)}$ under the pure projection (see Appendix~\ref{app:reduce} for more details).
In the lightcone limit we have $A^{(1)}_{[1,0,0]}=A^{(1)}_{[0,0,0]}$ and we only have two new coefficients $A^{(1)}_{[0,1,0]}$ and $A^{(1)}_{[0,0,1]}$. Comparing \eqref{eq:DD33N4} and \eqref{eq:DD33N2}, it seems that $A^{(1)}_{[0,1,0]}$ is missing. However, $A^{(1)}_{[0,1,0]}$ is actually the same as the $A^{(1)}_{[0,0,1]}$ after exchanging $3$ and $4$ due to the symmetry between $\mathcal{O}_3(x_3,y_3)$ and $\mathcal{O}_3(x_4,y_4)$. {The skeleton diagram is given by the left panel\footnote{In the Feynman diagrams of this paper, we use blue propagators to represent $d_{13}$ and $d_{14}$, and green propagators to represent $d_{23}$ and $d_{24}$. This color coding helps to highlight the planarity of the PCGG.} of Figure~\ref{dd33one}, after inserting the Lagrangian density, we obtain the corresponding one-loop Feynman diagram. To compute the coefficient $A^{(1)}_{[0,0,1]}$, we again identify $u_3=u_4$. Notice that the only skeleton diagram contain two propagators between $3$ and $4$. This can be seen as follows. Under pure projection, there are neither propagators between $1$ and $3$, nor between $2$ and $4$. On the other hand, planarity requires that, in the skeleton, if there are more than one propagators between operator 3 and the PCGG, a propagator between $1$ and $3$ must be adjacent to\footnote{Due to the form of the PCGG $\text{tr}(\tilde{q}_1q_2\tilde{q}_1 q_2\cdots)$.} a propagator between $2$ and $3$. This implies that there is exactly one propagator between $2$ and $3$, as well as one between $1$ and $4$. The remaining fields $\tilde{q}_3$ and $q_4$ contract with each other, resulting in two propagators connecting $3$ and $4$. At one-loop, the integrand only involves one inserted Lagrangian. Thus, there remains at least one free propagator carrying the factor $[34]$ after Lagrangian insertion, which vanishes upon harmonic identification. As a result, $A^{(1)}_{[0,0,1]}$ and $F^{(1)}_{[0,0,1]}$ are both vanishing. Due to the symmetry between $\mathcal{O}_3(x_3,y_3)$ and $\mathcal{O}_3(x_4,y_4)$, the coefficient $F^{(1)}_{[0,1,0]}$ is also vanishing.

Finally we come to the terms that vanish in the lightcone limit. The recursive procedure can not determine these terms. However, at one-loop level, there are actually no such terms. This can be proven by counting the conformal weight at each  point.} {If such a term exists, it must be proportional to $x_{34}^2$. By construction, the correlation function has conformal weight $p$ at each point.} For the coefficient $F_{[l,m,n]}^{(1)}$, the prefactor $\mathcal{R}_{\mathcal{N}=4}$ and $\mathcal{X}^{p-2-m-n}\mathcal{Y}^m \mathcal{Z}^n$ carry conformal weight $p-1$ at $3$ and $4$. Therefore, $F_{[l,m,n]}^{(1)}$ can only carry conformal weight one at each point. At one-loop order, each diagram has two T-blocks coming from the Lagrangian insertion. The product of these two T-blocks contain a term of the form $\prod_{r=1}^4 1/x_{r5}^2$, which already exhaust the conformal weights. As a result, the numerators of $F_{[l,m,n]}^{(1)}$ must be a constant and cannot contain the factor $x_{34}^2$. Therefore such missing terms do not exist.
\begin{figure}[h!]
    \centering  \includegraphics[width=0.7\linewidth]{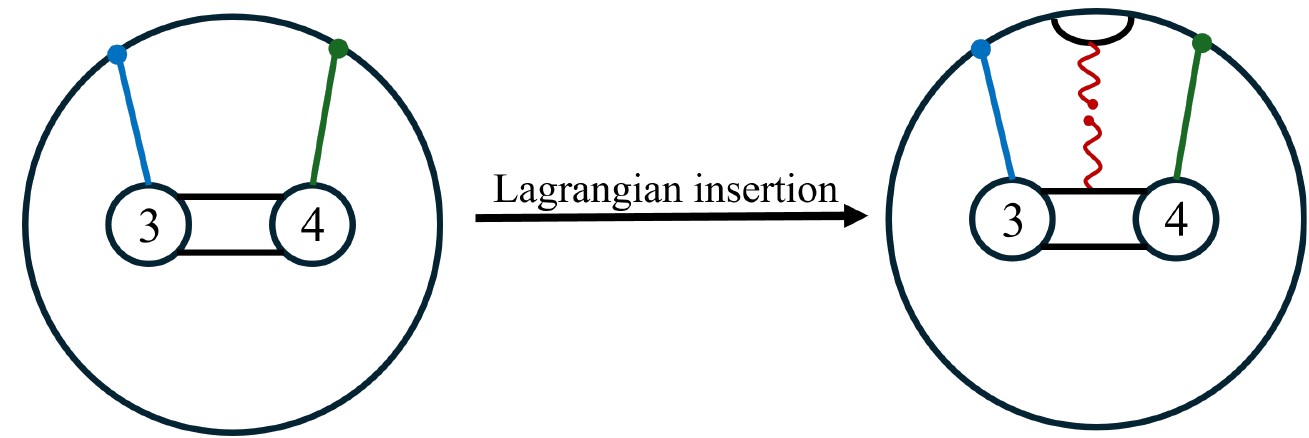}
    \caption{One-loop $\langle\tilde{\mathcal{D}}\mathcal{D}\mathcal{O}_3\mathcal{O}_3\mathcal{L}\rangle$ survival diagram }
    \label{dd33one}
\end{figure}

\begin{figure}[h!]
    \centering
    \includegraphics[width=0.25\linewidth]{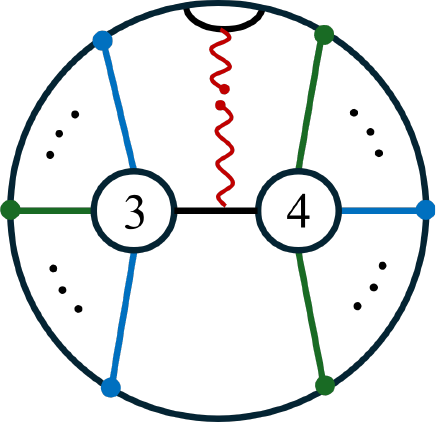}
    \caption{One-loop diagram that contributes to $F_{[0,\frac{p-2}{2},\frac{p-2}{2}]}^{(1)}$.}
    \label{dd44one}
\end{figure}

\paragraph{$p=4$ case} For $p=4$, the $\mathcal{N}=4$ correlator is given by\footnote{$F^{(1)}_{[0,0,1]}=F^{(1)}_{[0,1,0]}=0$ implies that $F^{(1)}_{[1,0,1]}=F^{(1)}_{[1,1,0]}=0$ by the lightcone OPE relation, therefore we have omitted the corresponding terms.} 
\begin{equation}
     \langle\mathcal{D}\mathcal{D}\mathcal{O}_4\mathcal{O}_4\rangle=\mathcal{R}_{\mathcal{N}=4}
d_{12}^{N-4}(\mathcal{X}^2\, F_{[2,0,0]}^{(1)}+\mathcal{Y}\mathcal{Z}\,F^{(1)}_{[0,1,1]}). 
\end{equation}
Since $F^{(1)}_{[2,0,0]}=F^{(1)}_{[0,0,0]}$ due to lightcone relation, which has been computed in the $p=2$ case, $F^{(1)}_{[0,1,1]}$ is the only new coefficient to be computed for $p=4$. We choose $m=1$ and the $\mathcal{N}=2$ correlator is 
\begin{equation}
\langle\tilde{\mathcal{D}}\mathcal{D}\mathcal{O}_3\mathcal{O}_3\mathcal{L}\rangle=\theta_5^4 \mathcal{R}_{\mathcal{N}=2}
\left(\frac{[12]}{x_{12}^2}\right)^{N-3}(X^2\, A_{[2,0,0]}^{(1)}+XZ\,A^{(1)}_{[1,0,1]}+XY\,A^{(1)}_{[1,1,0]}+YZ\,A^{(1)}_{[0,1,1]}).
\end{equation}
Upon the reduction procedure, $A^{(1)}_{[0,1,1]}$ corresponds to $F^{(1)}_{[0,1,1]}$. Similar to the previous case, we can firstly identify $u_3=u_4$ and find that there is only one non-vanishing diagram, given in Figure~\ref{dd44one}.
It is easy to see that the interaction part of this diagram is the same as Figure~\ref{dd22one}, which we have computed before. Therefore we can directly plug in the result and find that
\begin{equation}
    F^{(1)}_{[0,1,1]}=\frac{1}{x_{13}^2x_{24}^2}F^{(1)}(z,\bar{z}).
\end{equation}
By the same argument as for the $p=3$ case, the terms that vanish in the lightcone limit do not exist and the recursion relation give the complete result.

\paragraph{Generic $p$} Based on the above calculations, it is straightforward to generalize the result to higher $p$ and a clear pattern can be observed. As mentioned before, the recursive relation leaves only one coefficient to be determined for each $p$. For odd $p$, the coefficient to be determined is $A_{[0,\frac{p-3}{2},\frac{p-1}{2}]}^{(1)}$ and for even $p$, it is $A_{[0,\frac{p-2}{2},\frac{p-2}{2}]}^{(1)}$.

For each $p$, we can choose $m=\lfloor\frac{p-2}{2}\rfloor$ in \eqref{eq:oneloopProject}. Under this projection, the skeleton diagrams contain at least two propagators between $3$ and $4$ for odd $p$. The inserted Lagrangian can only turn one of these propagators into a T-block and the remaining free propagator vanishes upon harmonic identification $u_3=u_4$. Therefore, all $A_{[0,\frac{p-3}{2},\frac{p-1}{2}]}^{(1)}$ vanish for odd $p$.

Therefore we focus on even $p$ case. To extract the coefficient $A_{[0,\frac{p-2}{2},\frac{p-2}{2}]}^{(1)}$, we apply the following harmonic identification
\begin{equation}
\frac{1}{[13]^4}\left.\left(\left.\frac{\langle\tilde{\mathcal{D}}\mathcal{D}\mathcal{O}_p\mathcal{O}_p\mathcal{L}\rangle|_{u_3=u_4}}{[12]^{N-p}}\right|_{u_1=u_2}\right)\right|_{u_1=u_3}=\theta_5^4 x_{12}^2 x_{34}^2 A_{[0,\frac{p-2}{2},\frac{p-2}{2}]}^{(1)}.
\end{equation}

For each $p$, after the harmonic identification, only one diagram survives, which is given in Figure~\ref{dd44one}.
This is essentially the same diagram as Figure~\ref{dd22one}, the only difference being that there are more free propagators, which are encoded in the different powers of $X$, $Y$ and $Z$. Evaluating diagram leads to
\begin{equation}
   \theta_5^4 x_{12}^2x_{34}^2 \frac{1}{\prod_{r=1}^4 x_{r5}^2}
\end{equation}
and thus 
\begin{equation}
    A_{[0,\frac{p-2}{2},\frac{p-2}{2}]}^{(1)}=\frac{1}{\prod_{r=1}^4 x_{r5}^2},
\end{equation}
which turns out to be the same as the $A_{[0,0,0]}^{(1)}$ and $A_{[0,1,1]}^{(1)}$ and gives \begin{equation}
    F_{[0,\frac{p-2}{2},\frac{p-2}{2}]}^{(1)}=\frac{1}{x_{13}^2x_{24}^2}F^{(1)}(z,\bar{z}).
\end{equation} after integration over $x_5$. Summing up all channels, we obtain the result for general even $p$
\begin{align}
    G_{\{p,p\}}^{(1)}=d_{12}^{N-p}\tilde{\mathcal{R}}_{\mathcal{N}=4}F^{(1)}(z,
    \bar{z})\sum_{m=0}^{[\frac{p-2}{2}]}\mathcal{X}^{p-2-2m}(\mathcal{YZ})^{m}\label{DDpp1}. 
\end{align}
This matches nicely the one-loop result first derived in \cite{Jiang:2019xdz}.

\paragraph{Generalization to $G_{\{p,q\}}^{(1)}$}
We now compute the more general correlation function $G_{\{p,q\}}^{(1)}$ based on $G_{\{p,p\}}^{(1)}$. We first notice that $p$ and $q$ must satisfy the relationship $|p-q|=2k~(k\in\mathbb{N})$, otherwise the result is vanishing. This can be seen by using the PCGG method, where the $\mathcal{N}=4$ SYM correlation function takes the following form
\begin{align}
   G_{\{p,q\}}^{(1)}=\sum_{l=0}^{p}\Big(\frac{g^2 d_{12}}{8\pi^2} \Big)^{N-\ell} (d_{34})^{q-m} \langle\text{tr}(\Phi_1\Phi_2)^{\ell}  \text{tr}(\Phi_3)^{m+p-q}\text{tr}(\Phi_4)^{m}\rangle^{(1)}\,,
\end{align}
where $\Phi_j\equiv Y_j\cdot\Phi$ and we have assumed $p>q$. To have non-zero result, we must have 
\begin{align}
    p-q=2\ell-2m \quad (\ell,m\in\mathbb{N}).
\end{align}
For $p<q$, we have the same argument and therefore we have $|p-q|=2k~(k\in\mathbb{N})$.
\begin{figure}[htp]
    \centering
    \includegraphics[width=0.65\linewidth]{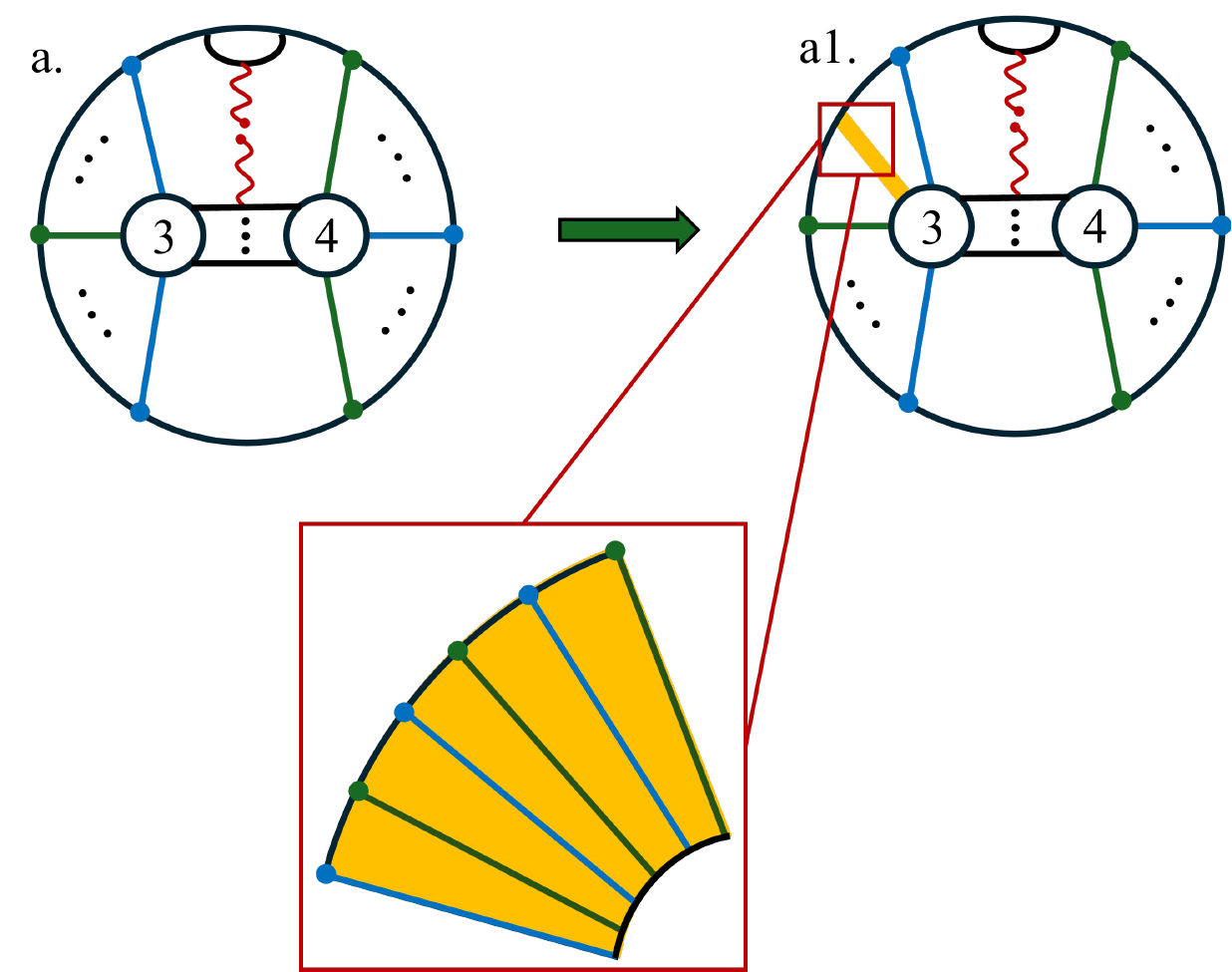}
    \caption{One-loop Feynman diagram for the transition from $G_{\{q,q\}}^{(1)}$ to $G_{\{q+2k,q\}}^{(1)}$. The red wavy lines are loop corrections and the yellow thick lines stand for $2k$ free propagators.
  }
    \label{Genericpq1}
\end{figure}

Next, assuming that $p=q+2k$ so that $\mathcal{O}_p(x_3)$ has extra free fields to contract with PCGG.
{At tree level, these $2 k$ extra free fields can only contract with in sequence due to the planarity of PCGG (See the illustration of the yellow part of $G_{q+2k,q}^{(1)}$ in Figure~\ref{Genericpq1}), thereby contributing a factor $(d_{13}d_{23})^k/(d_{12})^k$. The additional $2k$ Wick contractions do not lead to new loop corrections.} As shown in Figure~\ref{Genericpq1}, the left panel represents a one-loop Feynman diagram for $\langle \mathcal{D}\mathcal{D}\mathcal{O}_q\mathcal{O}_q\rangle$ while  the right panel is the one-loop Feynman diagram  for $\langle \mathcal{D}\mathcal{D}\mathcal{O}_{q+2k}\mathcal{O}_q\rangle$.  The differences between them only occurs in the free propagator part (the yellow part in Figure~\ref{Genericpq1}a1). As a result, we obtain the expression for $\langle \mathcal{D}\mathcal{D}\mathcal{O}_p\mathcal{O}_q\rangle$:
\begin{align}
G_{\{p,q\}}^{(1)}= 
\begin{cases} 
(\frac{d_{13}d_{23}}{d_{12}})^{k} G_{\{q,q\}}^{(1)}, & (p-q=2k,~k\in\mathbb{N})\\
\\
(\frac{d_{14}d_{24}}{d_{12}})^{k} G_{\{p,p\}}^{(1)}, & (q-p=2k,~k\in\mathbb{N}) 
\end{cases} 
\end{align}
where $G_{\{p,p\}}^{(1)}$ is given in \eqref{DDpp1}.

\section{Two-loop computations}
\label{sec:two}
In this section, we apply the harmonic PCGG method to compute the two-loop results of $G^{(2)}_{\{p,p\}}$. The results can be written in terms of the various two-loop conformal integrals are defined as:
\begin{align}
    F_z^{(2)} &= F^{(2)}(z,\bar{z}),\label{fz}\\
    F_{1-z}^{(2)} &= F^{(2)}(1-z,1-\bar{z}),\label{f1-z} \\
    F_{\frac{z}{z-1}}^{(2)} &= \frac{1}{(1-z)(1-\bar{z})} F^{(2)}\left( \frac{z}{z-1}, \frac{\bar{z}}{\bar{z}-1} \right),\label{f/(1-z)}
\end{align}
with the fundamental two-loop integral \( F^{(2)}(z,\bar{z}) \) given by
\begin{align}
\label{eq:fundFz}
    F^{(2)}_z(z,\bar{z}) &= \frac{x_{13}^2 x_{24}^2 x_{14}^2}{\pi^4} \int \frac{\mathrm{d}^4x_5 \, \mathrm{d}^4x_6}{x_{15}^2 x_{25}^2 x_{45}^2 x_{56}^2 x_{16}^2 x_{36}^2 x_{46}^2} \nonumber \\
    &= \frac{1}{z - \bar{z}} \Bigg[ \frac{\ln^2(z\bar{z})}{2} \Big( \mathrm{Li}_2(z) - \mathrm{Li}_2(\bar{z}) \Big) - 3\ln(z\bar{z}) \Big( \mathrm{Li}_3(z) - \mathrm{Li}_3(\bar{z}) \Big) + 6 \Big( \mathrm{Li}_4(z) - \mathrm{Li}_4(\bar{z}) \Big) \Bigg].
\end{align}
At two-loop order, the correlation function takes the form
\begin{align}
 G_{\{p,p\}}^{(2)}=\mathcal{R}_{\mathcal{N}=4}d_{12}^{N-p}\sum_{m+n+l=p-2}F^{(2)}_{[l,m,n]}(u,v)\mathcal{X}^l\mathcal{Y}^m\mathcal{Z}^n,
\end{align}
while $F^{(2)}_{[l,m,n]}=0$ for $|m-n|>1$ according to \textbf{Theorem 2}. Based on the lightcone OPE relation, the remaining $F^{(2)}_{[l,m,n]}$ can be deduced from $F^{(2)}_{[l-1,m,n]}$ and recursively from $F^{(2)}_{[0,m,n]}$. Further, $G_{\{p,p\}}^{(2)}$ is invariant under the exchange of $\mathcal{O}_p(x_3,y_3)$ and $\mathcal{O}_p(x_4,y_4)$. Taking into account these properties, it turns out to be sufficient to compute only two coefficients: $F^{(2)}_{[0,m,m]}$ and $F^{(2)}_{[0,m,m+1]}$.

\begin{figure}[h!]
    \centering
 \includegraphics[width=0.35\linewidth]{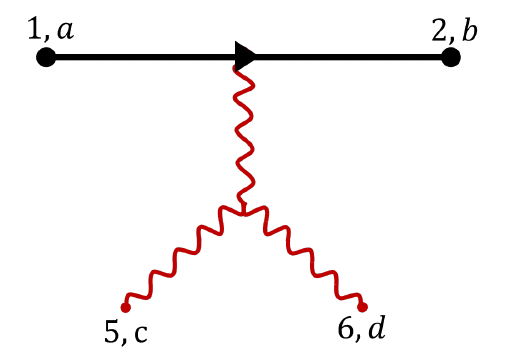}
    \caption{Interaction block involving self-interaction of gluons. Such diagrams vanish automatically in the frame $\theta_5=\theta_6=0$. }
    \label{non-abelian}
\end{figure}

\begin{figure}[h!]
    \centering
 \includegraphics[width=0.75\linewidth]{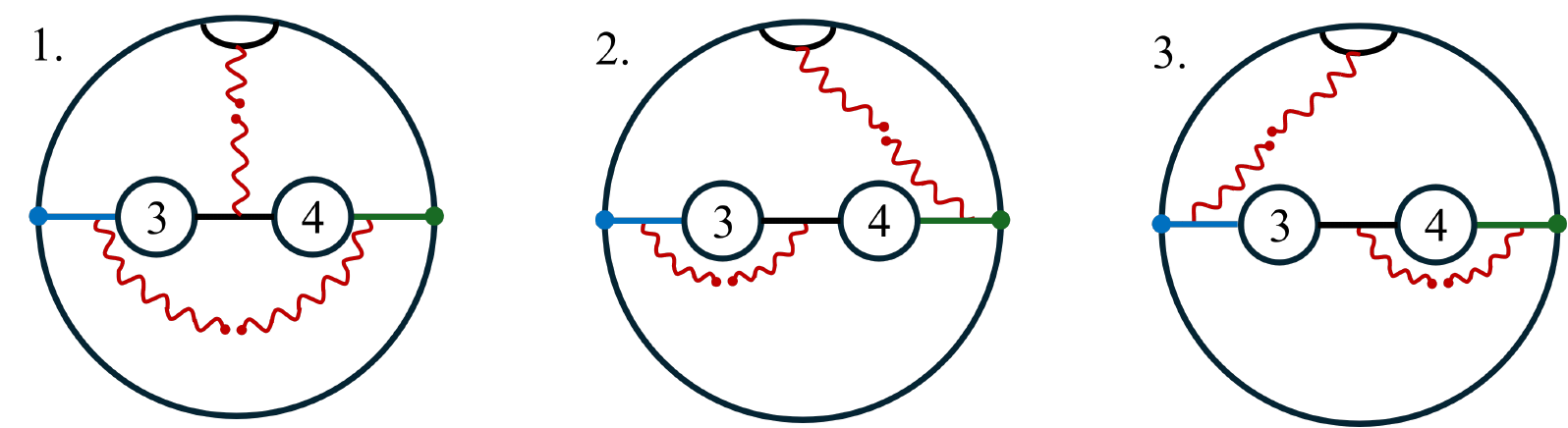}
    \caption{Two-loop $\langle\tilde{\mathcal{D}}\mathcal{D}\mathcal{O}_2\mathcal{O}_2\mathcal{L}\mathcal{L}\rangle$ diagrams with one T-block $\langle \tilde{q}_1 W q_2\rangle$. }
    \label{dd22two1}
\end{figure}

\begin{figure}[h!]
    \centering  \includegraphics[width=0.5\linewidth]{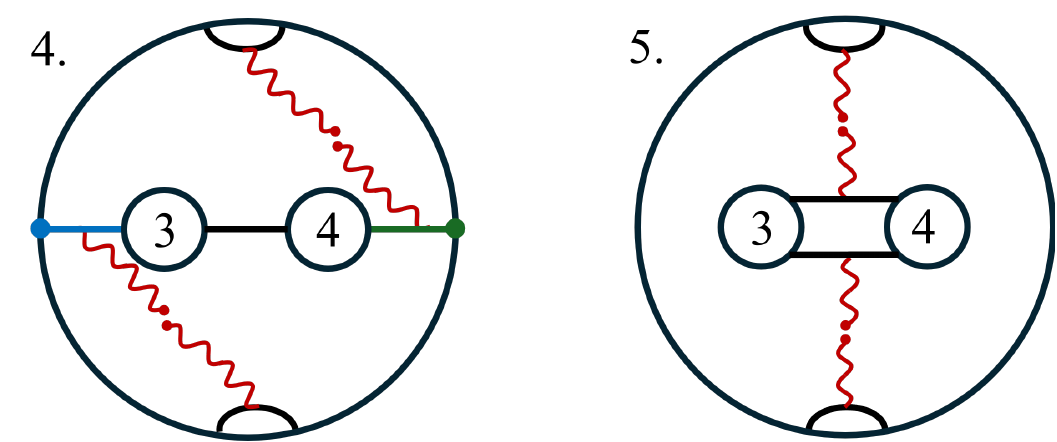}
    \caption{Two-loop $\langle\tilde{\mathcal{D}}\mathcal{D}\mathcal{O}_2\mathcal{O}_2\mathcal{L}\mathcal{L}\rangle$ diagrams with two T-blocks $\langle \tilde{q}_1 W q_2\rangle$. }
    \label{dd22two2}
\end{figure}

\begin{figure}[h!]
    \centering    \includegraphics[width=0.25\linewidth]{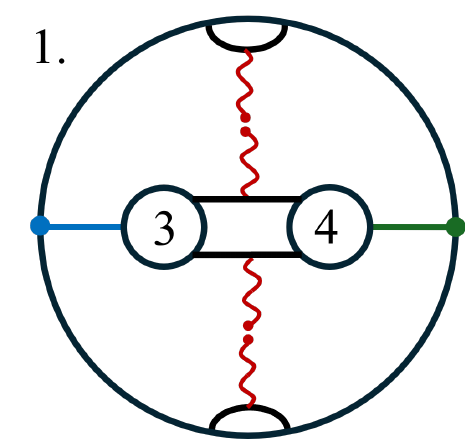}
    \caption{Two-loop diagram that contributes to $\langle\tilde{\mathcal{D}}\mathcal{D}\mathcal{O}_3\mathcal{O}_3\mathcal{L}\mathcal{L}\rangle$.}
    \label{dd33two}
\end{figure}

\begin{figure}[h!]
    \centering   \includegraphics[width=0.95\linewidth]{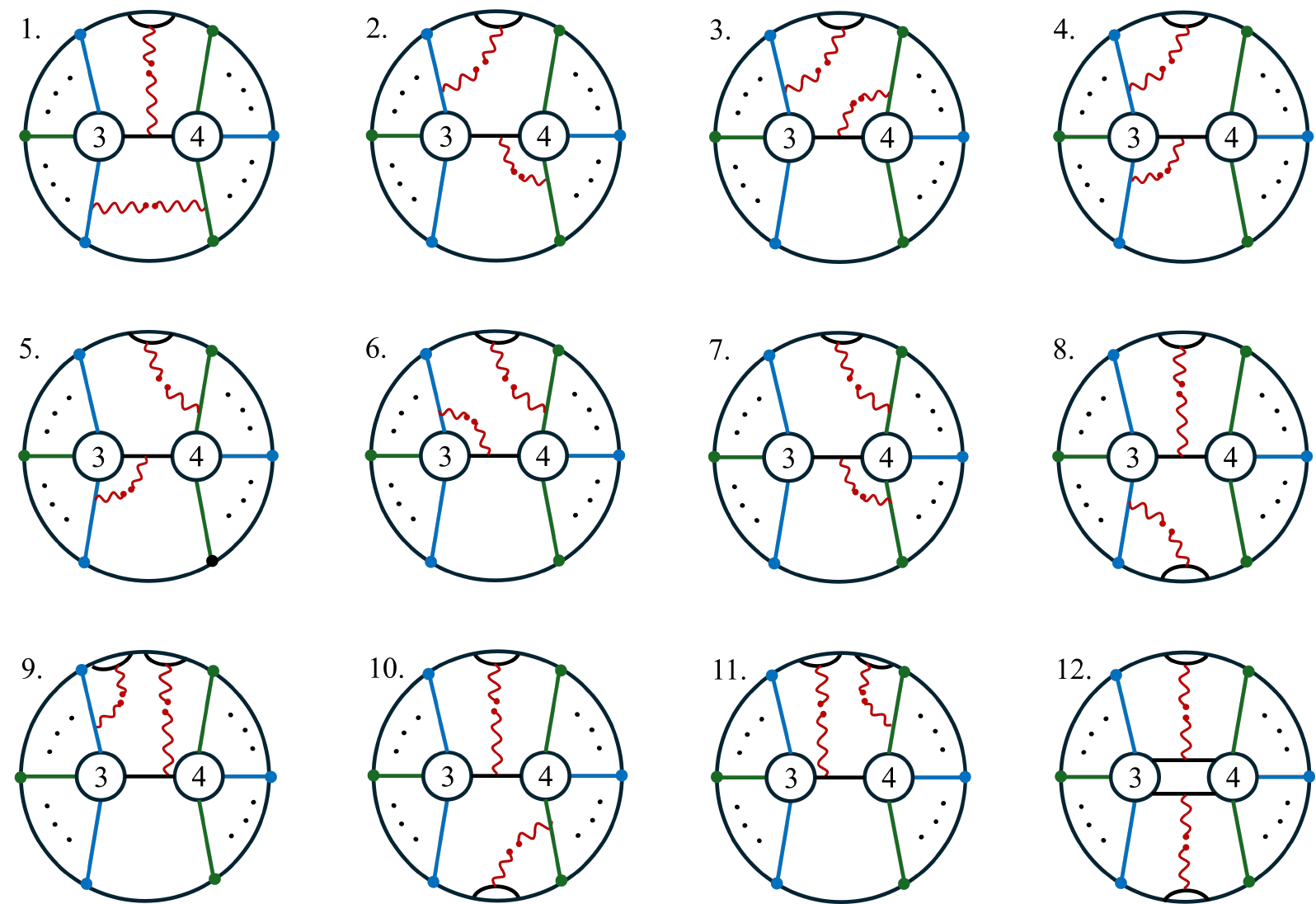}
    \caption{Two-loop diagrams that contribute to $\langle\tilde{\mathcal{D}}\mathcal{D}\mathcal{O}_p\mathcal{O}_p\mathcal{L}\mathcal{L}\rangle$. There are two different kinds of interactions; the three-point face  interactions and the four-point face interactions. Here, we list the Feynman diagrams that contain only four-point face interactions.}
    \label{dd44two1}
\end{figure}

\begin{figure}[h!]
    \centering    \includegraphics[width=0.95\linewidth]{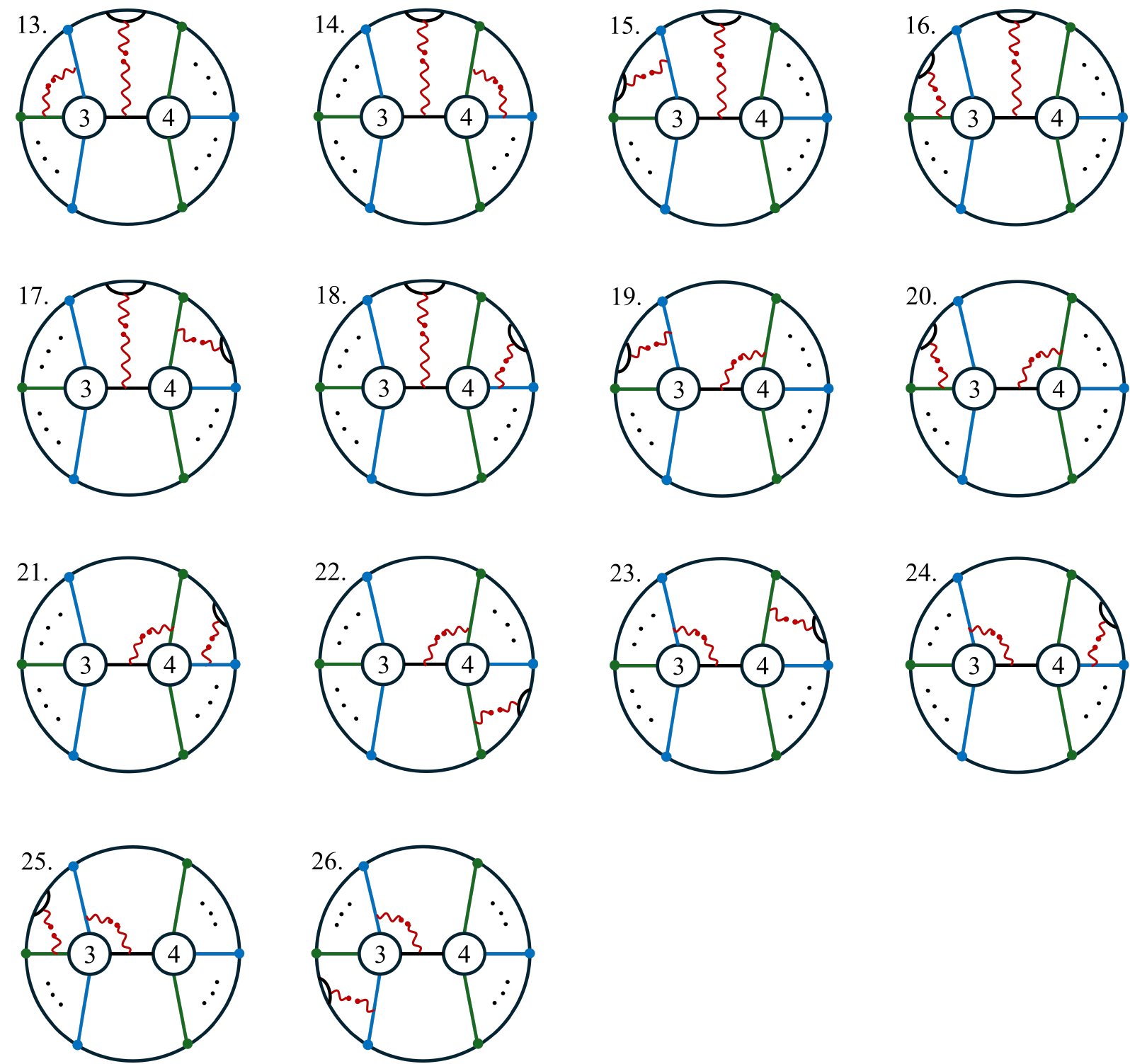}
    \caption{Two-loop diagrams that contribute to $\langle\tilde{\mathcal{D}}\mathcal{D}\mathcal{O}_p\mathcal{O}_p\mathcal{L}\mathcal{L}\rangle$. There are two different kinds of interactions; the three-point face  interactions and the four-point face interactions. Here, we list the Feynman diagrams that contain  three- and four-point face interactions }
    \label{dd44two2}
\end{figure}

In the Lagrangian insertion formalism, we need to compute the following $\mathcal{N}=2$ integrand with two inserted Lagrangian densities 
\begin{equation}
\langle\tilde{\mathcal{D}}\mathcal{D}\mathcal{O}_p\mathcal{O}_p\mathcal{L}\mathcal{L}\rangle=\Theta_{5,6}F(x,u),
\end{equation}
where the two-loop superconformal nilpotent invariant is
\begin{equation}
    \Theta_{5,6}=\left[\theta_5^4 \theta_6^4 R_{\mathcal{N}=2}+\ldots+\left(\theta_1^{+}\right)^2\left(\theta_2^{+}\right)^2\left(\theta_3^{+}\right)^2\left(\theta_4^{+}\right)^2 x_{56}^4\right] \frac{x_{56}^4}{\prod_{i=1}^4 x_{i 5}^2 x_{i 6}^2}.
\end{equation}
At two-loop order, it is more convenient to work in the frame $\theta_5=\theta_6=0$. One important advantage of this frame is that all Feynman diagrams containing gluon self-interaction, for instance the one shown in Figure~\ref{non-abelian}, vanish automatically. In this frame, the correlation function can be expressed as
\begin{equation}    \langle\tilde{\mathcal{D}}\mathcal{D}\mathcal{O}_p\mathcal{O}_p\mathcal{L}\mathcal{L}\rangle=\prod_{r=1}^4\theta^{+2}_r
\left(\frac{[12]}{x_{12}^2}\right)^{N-p}\sum_{l+m+n=p-2}X^lY^mZ^n A_{[l,m,n]}^{(2)},
\end{equation}
where the coefficients $A_{[l,m,n]}^{(2)}$ do not depend on harmonic variables.
Similar to the one-loop calculation, the correlation functions with different $p$ can be computed recursively. We will detail the computation for the first few values of $p$, from which it is clear to see the pattern for generic $p$.
\paragraph{$p=2$ case}
The general structure of $\langle\mathcal{D}\mathcal{D}\mathcal{O}_2\mathcal{O}_2\rangle^{(2)}$ in $\mathcal{N}=4$ SYM is
\begin{equation}
\langle\mathcal{D}\mathcal{D}\mathcal{O}_2\mathcal{O}_2\rangle^{(2)}=\mathcal{R}_{\mathcal{N}=4} d_{12}^{N-2}F_{[0,0,0]}^{(2)}.
\end{equation}
For $p=2$, we perform the pure projection
\begin{equation}
\begin{aligned}      \langle\tilde{\mathcal{D}}\mathcal{D}\mathcal{O}_2\mathcal{O}_2\mathcal{L}\mathcal{L}\rangle
        = \prod_{r=1}^4\theta^{+2}_r \left(\frac{[12]}{x_{12}^2}\right)^{N-2}A_{[0,0,0]}^{(2)}.
\end{aligned}
\end{equation}
Under this projection $F_{[0,0,0]}^{(2)}$ corresponds to $A_{[0,0,0]}^{(2)}$. We take the following harmonic identification:
\begin{equation}    \frac{1}{[12]^{N-2}}\left( \langle\tilde{\mathcal{D}}\mathcal{D}\mathcal{O}_2\mathcal{O}_2\mathcal{L}\mathcal{L}\rangle|_{u_1=u_4,u_2=u_3}\right)|_{u_1=u_2}=\prod_{r=1}^4\theta^{+2}_r A_{[0,0,0]}^{(2)}.
\end{equation}
This identification removes all Feynman diagrams containing free propagators carrying the factor $[14]$ and $[23]$. The contributing diagrams can be classified into two groups by the number of the T-block connecting $1$ and $2$, which are given in Figure~\ref{dd22two1} and Figure~\ref{dd22two2} respectively. 
The expressions of the building blocks can be computed straightforwardly in the frame $\theta_5=\theta_6=0$. The sum of these diagrams give the result
\begin{equation}
    A_{[0,0,0]}^{(2)}=\frac{1}{x_{12}^2x_{34}^2x_{14}^2x_{23}^2}\frac{x_{56}^2}{\prod_{r=1}^4x_{r5}^2x_{r6}^2} P,
\end{equation}
where
\begin{align}        P=&x_{13}^2x_{24}^2x_{56}^2+x_{14}^2x_{23}^2x_{56}^2-x_{12}^2x_{34}^2x_{56}^2+x_{13}^2x_{25}^2x_{46}^2+x_{13}^2x_{26}^2x_{45}^2\label{P22}\\&+x_{24}^2x_{15}^2x_{36}^2+x_{24}^2x_{16}^2x_{35}^2
    -x_{12}^2x_{35}^2x_{46}^2-x_{12}^2x_{36}^2x_{45}^2+x_{34}^2x_{15}^2x_{26}^2.\nonumber
\end{align}
Going back to the frame $\theta_r^+=0$ $(r=1,2,3,4)$ and integrating over inserted coordinates, we obtain
\begin{align}
    F_{[0,0,0]}^{(2)}&=\int\rd^4x_5\rd^4x_6\frac{1}{x_{56}^4}A_{[0,0,0]}\\\nonumber
    &=\frac{1}{x_{12}^2x_{34}^2x_{14}^2x_{23}^2}\left[(2-z-\bar{z})(F^{(1)}(z,\bar{z}))^2+4 F_{z}^{(2)}+4 F_{\frac{z}{1-z}}^{(2)}\right],
\end{align}
where the two-loop conformal integrals are defined in \eqref{fz}, \eqref{f/(1-z)} and \eqref{eq:fundFz}.
\paragraph{$p=3$ case}
The general structure of $\langle \mathcal{D}\mathcal{D}\mathcal{O}_3\mathcal{O}_3\rangle^{(2)}$ in $\mathcal{N}=4$ SYM reads
\begin{equation}
    \langle \mathcal{D}\mathcal{D}\mathcal{O}_3\mathcal{O}_3\rangle^{(2)}=\mathcal{R}_{\mathcal{N}=4} d_{12}^{N-3}\left[\mathcal{X}F_{[1,0,0]}^{(2)}+\mathcal{Y}F_{[0,1,0]}^{(2)}+\mathcal{Z}F_{[0,0,1]}^{(2)}\right],
\end{equation}
where 
\begin{equation}   F_{[1,0,0]}^{(2)}=F_{[0,0,0]}^{(2)}+\text{missing term}
\end{equation}
due to the lightcone OPE relation \eqref{lcr}. Here, ``missing term'' refers to the terms that vanish in the lightcone limit. We shall derive these missing terms for generic $p$ in what follows. In the current case, we take $p=3$ which gives
\begin{equation}
    \text{missing term}=-2F^{(2)}_{1-z}.
\end{equation}
$F_{[0,0,1]}^{(2)}$ and $F_{[0,1,0]}^{(2)}$ are related to each other by exchanging $3$ and $4$ and thus we can focus on the computation of $F_{[0,0,1]}^{(2)}$. We again choose the pure projection
\begin{equation}
\langle\tilde{\mathcal{D}}\mathcal{D}\mathcal{O}_3\mathcal{O}_3\mathcal{L}\mathcal{L}\rangle= \prod_{r=1}^4\theta^{+2}_r \left(\frac{[12]}{x_{12}^2}\right)^{N-3}(X A_{[1,0,0]}^{(2)}+Z A_{[0,0,1]}^{(2)}),
\end{equation}
where $A_{[0,0,1]}^{(2)}$ corresponds to $F_{[0,0,1]}^{(2)}$. To extract $A_{[0,0,1]}^{(2)}$ from the $\mathcal{N}=2$ correlator, we perform the following harmonic identification
\begin{equation}  \frac{1}{[13]^2}\left.\left(\left.\frac{\langle\tilde{\mathcal{D}}\mathcal{D}\mathcal{O}_3\mathcal{O}_3\mathcal{L}\mathcal{L}\rangle|_{u_3=u_4}}{[12]^{N-3}}\right|_{u_1=u_2} \right)\right|_{u_1=u_3}=\prod_{r=1}^4\theta^{+2}_r A_{[0,0,1]}^{(2)}\,,
\end{equation}
under which there is only one non-vanishing Feynman diagram, given in Figure~\ref{dd33two}.
Using the simplified Feynman rules, this diagram evaluates to 
\begin{align}
       &\frac{[14]}{x_{14}^2} \frac{[23]}{x_{23}^2}\langle\tilde{q}_1 W_5q_2\rangle\langle\tilde{q}_3 W_5q_4\rangle\langle\tilde{q}_1 W_6q_2\rangle\langle\tilde{q}_3 W_6q_4\rangle+(5\leftrightarrow6)\\\nonumber
       =&\prod_{r=1}^4\theta^{+2}_r\frac{[14]}{x_{14}^2} \frac{[23]}{x_{23}^2}\frac{1}{x_{12}^2 x_{34}^2}\frac{x_{56}^4}{\prod_{r=1}^4 x_{r5}^2x_{r6}^2}\,.
\end{align}
It is then simple to go back to the frame $\theta_r^+=0$ ($r=1,\ldots,4$) and read off $A_{[0,0,1]}^{(2)}$ 
\begin{equation}
 A_{[0,0,1]}^{(2)}=\frac{x_{14}^2x_{23}^2}{x_{13}^2x_{24}^2}  \frac{x_{13}^2x_{24}^2}{\prod_{r=1}^4 x_{r5}^2x_{r6}^2}.
\end{equation}
After integration over $x_5$ and $x_6$, we obtain
\begin{equation}
    F^{(2)}_{[0,0,1]}=\int \rd^4 x_5 \rd^4 x_6 A_{[0,0,1]}^{(2)}=(1-z)(1-\bar{z})(F^{(1)}(z,\bar{z}))^2.
\end{equation}
{Combining the previous results, we obtain
\begin{equation}
\begin{aligned}
    \langle \mathcal{D}\mathcal{D}\mathcal{O}_3\mathcal{O}_3\rangle^{(2)} =\tilde{\mathcal{R}}_{\mathcal{N}=4}d_{12}^{N-3}&\{\mathcal{X} [(2-z-\bar{z})(F^{(1)}(z,\bar{z}))^2+4 F_{z}^{(2)}+4 F_{\frac{z}{1-z}}^{(2)}  ]\\
    &+\mathcal{Y}(1-z)(1-\bar{z})(F^{(1)}(z,\bar{z}))^2\\
    &+\mathcal{Z}(F^{(1)}(z,\bar{z}))^2-2\mathcal{X}F^{(2)}_{1-z}\}\,.
\end{aligned}
\end{equation}
}
\paragraph{$p=4$ case}
The $\mathcal{N}=4$ correlator $\langle \mathcal{D}\mathcal{D}\mathcal{O}_4 \mathcal{O}_4\rangle^{(2)}$ has the following structure:
\begin{equation}
   \langle \mathcal{D}\mathcal{D}\mathcal{O}_4 \mathcal{O}_4\rangle^{(2)}=\mathcal{R}_{\mathcal{N}=4}d_{12}^{N-4}\left[\mathcal{X}^2 F_{[2,0,0]}^{(2)}+\mathcal{X}\mathcal{Y}F_{[1,1,0]}^{(2)}+\mathcal{X}\mathcal{Z}F_{[1,0,1]}^{(2)}+\mathcal{Y}\mathcal{Z}F_{[0,1,1]}^{(2)}\right]\,, 
\end{equation}
where we have dropped $F_{[0,0,2]}^{(2)}$ and $F_{[0,2,0]}^{(2)}$ due to the \textbf{Theorem 2}. The coefficients $F_{[2,0,0]}^{(2)}$, $F_{[1,1,0]}^{(2)}$ and $F_{[1,0,1]}^{(2)}$ are partially determined by the lightcone OPE relation up to some missing terms in the lightcone limit. Again, we only need to focus on the computation of one term, $F_{[0,1,1]}^{(2)}$. The simplest $\mathcal{N}=2$ correlator containing $F_{[0,1,1]}^{(2)}$ corresponds to taking $p=4$, $r=1$ in \eqref{eq:N2reductionr} and the correlator takes the form
\begin{equation}     \langle\tilde{\mathcal{D}}\mathcal{D}\mathcal{O}_4\mathcal{O}_4\mathcal{L}\mathcal{L}\rangle= \prod_{r=1}^4\theta^{+2}_r \left(\frac{[12]}{x_{12}^2}\right)^{N-3}(X^2A_{[2,0,0]}^{(2)}+XYA_{[1,1,0]}^{(2)}+XZA_{[1,0,1]}^{(2)}+YZA_{[0,1,1]}^{(2)})\,.
\end{equation}
In the $\mathcal{N}=2$ reduction, $A^{(2)}_{[0,1,1]}$ corresponds to $F^{(2)}_{[0,1,1]}$.
We perform the following harmonic identification 
\begin{equation}
 \frac{1}{[13]^4}\left.\left(\left. \frac{\langle\tilde{\mathcal{D}}\mathcal{D}\mathcal{O}_4\mathcal{O}_4\mathcal{L}\mathcal{L}\rangle|_{u_3=u_4}}{[12]^{N-4}}\right|_{u_1=u_2}\right)\right|_{u_1=u_3}=\prod_{r=1}^4\theta^{+2}_r A_{[0,1,1]}^{(2)}\,.
\end{equation}
The contributing diagrams are listed in Figure~\ref{dd44two1} and Figure~\ref{dd44two2}\footnote{More precisely, we need to take $p = 4$ and remove the ellipsis from the Feynman diagrams in Figure~\ref{dd44two1} and Figure~\ref{dd44two2}.}.

Here several points deserve more discussions. First, we organize these Feynman diagrams into two groups, classified by the positions of the inserted Lagrangian densities. The first group consists of diagrams in which both Lagrangians are inserted in four-point faces and the second group consists of diagrams containing two Lagrangians inserted in a three-point face and a four-point face individually. 

Second, we need to multiply a symmetry factor to each diagram. For example, Diagram-$1$ in Figure~\ref{dd44two1} has a symmetry factor $6$ which originates from the fact that the length-6 PCGG provides us 6 possible choices to construct the factor $\langle\tilde{q}_1 W q_2\rangle$. The symmetry factor of the Diagram-$13$ is $6\times2$, where the factor $6$ has the same origin of Diagram-1 and the factor $2$ comes from the fact that there are two three-point faces to insert the Lagrangian densities. The explicit formula of these diagrams for general $p$, along with the proper symmetry factors are listed in Table~\ref{tab:my_label1} and Table~\ref{tab:my_label2}. Setting $p=4$ gives the symmetry factors of the present case.

Third, in the harmonic identification procedure, we need to factor out $[12]^{N-4}$ and then set $u_1=u_2$. Some diagrams, for instance Diagram-$8$ and Diagram-$10$, carry a factor $[12]^{N-5}$ separately while their sum contributes an additional $[12]$ factor due to the harmonic identity
\begin{equation}
    [13^-][23]+[13][23^-]=[12].
\end{equation}
For simplicity of counting, we multiply a factor $\frac{1}{2}$ to this type of diagrams, which is indicated in red in the tables.

After summing over all contributing diagrams and performing the integration over the spacetime points and Grassmann variables, we obtain
\begin{equation}
    F_{[0,1,1]}^{(2)}=(2-z-\bar{z})(F^{(1)}(z,\bar{z}))^2+4 F_{z}^{(2)}+4 F_{\frac{z}{1-z}}^{(2)},
\end{equation}
which turns out to be the same as the $F_{[0,0,0]}^{(2)}$. To obtain the full result, we need to compute $F_{[2,0,0]}^{(2)}$, $F_{[1,1,0]}^{(2)}$ and $F_{[1,0,1]}^{(2)}$. As mentioned in Section~\ref{light-cone}, these coefficients can be deduced from the correlator of $\langle \mathcal{D}\mathcal{D}\mathcal{O}_3 \mathcal{O}_3\rangle$ using the lightcone OPE relations, up to some missing terms to be determined. To calculate these terms, notice that we have
\begin{equation}
    \mathcal{X}^2 F_{[2,0,0]}=\mathcal{X}^2(F_{[1,0,0]}+\text{missing term})
\end{equation}
\begin{equation}
    \begin{aligned}
          \mathcal{X}^2 F_{[2,0,0]}&=\mathcal{X}^2(F_{[1,0,0]}+\text{missing term}),  \\
           \mathcal{X}\mathcal{Y} F_{[1,1,0]}&=\mathcal{X}\mathcal{Y}(F_{[0,1,0]}+\text{missing term}),  \\
          \mathcal{X}\mathcal{Z} F_{[1,0,1]}&=\mathcal{X}\mathcal{Z}(F_{[0,0,1]}+\text{missing term}). 
    \end{aligned}
\end{equation}
Taking the sum of the above expressions, we find that the complete missing term takes the following form
\begin{equation}
    \text{missing terms}= \mathcal{X}^2 a+\mathcal{X}\mathcal{Y}b+\mathcal{X}\mathcal{Z}c,
\end{equation}
where $a$, $b$ and $c$ are some coefficients which depend on cross ratios. We will show below that $b=c=0$ and
\begin{equation}
    a=-2 F^{(2)}_{1-z}.
\end{equation}
Therefore, the full two-loop result of $\langle \mathcal{D}\mathcal{D}\mathcal{O}_4 \mathcal{O}_4\rangle$ is given by 
\begin{equation}
  \begin{aligned}
      \langle \mathcal{D}\mathcal{D}\mathcal{O}_4 \mathcal{O}_4\rangle^{(2)}=&\tilde{\mathcal{R}}_{\mathcal{N}=4}\{(\mathcal{X}^2+\mathcal{Y}\mathcal{Z})[(2-z-\bar{z})(F^{(1)}(z,\bar{z}))^2+4 F_{z}^{(2)}+4 F_{\frac{z}{1-z}}^{(2)}  ]\\
      &+\mathcal{X}\mathcal{Y}(1-z)(1-\bar{z})(F^{(1)}(z,\bar{z}))^2+\mathcal{X}\mathcal{Z}(F^{(1)}(z,\bar{z}))^2\\
      &-2\mathcal{X}^2 F^{(2)}_{1-z}\}.
  \end{aligned}  
\end{equation}

\paragraph{Generic $p$}
From previous examples, we find a recursive structure of the correlators. For generic $p$, the recursive structure and \textbf{Theorem 2} guarantee that we only need to compute $F_{[0,\lfloor\frac{p-2}{2}\rfloor,\lfloor\frac{p-1}{2}\rfloor]}^{(2)}$ up to some missing terms in the lightcone limit. We choose the following $\mathcal{N}=2$ projection
\begin{equation}
     \langle\det\tilde{q}_1\det q_2\text{tr}(\tilde{q}_3^{\lfloor\frac{p-2}{2}\rfloor} q_3^{p-\lfloor\frac{p-2}{2}\rfloor})\text{tr}(q_4^{\lfloor\frac{p-2}{2}\rfloor} \tilde{q}_4^{p-\lfloor\frac{p-2}{2}\rfloor})\rangle= \prod_{r=1}^4\theta^+_r \sum_{l+m+n=p-2} X^l Y^m Z^n f_{[l,m,n]}^{(2)}.
\end{equation}

We consider even $p$ first. Under harmonic identification, 26 independent diagrams survive, which are listed in Figure~\ref{dd44two1} and Figure~\ref{dd44two2}. For arbitrary $p>4$, the interaction parts of the Feynman diagrams are the same, the differences between different $p$ are the symmetry factors and the number of free propagators.

The 26 Feynman diagrams are classified into two groups. The first group, consisting of 12 diagrams, contain diagrams with only four-point face interactions. The second group consists of 14 diagrams involving both three- and four-point face interactions. The color factor, symmetry factor, and kinematic factor of these diagrams are listed in Table~\ref{tab:my_label1} and Table~\ref{tab:my_label2}.

\begin{table}[htbp]
    \centering
    \renewcommand{\arraystretch}{2.5}  
    \setlength{\tabcolsep}{6pt}  
    \begin{tabular}{c|c|c|c}
     \hline
        Diagram & Color Factor & Symmetry Factor & Kinematic Factor \\
        \hline
        1 & $-\frac{N_c^{p+5}}{2^{p+2}}$ & $2p$ &\makecell[c]{$ (YZ)^{\frac{p-2}{2}} 
                \Big[  \tau_{13}\tau_{24} - \tau_{12}\tau_{34} - \tau_{14}\tau_{23} 
                      $\\[3mm]$ -\tau_{13}(\rho_2^2\sigma_4^2 + \rho_4^2\sigma_2^2) - \tau_{24}(\rho_1^2\sigma_3^2 + \rho_3^2\sigma_1^2) 
                      $\\[3mm]$ +\tau_{12}(\rho_3^2\sigma_4^2 + \rho_4^2\sigma_3^2) + \tau_{34}(\rho_1^2\sigma_2^2 + \rho_2^2\sigma_1^2) 
                       $\\[3mm]$+\tau_{14}(\rho_2^2\sigma_3^2 + \rho_3^2\sigma_2^2) + \tau_{23}(\rho_1^2\sigma_4^2 + \rho_4^2\sigma_1^2) \Big]
           $} \\
        \hline
        2 & $-\frac{N_c^{p+5}}{2^{p+2}}$ & $2p$ & $(YZ)^{\frac{p-2}{2}}\tau_{13}(\rho_2^2\sigma_4^2+\rho_4^2\sigma_2^2)$ \\
        \hline
        3 & $-\frac{N_c^{p+5}}{2^{p+2}}$ & $2p$ & $(YZ)^{\frac{p-2}{2}} \tau_{13}(\rho_2^2\sigma_4^2+\rho_4^2\sigma_2^2)$ \\
        \hline
        4 & $-\frac{N_c^{p+5}}{2^{p+2}}$ & $2p$ & $(YZ)^{\frac{p-4}{2}}\Big[-\frac{[13][24][14]^2}{x_{13}^2 x_{24}^2 x_{14}^4}\Big]\tau_{23}(\rho_1^2\sigma_4^2+\rho_4^2\sigma_1^2)$\\
        \hline
        5 & $-\frac{N_c^{p+5}}{2^{p+2}}$ & $2p$ & $(YZ)^{\frac{p-2}{2}} \tau_{24} (\rho_1^2\sigma_3^2+\rho_3^2\sigma_1^2)$ \\
        \hline
        6 & $-\frac{N_c^{p+5}}{2^{p+2}}$ & $2p$ & $(YZ)^{\frac{p-2}{2}} \tau_{24} (\rho_1^2\sigma_3^2+\rho_3^2\sigma_1^2)$ \\
        \hline
        7 & $-\frac{N_c^{p+5}}{2^{p+2}}$ & $2p$ & $(YZ)^{\frac{p-4}{2}} \left[ -\frac{[13][24][23]^2}{x_{13}^2 x_{24}^2 x_{23}^4} \right] \tau_{14}(\rho_2^2\sigma_3^2+\rho_3^2\sigma_2^2)$ \\
        \hline
        8 & $-\frac{N_c^{p+6}}{2^{p+3}}$ & $2(p+1)$ & $\textcolor{red}{\frac{1}{2}}(YZ)^{\frac{p-2}{2}} \frac{[14]}{x_{14}^2} \tau_{12}(\rho_3^2\sigma_4^2 + \rho_4^2\sigma_3^2)$ \\
        \hline
        9 & $-\frac{N_c^{p+6}}{2^{p+3}}$ & $2(p+1)$ & $\textcolor{red}{\frac{1}{2}}(YZ)^{\frac{p-2}{2}} \frac{[14]}{x_{14}^2} \tau_{12}(\rho_3^2\sigma_4^2 + \rho_4^2\sigma_3^2)$ \\
        \hline
        10 & $-\frac{N_c^{p+6}}{2^{p+3}}$ & $2(p+1)$ & $-\textcolor{red}{\frac{1}{2}}(YZ)^{\frac{p-2}{2}} \frac{[23]}{x_{23}^2} \tau_{12}(\rho_3^2\sigma_4^2 + \rho_4^2\sigma_3^2)$ \\
        \hline
        11 & $-\frac{N_c^{p+6}}{2^{p+3}}$ & $2(p+1)$ & $-\textcolor{red}{\frac{1}{2}}(YZ)^{\frac{p-2}{2}} \frac{[23]}{x_{23}^2} \tau_{12}(\rho_3^2\sigma_4^2 + \rho_4^2\sigma_3^2)$ \\
        \hline
        12 & $-\frac{N_c^{p+5}}{2^{p+2}}$ & $4p$ & $2(YZ)^{\frac{p-2}{2}} \tau_{12} \tau_{34}$ \\
        \hline
    \end{tabular}
    \caption {Color factors, symmetry factors and kinematic factors of Feynman diagrams that contain only four-point face interactions.}
    \label{tab:my_label1}
\end{table}

\begin{table}[htbp]
    \centering
    \renewcommand{\arraystretch}{2.5}  
    \setlength{\tabcolsep}{6pt}     
    \begin{tabular}{c|c|c|c}
        \hline
        Diagram & Color Factor & Symmetry Factor & Kinematic Factor \\
        \hline
        13 & $-\frac{N_c^{p+5}}{2^{p+2}}$ & $2p(p-2)$ & $(YZ)^{\frac{p-4}{2}}\Big[-\frac{[23][14][14]^2}{x_{23}^2 x_{14}^2 x_{14}^4}\Big]\tau_{12}(\rho_3^2\sigma_4^2+\rho_4^2\sigma_3^2)$ \\
        \hline
        14 & $-\frac{N_c^{p+5}}{2^{p+2}}$ & $2p(p-2)$ & $(YZ)^{\frac{p-4}{2}}\Big[-\frac{[14][24][23]^2}{x_{14}^2 x_{24}^2 x_{23}^4}\Big]\tau_{12}(\rho_3^2\sigma_4^2+\rho_4^2\sigma_3^2)$ \\
        \hline
        15 & $-\frac{N_c^{p+6}}{2^{p+3}}$ & $2(p+1)(p-2)$ & $\textcolor{red}{\frac{1}{2}}(YZ)^{\frac{p-2}{2}}\frac{[14]}{x_{14}^2}\tau_{12}(\rho_3^2\sigma_4^2+\rho_4^2\sigma_3^2)$ \\
        \hline
        16 & $-\frac{N_c^{p+6}}{2^{p+3}}$ & $2(p+1)(p-2)$ & $\textcolor{red}{\frac{1}{2}}(YZ)^{\frac{p-4}{2}}\Big[-\frac{[14]^2[23]^2[24]}{x_{14}^4 x_{23}^4 x_{24}^2}\Big]\tau_{12}(\rho_3^2\sigma_4^2+\rho_4^2\sigma_3^2)$ \\
        \hline
        17 & $-\frac{N_c^{p+6}}{2^{p+3}}$ & $2(p+1)(p-2)$ & $-\textcolor{red}{\frac{1}{2}}(YZ)^{\frac{p-2}{2}}\frac{[23]}{x_{23}^2}\tau_{12}(\rho_3^2\sigma_4^2+\rho_4^2\sigma_3^2)$ \\
        \hline
        18 & $-\frac{N_c^{p+6}}{2^{p+3}}$ & $2(p+1)(p-2)$ & $\textcolor{red}{\frac{1}{2}}(YZ)^{\frac{p-4}{2}}\Big[\frac{[14]^2[23]^2[24]}{x_{14}^4 x_{23}^4 x_{24}^2}\Big]\tau_{12}(\rho_3^2\sigma_4^2+\rho_4^2\sigma_3^2)$ \\
        \hline
        19 & $-\frac{N_c^{p+5}}{2^{p+2}}$ & $2p(p-2)$ & $(YZ)^{\frac{p-2}{2}}\tau_{13}(\rho_2^2\sigma_4^2+\rho_4^2\sigma_2^2)$ \\
        \hline
        20 & $-\frac{N_c^{p+5}}{2^{p+2}}$ & $2p(p-2)$ & $(YZ)^{\frac{p-4}{2}}\Big[-\frac{[14][24][23]^2}{x_{14}^2 x_{24}^2 x_{23}^4}\Big]\tau_{13}(\rho_2^2\sigma_4^2+\rho_4^2\sigma_2^2)$ \\
        \hline
        21 & $-\frac{N_c^{p+5}}{2^{p+2}}$ & $2p(p-2)$ & $(YZ)^{\frac{p-4}{2}}\left[-\frac{[13][14][23]^2}{x_{13}^2 x_{14}^2 x_{23}^4}\right]\tau_{14}(\rho_2^2\sigma_3^2+\rho_3^2\sigma_2^2)$ \\
        \hline
        22 & $-\frac{N_c^{p+5}}{2^{p+2}}$ & $2p(p-3)$ & $(YZ)^{\frac{p-4}{2}}\Big[-\frac{[13][24][23]^2}{x_{13}^2 x_{24}^2 x_{23}^4}\Big]\tau_{14}(\rho_2^2\sigma_3^2+\rho_3^2\sigma_2^2)$ \\
        \hline
        23 & $-\frac{N_c^{p+5}}{2^{p+2}}$ & $2p(p-2)$ & $(YZ)^{\frac{p-2}{2}}\tau_{24}(\rho_1^2\sigma_3^2+\rho_3^2\sigma_1^2)$ \\
        \hline
        24 & $-\frac{N_c^{p+5}}{2^{p+2}}$ & $2p(p-2)$ & $(YZ)^{\frac{p-4}{2}}\Big[\frac{[13][23][14]^2}{x_{13}^2 x_{23}^2 x_{14}^4}\Big]\tau_{24}(\rho_1^2\sigma_3^2+\rho_3^2\sigma_1^2)$ \\
        \hline
        25 & $-\frac{N_c^{p+5}}{2^{p+2}}$ & $2p(p-2)$ & $(YZ)^{\frac{p-4}{2}}\left[\frac{[23][24][14]^2}{x_{23}^2 x_{24}^2 x_{14}^4}\right]\tau_{23}(\rho_1^2\sigma_4^2+\rho_4^2\sigma_1^2)$ \\
        \hline
        26 & $-\frac{N_c^{p+5}}{2^{p+2}}$ & $2p(p-3)$ & $(YZ)^{\frac{p-4}{2}}\left[-\frac{[13][24][14]^2}{x_{13}^2 x_{24}^2 x_{14}^4}\right]\tau_{23}(\rho_1^2\sigma_4^2+\rho_4^2\sigma_1^2)$ \\
        \hline
    \end{tabular}
    \caption{Color factors, symmetry factors and kinematic factors of Feynman diagrams that involve both three- and four-point face interactions.}
    \label{tab:my_label2}
\end{table}

The sum of all diagrams in Table~\ref{tab:my_label1}  and  Table~\ref{tab:my_label2} gives
     \begin{align}
            \frac{N_c^5 N_c!  (d_{12})^{N-p} X^{N-p-2}(YZ)^{\frac{p-2}{2}}}{x_{12}^2 x_{34}^2 x_{13}^2 x_{24}^2}&[\tau_{13}\tau_{24}+\tau_{14}\tau_{23}-\tau_{12}\tau_{34}+\tau_{13}(\rho_2^2\sigma_4^2+\rho_4^2\sigma_2^2)\label{two}\\
    &+\tau_{24}(\rho_1^2\sigma_3^2+\rho_3^2\sigma_1^2)
    -\tau_{12}(\rho_3^2\sigma_4^2+\rho_4^2\sigma_3^2)+\tau_{34}(\rho_1^2\sigma_2^2+\rho_2^2\sigma_1^2)\nonumber\\
    &+\tau_{14}(\rho_2^2\sigma_3^2+\rho_3^2\sigma_2^2)+\tau_{23}(\rho_1^2\sigma_4^2+\rho_4^2\sigma_1^2)].\nonumber
\end{align}
In the $\theta_5=\theta_6=0$ frame, (\ref{two}) simplifies to     
    \begin{align}
         N_c^5 N_c!\frac{(\theta_1^{+})^2(\theta_2^{+})^2(\theta_3^{+})^2(\theta_4^{+})^2x_{56}^2}{x_{12}^2x_{34}^2x_{13}^2x_{24}^2}(d_{12})^{N-p} X^{N-p}(YZ)^{\frac{p-2}{2}}\frac{P_1}{x_{15}^2x_{25}^2x_{35}^2x_{45}^2x_{16}^2x_{26}^2x_{36}^2x_{46}^2}\,,
\end{align}
where
\begin{align}   P_1&=x_{13}^2x_{24}^2x_{56}^2+x_{14}^2x_{23}^2x_{56}^2-x_{12}^2x_{34}^2x_{56}^2+x_{13}^2x_{25}^2x_{46}^2+x_{13}^2x_{26}^2x_{45}^2\label{P1}\\&+x_{24}^2x_{15}^2x_{36}^2+x_{24}^2x_{16}^2x_{35}^2
    -x_{12}^2x_{35}^2x_{46}^2-x_{12}^2x_{36}^2x_{45}^2+x_{34}^2x_{15}^2x_{26}^2\nonumber\\&
    +x_{34}^2x_{16}^2x_{25}^2
+x_{14}^2x_{25}^2x_{36}^2+x_{14}^2x_{26}^2x_{35}^2+x_{23}^2x_{15}^2x_{46}^2+x_{23}^2x_{16}^2x_{45}^2.\nonumber
\end{align}
    
Going back to the $\theta_r=\bar{\theta}_r=0~(r=1,\cdots,4)$ frame, we obtain
\begin{align}
 A_{[0,\frac{p-2}{2},\frac{p-2}{2}]}^{(2)}=&\,N_c^5 N_c!\theta_5^4\theta_6^4\mathcal{R}_{\mathcal{N}=2}(d_{12})^{N-p} X^{N-p}(YZ)^{\frac{p-2}{2}}\\\nonumber
 &\times\frac{P_1}{x_{15}^2x_{25}^2x_{35}^2x_{45}^2x_{16}^2x_{26}^2x_{36}^2x_{46}^2x_{56}^2}.
\end{align}

Notice that although individual diagram in the tables depends on $p$, the summation is independent of $p$. This property already hints at the 10-dimention hidden conformal symmetry that will be discussed in next section.

\begin{figure}
    \centering
    \includegraphics[width=0.55\linewidth]{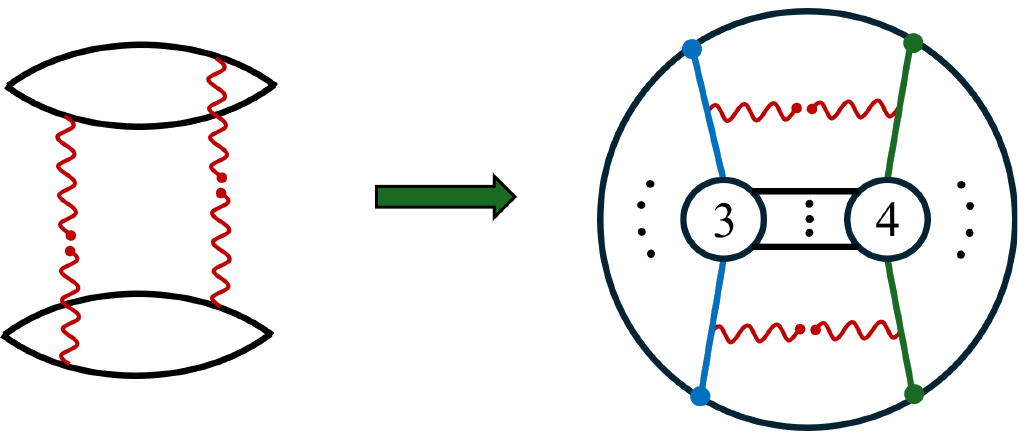}
    \caption{The Feynman diagrams that contribute to $x_{12}^2x_{34}^2x_{56}^2$.}
    \label{missingex}
\end{figure}

\begin{figure}
    \centering
    \includegraphics[width=0.25\linewidth]{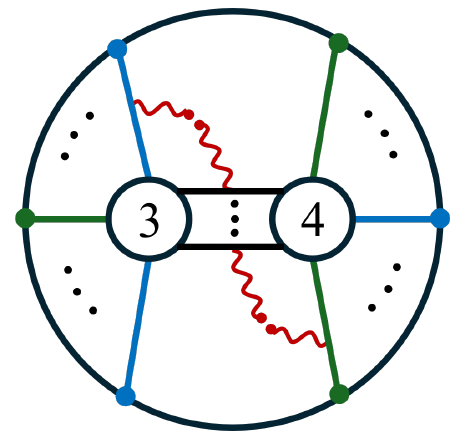}
    \caption{The Feynman diagrams that contribute to the missing terms.}
    \label{missingterm}
\end{figure}

For odd $p$, there is only one diagram contributing to the $A_{[0,\frac{p-3}{2},\frac{p-1}{2}]}^{(2)}$, which is similar to the $p=3$ case. Thus we can determine 
\begin{equation}
    A_{[0,\frac{p-3}{2},\frac{p-1}{2}]}^{(2)}\propto A_{[0,0,1]}^{(2)}.
\end{equation}
Using the lightcone OPE relation, the integrand of $\langle \mathcal{D}\mathcal{D}\mathcal{O}_p\mathcal{O}_p\rangle^{(2)}$ can be written as
\begin{align}     \langle\tilde{\mathcal{D}}\mathcal{D}\mathcal{O}_p\mathcal{O}_p\mathcal{L}\mathcal{L}\rangle=&\frac{\theta_5^4\theta_6^4\mathcal{R}_{\mathcal{N}=2}}{x_{15}^2x_{25}^2x_{35}^2x_{45}^2x_{16}^2x_{26}^2x_{36}^2x_{46}^2x_{56}^2} \Big[P_1 (d_{12})^{N-p} \sum_{m=0}^{\lfloor\frac{p-2}{2}\rfloor} X^{p-2-2 m}(Y Z)^m \\
& +P_2Y(d_{12})^{N-p} \sum_{m=0}^{\lfloor\frac{p-3}{2}\rfloor} X^{p-3-2 m}(Y Z)^m\nonumber \\
& +P_3 Z(d_{12})^{N-p} \sum_{m=0}^{\lfloor\frac{p-3}{2}\rfloor} X^{p-3-2 m}(Y Z)^m\nonumber\\
    &+\text{missing terms}\Big],\nonumber
\end{align}
where
\begin{align}
    P_2=x_{14}^2x_{23}^2x_{56}^2\,,\qquad
    P_3=x_{13}^2x_{24}^2x_{56}^2\,.
\end{align}
Finally we need to determine the missing terms.
{Notice that the missing terms come from the lightcone OPE relation
\begin{equation}
\label{eq:recursive}
    \begin{aligned}
       \mathcal{X}^l \mathcal{Y}^m \mathcal{Z}^n F_{[l,m,n]}=\mathcal{X}^{l} \mathcal{Y}^m \mathcal{Z}^n (F_{[l-1,m,n]}+\tilde{f}_{[l,m,n]}),
    \end{aligned}
\end{equation}
where $\tilde{f}_{[l,m,n]}$ only depends on cross ratios. Therefore each time when we apply the lightcone OPE relation as in \eqref{eq:recursive}, we generate a potential missing term $d_{12}^{N-p}\mathcal{X}^l \mathcal{Y}^m \mathcal{Z}^n\tilde{f}_{[l,m,n]}$. Further taking into account the restriction $|m-n|\le 1$ due to \textbf{Theorem 2}, we can write down the following ansatz
\begin{equation}
    \text{missing terms}=d_{12}^{N-p}\sum_{m=0}^{\lfloor\frac{p-3}{2}\rfloor} \mathcal{X}^{p-3-2 m}(\mathcal{Y}\mathcal{Z})^m(A_m \mathcal{X} +B_m \mathcal{Y}+C_m \mathcal{Z}),
\end{equation}
where $A_m$, $B_m$ and $C_m$ are functions depend on the coordinates $x_j$.}
According to (\ref{lcr}), such terms are proportional to $x_{34}^2$. To compensate for the conformal weight of the correlator, the {coefficients $A_m$, $B_m$ and $C_m$ should carry conformal weight one at each point, from $x_1$ to $x_6$. In addition, the missing terms should be symmetric under the exchange of $1$ and $2$, as well as $3$ and $4$. Meanwhile, these coefficients themselves are symmetric under the exchange of $3$ and $4$ because their dependence on $x_3$ and $x_4$ appears only through $x_{34}^2$. Therefore the most general ansatz for these coefficients are
\begin{equation}
\begin{aligned}
        &B_m=C_m=c_1x_{34}^2 (x_{15}^2x_{26}^2+x_{16}^2x_{25}^2)+c_2x_{12}^2x_{34}^2x_{56}^2,\\
        &A_m=c_3x_{34}^2 (x_{15}^2x_{26}^2+x_{16}^2x_{25}^2)+c_4x_{12}^2x_{34}^2x_{56}^2,
\end{aligned}
\end{equation}
where $c_i$ are constant numbers.} The terms with coefficient $c_2$ and $c_4$ can only come from the diagram in Figure~\ref{missingex}, which are excluded by the previous analysis of harmonic identification. So we can further simplify the ansatz to be
\begin{equation}
    \text{missing terms}=x_{34}^2 (x_{15}^2x_{26}^2+x_{16}^2x_{25}^2)d_{12}^{N-p}\sum_{m=0}^{\lfloor\frac{p-3}{2}\rfloor} \mathcal{X}^{p-3-2 m}(\mathcal{Y}\mathcal{Z})^m(a_m \mathcal{X} +b_m \mathcal{Y}+c_m \mathcal{Z}),
\end{equation}
where $a_m$, $b_m$ and $c_m$ are constants and $b_m=c_m$.
Since the factor $x_{34}^2(x_{15}^2x_{26}^2)$ can only come from the products of T-blocks, it is possible to specify the interacting subgraph that contributes this factor. In fact, all interacting subgraphs at two-loops have been listed in the Figure 7 in \cite{Arutyunov:2003ad} and we find that the only required subgraph is 
\begin{equation}
    \left(T_{145} T_{236}+(5 \leftrightarrow 6)\right) T_{345} T_{346}=\frac{\tau_{34}\left(\rho_1^2 \sigma_2^2+\rho_2^2 \sigma_1^2\right)}{x_{14}^2 x_{23}^2 x_{34}^4}.
\end{equation}
To determine $a_m$, $b_m$ and $c_m$, it is convenient to choose the projection $\langle \det \tilde{q}_1 \det q_2 \\  \text{tr}(\tilde{q}_3^{p-m} q_3^{m})\text{tr} (q_4^{p-m} \tilde{q}_4^{m})\rangle$. Among the diagrams listed in \cite{Arutyunov:2003ad}, only one of them contains the factor $x_{34}^2$ after taking into account planarity, this is the Feynman diagrams of Figure~\ref{missingterm}.
The contribution can be extracted from the complete result by identifying $u_3=u_4$ after factorizing out $[34]^m$ and this gives
\begin{equation}
    a_m=-1\,,\qquad b_m=c_m=0.
\end{equation}
Therefore the missing term is fixed to be
\begin{equation}
    \text{missing terms}=-(x_{34}^2x_{15}^2x_{26}^2+x_{34}^2x_{15}^2x_{26}^2)\sum_{m=0}^{\lfloor\frac{p-3}{2}\rfloor} \mathcal{X}^{p-2-2 m}(\mathcal{Y}\mathcal{Z})^m.
\end{equation}
After performing the integrals over the spacetime points $x_5,x_6$ and the Grassmann variables $\theta_5,\theta_6$ and uplifting the results to the $\mathcal{N}=4$ SYM theory, the two-loop result for generic $p$ is given by
\begin{align}
          G_{\{p,p\}}^{(2)}=&\tilde{\mathcal{R}}_{\mathcal{N}=4}d_{12}^{N-p}\Big[H(z,\bar{z}) \sum_{m=0}^{[\frac{p-2}{2}]} \mathcal{X}^{p-2-2m}(\mathcal{Y}\mathcal{Z})^m \label{DDpp2}\\
    &+ (1-z)(1-\bar{z}) \Big(F^{(1)}(z,\bar{z})\Big)^2 \mathcal{Y}  \sum_{m}^{[\frac{p-3}{2}]} \mathcal{X}^{p-3-2m}(\mathcal{Y}\mathcal{Z})^m\nonumber\\
    &+ \Big(F^{(1)}(z,\bar{z})\Big)^2 \mathcal{Z} \sum_{m=0}^{[\frac{p-3}{2}]} \mathcal{X}^{p-3-2m}(\mathcal{Y}\mathcal{Z})^m-2F_{1-z}^{(2)}\sum_{m=0}^{[\frac{p-3}{2}]} \mathcal{X}^{p-2-2 m}(\mathcal{Y}\mathcal{Z})^m\Big],\nonumber
\end{align}
where
\begin{equation}
    H(z,\bar{z})=(2-z-\bar{z})(F^{(1)}(z,\bar{z}))^2+4 F_{z}^{(2)}+4 F_{\frac{z}{1-z}}^{(2)}.
\end{equation}

\begin{figure}[htp]
    \centering
    \includegraphics[width=0.9\linewidth]{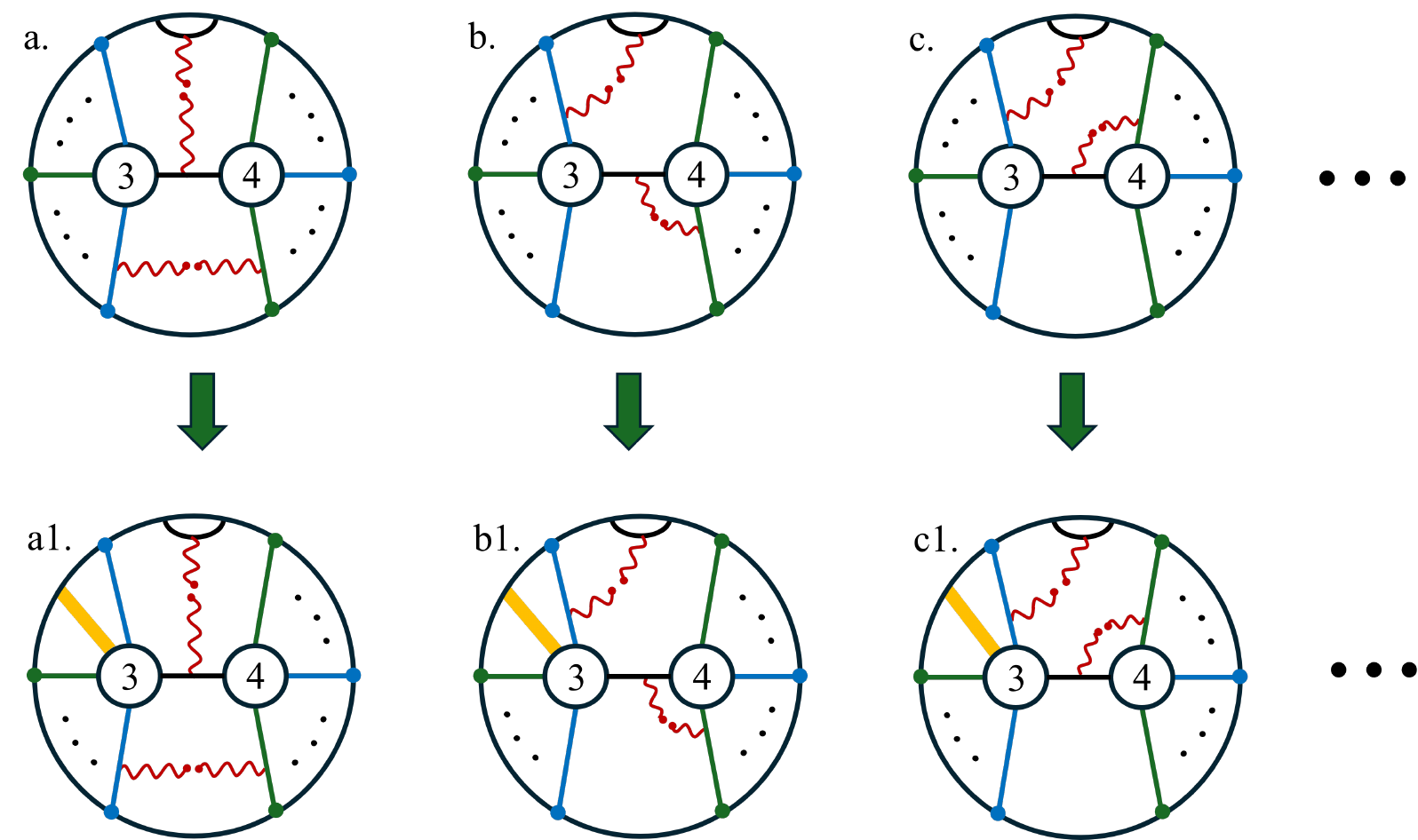}
    \caption{Two-loop Feynman diagram for the transition from $G_{\{q,q\}}^{(2)}$ to $G_{\{q+2k,q\}}^{(2)}$. The red wavy lines are loop corrections and the yellow thick lines stand for $2k$ free propagators.}
    \label{Genericpq2}
\end{figure}

\paragraph{Generalization to $G_{\{p,q\}}^{(2)}$}
The derivation of the two-loop contribution for $G_{\{p,q\}}^{(2)}$ is analogous to the one-loop case. As shown in Figure \ref{Genericpq2}, all diagrams contributing to the two-loop correction of $G_{\{p,q\}}^{(2)}$ differ from those of $G_{\{p,p\}}^{(2)}$ only in the free-propagator parts (highlighted in yellow), they introduce no additional corrections to the loop integrals. Consequently, the two-loop result for $G_{\{p,q\}}^{(2)}$ can be written as
\begin{align}
G_{\{p,q\}}^{(2)}= 
\begin{cases} 
(\frac{d_{13}d_{23}}{d_{12}})^{k} G_{\{q,q\}}^{(2)}, & (p-q=2k,~k\in \mathbb{N})\\
\\
(\frac{d_{14}d_{24}}{d_{12}})^{k} G_{\{p,p\}}^{(2)}, & (q-p=2k,~k\in \mathbb{N}) 
\end{cases} 
\end{align}
where $G_{\{p,p\}}^{(2)}$ is given in \eqref{DDpp2}.

\section{Hidden higher dimensional conformal symmetry}
\label{sec:hidden}
In \cite{Chen:2025yxg}, it was discovered that the giant graviton two-point function at strong coupling exhibits an intriguing hidden structure, enabling the calculation of all $\langle\mathcal{D}\mathcal{D}\mathcal{O}_p\mathcal{O}_q\rangle$ correlators from a single generating function constructed from $\langle\mathcal{D}\mathcal{D}\mathcal{O}_2\mathcal{O}_2\rangle$. This structure reflects a 10D hidden conformal symmetry, first observed in four-point functions of single-trace operators $\langle\mathcal{O}_{k_1}\mathcal{O}_{k_2}\mathcal{O}_{k_3}\mathcal{O}_{k_4}\rangle$ \cite{Caron-Huot:2018kta}. A natural question is whether this structure persists at weak coupling, or even at finite coupling?

For four-point functions of single-trace operators, this hidden 10D conformal symmetry also manifests at weak coupling, albeit in a slightly different form. Within the Lagrangian insertion approach, the integrand of $\langle\mathcal{O}_{k_1}\mathcal{O}_{k_2}\mathcal{O}_{k_3}\mathcal{O}_{k_4}\rangle$ at the first few loop orders can be obtained from a similar generating function based on the simplest four-point function, $\langle\mathcal{O}_{2}\mathcal{O}_{2}\mathcal{O}_{2}\mathcal{O}_{2}\rangle$ \cite{Caron-Huot:2021usw}. This observation suggests the possibility of an analogous structure in the present case at the integrand level. Indeed, we identify the hidden higher-dimensional symmetry in the integrand up to two-loop order. Specifically, the integrand of $\langle\mathcal{D}\mathcal{D}\mathcal{O}_{p}\mathcal{O}_{q}\rangle$ can be derived from generating functions constructed from $\langle\mathcal{D}\mathcal{D}\mathcal{O}_{2}\mathcal{O}_{2}\rangle$, employing the same replacement rules for $x_{ij}^2$ as in the strong coupling regime \cite{Chen:2025yxg}.

More explicitly, we define the generating function as follows:
\begin{align}
    \mathbf{I}^{(n)}(\lambda_1,\lambda_2)=\Big(\frac{x_{12}^6}{x_{13}^2 x_{14}^2 x_{23}^2 x_{24}^2x_{34}^2 }\times H_{\{2,2\}}^{(n)}\Big|_{x_{ij}^2\rightarrow \hat{x}_{ij}^2}\Big)\Big|_{Y_3\rightarrow\frac{\lambda_1}{\sqrt{2}}Y_3, Y_4\rightarrow\frac{\lambda_2}{\sqrt{2}}Y_4}\,,
 \label{gfunction}
\end{align}
where $x_{ij}^2\to \hat{x}_{ij}^2$ are the replacement rules which uplift specific combinations of distances in 4D to their counterparts in a higher dimensional space, whose explicit form will be given in \eqref{eq:replacementRule}. Here $H_{\{2,2\}}^{(n)}$ are related to the integrands of $G_{\{2,2\}}^{(n)}$ whose explicit forms are
\begin{align}
    H^{(1)}_{\{2,2\}}&=\frac{-1}{\pi^2 x_{15}^2x_{25}^2x_{35}^2x_{45}^2}\,,\\\nonumber
    H^{(2)}_{\{2,2\}}&=\frac{P_1}{\pi^4 x_{15}^2x_{25}^2x_{35}^2x_{45}^2x_{16}^2x_{26}^2x_{36}^2x_{46}^2x_{56}^2}\,.
\end{align}
where $P_1$ has been defined in (\ref{P1}).

\paragraph{Replacement rule} The replacement $x_{ij}^2 \rightarrow \hat{x}_{ij}^2$ uplifts distances in a 4D space to a higher dimensional space, combining the 6D embedding space vectors $P_A \in \mathbb{R}^{5,1}$
\begin{align}
    P_A=\left(\frac{1-x^2}{2},\frac{1+x^2}{2},\Vec{x}\right)
\end{align}
and the  6D null vectors $Y_j$ into the 12D framework. 
\begin{align}
Z_i=(P_{i,A},Y_{i,I}). \quad A=1,\dots,6,\quad I=1,\dots 6,
\end{align}
where the index $i$ labels distinct fields  .We now use the following replacement rules
\begin{align}
   P_{3}\cdot P_{4}&\to Z_{3}\cdot Z_{4},\\
   P_{3}\cdot\mathbb{N}\cdot P_{4}&\to Z_{3}\cdot(\mathbb{N}+\mathbb{M})\cdot Z_{4}
\end{align}
where $\mathbb{N}_{AB}$, $\mathbb{M}_{ij}$ are spacetime and $R$-symmetry projectors respectively
\begin{align}
    \mathbb{N}_{AB}&=\frac{P_{1,A}P_{2,B}+P_{2,A}P_{1,B}}{P_1\cdot P_2},\nonumber\\
    \mathbb{M}_{IJ}&=\delta_{IJ}-\frac{Y_{1,I}Y_{2,J}+Y_{2,I}Y_{1,J}}{Y_1\cdot Y_2}\,.
\end{align}
This rule  originates from the defect picture in the bulk. In AdS space, the giant gravitons can be seen as a one-dimensional defect, which breaks the symmetry from $SO(5,1)\times SO_R(6)$ down to $SO(4)\times SO_R(4)$. Here, $\mathbb{N}$ projects any vector onto the subspace spanned by $P_{1,A}$ and $P_{2,A}$, while $\mathbb{M}$ projects it onto the subspace orthogonal to  $Y_{1,I}$ and $Y_{2,I}$.
Explicitly,  the replacement rules can be expressed in terms of the variables $x_{ij}^2$ and $y_{ij}^2=Y_i\cdot Y_j$ as
\begin{align}
\label{eq:replacementRule}
    x_{34}^2 &\rightarrow x_{34}^2- 2y_{34}^2\,,\\\nonumber
    \frac{x_{13}^2x_{23}^2}{x_{12}^2}&\rightarrow \frac{x_{13}^2x_{23}^2}{x_{12}^2}+\frac{2 y_{13}^2y_{23}^2}{y_{12}^2}\,,\nonumber\\
    \frac{x_{14}^2x_{24}^2}{x_{12}^2}&\rightarrow \frac{x_{14}^2x_{24}^2}{x_{12}^2}+\frac{2 y_{14}^2y_{24}^2}{y_{12}^2}\,,\nonumber\\  \frac{x_{13}^2x_{24}^2+x_{14}^2x_{23}^2}{2x_{12}^2}&\rightarrow  \frac{x_{13}^2x_{24}^2+x_{14}^2x_{23}^2}{2x_{12}^2}+\frac{y_{13}^2y_{24}^2+y_{14}^2y_{23}^2-y_{12}^2y_{34}^2}{y_{12}^2}
\end{align}
with other $x_{ij}^2$ remain invariant. After performing the rescaling
\begin{align} 
Y_3\rightarrow\frac{\lambda_1}{\sqrt{2}}Y_3,\qquad  Y_4\rightarrow\frac{\lambda_2}{\sqrt{2}}Y_4, 
\end{align}
we obtain the generating function $\mathbf{I}^{(n)}(\lambda_1,\lambda_2)$. Expanding this function in $\lambda_1$, $\lambda_2$ yields
\begin{align}
\mathbf{I}^{(n)}(\lambda_1,\lambda_2)=\sum_{p,q=2}\mathcal{H}_{\{p,q\}}^{(n)}\lambda_1^{p-2}\lambda_2^{q-2}\,.
\end{align}  
    Here $\mathcal{H}_{\{p,q\}}^{(n)}$ are the integrand of $G_{\{p,q\}}^{(n)}$. This construction allows us to systematically derive the integrands for all $G_{\{p,q\}}^{(n)}$.
\begin{align}
G_{\{p,q\}}^{(n)}=\mathcal{R}_{\mathcal{N}=4}\prod_{i=1}^n\int dx_{4+i}^4\frac{x_{13}^2x_{14}^2x_{23}^2x_{24}^2x_{34}^2}{x_{12}^6}\mathcal{H}_{\{p,q\}}^{(n)}.
\end{align}
We present the explicit forms of $\mathbf{I}^{(n)}(\lambda_1,\lambda_2)$ for $n=1,2$.

\paragraph{One-loop}
At one loop, the generating function takes the explicit form
\begin{align}
  \mathbf{I}^{(1)}(\lambda_1,\lambda_2)=-\frac{(\frac{x_{13}^2x_{23}^2}{x_{12}^2}+\frac{\lambda_1^2y_{13}^2y_{23}^2}{y_{12}^2})^{-1}(\frac{x_{14}^2x_{24}^2}{x_{12}^2}+\frac{\lambda_2^2 y_{14}^2y_{24}^2}{y_{12}^2})^{-1}x_{12}^2}{\pi^2(x_{34}^2-\lambda_1\lambda_2y_{34}^2)}\times\frac{1}{x_{15}^2x_{25}^2x_{35}^2x_{45}^2}.
\end{align}
It is straightforward to verify that for $p-q=2k+1$ (assuming $p>q$), the coefficient of $\lambda_1^{q+2k-1}\lambda_2^{q-2}$ is vanishing; for $p-q=2k$, the coefficient of the $\lambda_1^{q+2k-2}\lambda_2^{q-2}$ is exactly the integrands of $G_{\{q+2k,q\}}^{(1)}$,  as confirmed by explicit calculations.

\paragraph{Two-loop}
At two loops, the generating function is given by
\begin{align}
    \mathbf{I}^{(2)}(\lambda_1,\lambda_2)=&-\frac{(\frac{x_{13}^2x_{23}^2}{x_{12}^2}+\frac{\lambda_1^2y_{13}^2y_{23}^2}{y_{12}^2})^{-1}(\frac{x_{14}^2x_{24}^2}{x_{12}^2}+\frac{\lambda_2^2y_{14}^2y_{24}^2}{y_{12}^2})^{-1}x_{12}^2}{(x_{34}^2-\lambda_1\lambda_2y_{34}^2)}\\
    &\times\frac{\mathbf{P}_1(\lambda_1,\lambda_2)}{\pi^4 x_{15}^2x_{25}^2x_{35}^2x_{45}^2x_{16}^2x_{26}^2x_{36}^2x_{46}^2x_{56}^2}\nonumber
\end{align}
where 
\begin{align}  \mathbf{P}_1(\lambda_1,\lambda_2)=&2x_{12}^2x_{56}^2\Big(\frac{x_{13}^2x_{24}^2}{2x_{12}^2}+\frac{x_{14}^2x_{23}^2}{2x_{12}^2}+\lambda_1\lambda_2\frac{y_{13}^2y_{24}^2+y_{14}^2y_{23}^2-y_{12}^2y_{34}^2}{2y_{12}^2}\Big)\\
&-x_{12}^2(x_{34}^2-\lambda_1\lambda_2 y_{34}^2)x_{56}^2
    +x_{13}^2x_{25}^2x_{46}^2
+x_{13}^2x_{26}^2x_{45}^2\nonumber\\
&+x_{24}^2x_{15}^2x_{36}^2+x_{24}^2x_{16}^2x_{35}^2+x_{14}^2x_{25}^2x_{36}^2+x_{14}^2x_{26}^2x_{35}^2\nonumber\\
&+x_{23}^2x_{15}^2x_{46}^2+x_{23}^2x_{16}^2x_{45}^2 -x_{12}^2x_{35}^2x_{46}^2-x_{12}^2x_{36}^2x_{45}^2\nonumber\\
&+(x_{34}^2-\lambda_1\lambda_2 y_{34}^2)(x_{15}^2x_{26}^2+x_{16}^2x_{25}^2)\nonumber
\end{align}
Again, we observe that for $p-q=2k+1$, the coefficient of the $\lambda_1^{q+2k-1}\lambda_2^{q-2}$ is vanishing while for $p-q=2k$, the coefficient of $\lambda_1^{q+2k-2}\lambda_2^{q-2}$ coincides with the integrand of $G_{\{q+2k,q\}}^{(2)}$.\par

It is natural to conjecture that the same structure holds at even higher loop orders, though explicit verification would require computations beyond two-loop order and we leave it for future investigations.

\section{Conformal data and integrability check}
\label{sec:integrbility}

In this section we test the two-loop result $\langle\mathcal{D}\mathcal{D}\mathcal{O}_p\mathcal{O}_p\rangle^{(2)}$ with integrability predictions. 

\subsection{Sum rules from conformal block decomposition}
We firstly express the two-loop result in a form which is more suitable for obtaining the OPE data
\begin{align}
    G_{\{p,p\}}^{(2)}=\tilde{\mathcal{R}}_{\mathcal{N}=4}d_{12}^{N-2}d_{34}^{p-2}P^{(2)}(u,v;\sigma,\tau),
\end{align}
where
\begin{align}
    P^{(2)}=&\,
        \Bigg[H(z,\bar{z}) \sum_{m=0}^{[\frac{p-2}{2}]} \left(\sigma u\frac{\tau u}{v}\right)^m 
    + (1-z)(1-\bar{z}) \Big(F^{(1)}(z,\bar{z})\Big)^2 \sigma u  \sum_{m}^{[\frac{p-3}{2}]}\left(\sigma u\frac{\tau u}{v}\right)^m\label{NNpp1reduced}\\
    &+ \Big(F^{(1)}(z,\bar{z})\Big)^2 \frac{\tau u}{v} \sum_{m=0}^{[\frac{p-3}{2}]} \left(\sigma u\frac{\tau u}{v}\right)^m-2F_{1-z}^{(2)}\sum_{m=0}^{[\frac{p-3}{2}]} \left(\sigma u\frac{\tau u}{v}\right)^m\Bigg],\nonumber
\end{align}
where $u,v$ and $\sigma,\tau$ are the conformal and harmonic cross ratios defined in \eqref{uv} and \eqref{sigmtau} respectively. We have use the relations
    $\mathcal{Y}/\mathcal{X}=\sigma u$ and $\mathcal{Z}/\mathcal{X}={\tau u}/{v}$
to write $\mathcal{X},\mathcal{Y},\mathcal{Z}$ in terms of the cross ratios.
To extract OPE data, we write the correlation function as \cite{Chicherin:2015edu}
\begin{align}
G_{\{p,p\}}=G_{\{p,p\}}^{(0)}+ C_{NNpp}d_{12}^{N}d_{34}^{p}\mathcal{S}(u,v;\sigma,\tau)\mathcal{H}(u,v;\sigma,\tau)\label{NNpp2}
\end{align}
where $\mathcal{S}(u,v;\sigma,\tau)$ is a polynomial of the cross ratios and $C_{NNpp}$ is a normalization factor.
\begin{align}
    \mathcal{S}(u,v;\sigma,\tau)&=\mathcal{R}_{\mathcal{N}=4}\frac{x_{12}^2 x_{34}^2 x_{14}^2 x_{23}^2}{x_{13}^2 x_{24}^2 y_{12}^4 y_{34}^4}\\
    &=v+\sigma^2 u v+\sigma v(v-1-u)+\tau(1-u-v)+\sigma\tau(u-1-v).\nonumber
\end{align}
The dynamical information is encoded in $\mathcal{H}(u, v; \sigma, \tau)$. Comparing (\ref{NNpp1reduced}) and (\ref{NNpp2}), we find
\begin{align}
    P^{(2)}(u,v;\sigma,\tau)&=\frac{x_{12}^{2}x_{34}^{2}x_{14}^{2}x_{23}^{2}}{y_{12}^{4}y_{34}^{4}}d_{12}^{2}d_{34}^{2}\mathcal{H}(u,v;\sigma,\tau)\Big|_{\mathcal{O}(g^4)}=\frac{v}{u}\mathcal{H}(u,v;\sigma,\tau)\Big|_{\mathcal{O}(g^4)}.\label{NNpp3}
\end{align}
To proceed, we expand the function $\mathcal{H}$ into eigenfunctions of the $SU(4)$ Casimir operator 
\begin{align}
    \mathcal{H}(u,v;\sigma,\tau)=\sum_{0\leq m<n\leq p-2}A_{n,m}(u,v)Y_{nm}(\sigma,\tau)
\end{align}
where the channel with indices $n, m$ corresponds to an exchanged supermultiplet in the $SU(4)$ representation with Dynkin label $[n-m,2m,n-m]$. The functions $Y_{nm}(\sigma,\tau)$ are given in terms of Jacobi polynomials
\begin{align}
    Y_{nm}(\sigma,\tau)=\frac{2(m!)^2((n+1)!)^2}{(2m)!(2n+2)!}\frac{P_{n+1}^{(0,0)}(y)P_m^{(0,0)}(\bar{y})-P_m^{(0,0)}(y)P_{n+1}^{0,0}(\bar{y})}{y-\bar{y}}\,,
\end{align}
where the variables \( y \) and \( \bar{y} \) are defined as
\begin{align}
    \sigma=\frac{1}{4}(1+y)(1+\bar{y}),\quad \tau=\frac{1}{4}(1-y)(1-\bar{y}).
\end{align}

We recall that the Jacobi polynomials $P_n^{(\alpha,\beta)}$ are defined by a finite hypergeometric series, which can be expressed using Rodrigues' formula as
\begin{align}
    P_n^{(\alpha,\beta)}(z)=\frac{(-1)^n}{2^nn!}(1-z)^{-\alpha}(1+z)^{-\beta}\frac{d^n}{dz^n}\left[(1-z)^\alpha(1+z)^\beta(1-z^2)^n\right].
\end{align}


The function $A_{n,m}$ can be further expanded in terms of conformal blocks that encode loop corrections from conformal operators of conformal dimension $\Delta$ and spin $S$. Explicitly, we write
\begin{align}
A_{nm}^{(0)}(u,v)+A_{nm}(u,v)=\sum_{\Delta,S}C_{nm}^{\Delta ,S}G_{\Delta}^{(S)}(u,v).
\end{align}
The conformal blocks $G_{\Delta}^{(S)}(u,v)$ have been defined in \cite{Dolan:2000ut,Nirschl:2004pa,Bissi:2015qoa,Doobary:2015gia}
\begin{align}
    G_{\Delta}^{(S)}(u,v)=\frac{u^{\frac{1}{2}(\Delta-S)}}{z-\bar{z}}\left(z\left(-z\right)^{S}f_{\Delta+S}(z)f_{\Delta-S-2}(\bar{z})-\bar{z}\left(-\bar{z}\right)^S f_{\Delta+S}(\bar{z})f_{\Delta-S-2}(z)\right),
\end{align}
with
\begin{align}
    f_\rho(z)={}_2F_1\left(\frac{1}{2}\rho,\frac{1}{2}\rho;\rho;z\right).
\end{align}
The conformal dimension of the superconformal primary can be written as  $\Delta=\Delta_0+\delta\Delta$, where $\delta\Delta$ is the anomalous dimension. In what follows, we will label an operator by its twist  $L=\Delta_0-S$ and spin. Correspondingly, we denote the anomalous dimension and the OPE coefficients of the operator with twist $L$ and spin $S$ by $\delta \Delta_{nm,L,S}$ and $C_{nm,L,S}$ respectively.

In order to obtain the conformal data, we take the OPE limit $z\rightarrow0, \bar{z}\rightarrow 0$ which corresponds to the small $u$ expansion of the conformal blocks. We expand the $A_{nm}$, $\delta\Delta$, $C_{nm}^{L,S}$  perturbatively
\begin{align*}
    A_{nm}(u,v)=\sum_{a=1}^{\infty}g^{2a} A_{nm}^{(a)}(u,v),\quad\delta\Delta_{nm,L,S}=\sum_{a=1}^{\infty}g^{2a}\delta\Delta^{(a)}_{nm,L,S},\quad C_{nm}^{L,S}=\sum_{r=1}^{\infty}g^{2a}C_{nm,L,S}^{(a)}.
\end{align*}
The perturbative expansion of $\mathcal{H}(u,v;\tau,\sigma)$ reads
\begin{align}
\label{NNpp4}
\mathcal{H}(u,v;\tau,\sigma)=&\, \sum_{a=0}^{\infty}g^{2a}A_{nm}^{(a)}Y_{nm}(\sigma,\tau)\\\nonumber
=&\,\sum_{a=0}^\infty g^{2a}\sum_{L=1+n}^{2p-2}\sum_{b=0}^a\sum_{S=0}^\infty\mathcal{P}_{nm,L,S}^{(a,b)}f_{L,S}^{(b)}(z,\bar{z},\mu)Y_{nm}(\sigma,\tau),
\end{align}
where $
\mu=\frac{1}{2}\log(z\bar{z})$ and the function $f_{L,S}^{(b)}$ is a linear combination of the derivatives of the conformal blocks,
\begin{align}
    f_{L,S}^{(b)}(z,\bar{z},\mu)=\frac{\partial^b}{\partial^b\gamma}\Big[e^{\gamma\mu}G_{L+\gamma}^{(S)}(u,v)\Big]\Bigg|_{\gamma\rightarrow0}.
\end{align}
In this paper, we only focus on twist\footnote{At twist $L = 2p$, double-trace operators mix into the OPE, making it difficult to isolate the single-trace contribution against which integrability can be tested.} $L\leq2p-2$. We call the numbers $\mathcal{P}_{nm,L,S}^{(a,b)}$, which can be extracted by  reading off the coefficient of $g^{2a}\mu^{b}$, the sum rules. Up to two loops, the relations among the sum rules $P_{nm,L,S}^{(a,b)}$, structure constants $C_{nm,L,S}$ and anomalous dimensions $\delta\Delta_{nm,L,S}$ are given in \eqref{eq:PCD}, where for brevity we omit the $SU(4)$ indices $n$, $m$ and the index $I$ labels the set of operators with the same twist $L$ and spin $S$
\begin{align}
\label{eq:PCD}    
\mathcal{P}_{L,S}^{(0,0)}&=\sum_{I}\left(C_{L,S,I}^{(0)}\right)^2,\\\nonumber
\mathcal{P}_{L,S}^{(1,0)}&=\sum_{I}2C_{L,S,I}^{(0)}C_{L,S,I}^{(1)},\\\nonumber    
\mathcal{P}_{L,S}^{(1,1)}&=\sum_{I}\delta\Delta_{L,S,I}^{(1)}\left(C_{L,S,I}^{(0)}\right)^2,\\\nonumber 
\mathcal{P}_{L,S}^{(2,0)}&=\sum_{I}\left[\left(C_{L,S,I}^{(1)}\right)^2+2C_{L,S,I}^{(0)}C_{L,S,I}^{(2)}\right],\\\nonumber   
\mathcal{P}_{L,S}^{(2,1)}&=\sum_{I}\left[2\delta\Delta_{L,S,I}^{(1)}C_{L,S,I}^{(1)}C_{L,S,I}^{(0)}+\delta\Delta_{L,S,I}^{(2)}\left(C_{L,S,I}^{(0)}\right)^2\right],\\\nonumber
\mathcal{P}_{L,S}^{(2,2)}&=\sum_{I}\frac{1}{2}\left(\delta\Delta_{L,S,I}^{(1)}\right)^2\left(C_{L,S,I}^{(0)}\right)^2.
\end{align}

To extract the sum rules, we use the relations \eqref{NNpp3} and \eqref{NNpp4}, expand $P^{(2)}$ and $f_{L,S}^{(b)}$ in $z$ and $\bar {z}$, then read off the coefficients of $\{z,\bar {z},\mu\}$. The results are collected in Appendix~\ref{sumruledata}.

\subsection{Sum rules from integrability}
 \begin{figure}[h!]
    \centering
    \includegraphics[width=0.6\linewidth]{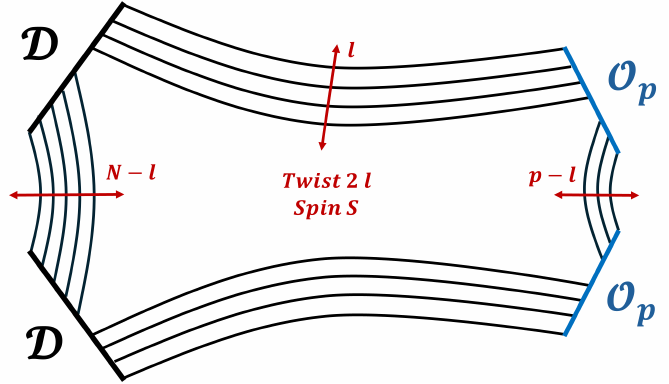}
    \caption{$\mathcal{D}$ and  $\mathcal{O}_p$  are all BPS operators, but the operators propagating between giant gravitons and single-trace operators are generally non-protected operators. If $l$ is the the number of Wick contractions between $\mathcal{D}$ and $\mathcal{O}_p$, the twist of the corresponding non-BPS operators is $2 l$.}
    \label{sumrulep}
\end{figure}

In this subsection, we compute the sum rule $\mathcal{P}_{nm,L,S}^{(a,b)}$ in the $SL(2)$ sector up to twist $6$ at two-loop order. The two-loop field theoretic results also give predictions for the sum rules beyond $SL(2)$ sectors. For the purpose of the current work, the $SL(2)$ predictions already provide a stringent test of our two-loop results. As $SU(4)$ indices satisfy  $n = m$ in the $SL(2)$ sector, we will omit these indices in the following subsection. 

In the planar $\mathcal{N}=4$ SYM theory, integrability results for conformal data typically consists of an asymptotic part and wrapping corrections, schematically \cite{Minahan:2002ve,Bena:2003wd,Minahan:2010js,Beisert:2010jr} 
\begin{align}    
\textbf{(Observables)}=\textbf{(Asymptotic part)}+\textbf{(Wrapping correction)}.
\end{align}
\paragraph{Asymptotic part} We first consider the asymptotic part of $SL(2)$ sector. The $SL(2)$ sector consists of a complex scalar field
and a  covariant derivative. It is the simplest non-compact subsector. The operators in the subsector are of type 
\begin{align}
    \hat{\mathcal{O}}_{L,S}=\text{tr}(D^SZ^L)+\cdots
\end{align}
where the twist $L$ counts the number of complex scalar fields $Z$, while the spin $S$ counts the number of covariant derivatives $D$.

In the $SL(2)$ sector, the asymptotic results for OPE data are given in terms of Bethe roots $\mathbf{u}=\{u_1,\dots,u_S\}$
\begin{align}
\label{sumrule}
\sum_{a=0}^{\infty}g^{2a}\sum_{b=0}^{a}\mu^b~\mathcal{P}_{L,S}^{(a,b)}=
\sum_{\mathbf{u}}\mathsf{d}_{L,S}(\mathbf{u})\mathsf{c}_{L,S}(\mathbf{u})e^{\delta\Delta_{L,S}(\mathbf{u})\,\mu}\,,
\end{align}
where $\mathsf{d}_{L,S}$ and $\mathsf{c}_{L,S}$ are the structure constants  between two BPS operators and a non-BPS operator,
\begin{align}
\langle\mathcal{D}(x_1)\mathcal{D}(x_2)\hat{\mathcal{O}}_{L,S}(x_3)\rangle&=\mathfrak{n}_{\mathcal{D}}\hat{\mathfrak{n}}_{L}^{1/2}\times\mathsf{d}_{L,S}\times d_{12}^{N-L/2}d_{13}^{L/2}d_{23}^{L/2},\\\nonumber 
\langle\mathcal{O}_p(x_1)\mathcal{O}_p(x_2)\hat{\mathcal{O}}_{L,S}(x_3)\rangle&=\mathfrak{n}_{p}\hat{\mathfrak{n}}_{L}^{1/2}\times\mathsf{c}_{L,S}\times d_{12}^{p-L/2}d_{13}^{L/2}d_{23}^{L/2},
\end{align}
where the terms $\mathfrak{n}_{\mathcal{D}}$,  $\mathfrak{n}_{p}$ and $\hat{\mathfrak{n}}_{L}$ represent the normalization of the operators. The asymptotic formulas of $\mathsf{d}_{L,S}$, $\mathsf{c}_{L,S}$ and $\delta\Delta_{L,S}$ are given in Appendix~\ref{Asymptotic formula}.

The Bethe roots are solutions of the asymptotic Bethe ansatz equations (BAE) and the zero momentum condition
\begin{align}  
    e^{i p(u_j) L} \prod_{k \neq j}^{S} S(u_j, u_k) = 1, \qquad \prod_{j=1}^{S} e^{i p(u_j) L} = 1, \label{asyBAE}  
\end{align}  
where the momentum \( p(u) \) and the $S$-matrix  are parametrized by the Zhukovsky variable \( x(u) \)
\begin{align}  
    x(u) = \frac{u + \sqrt{u^2 - 4g^2}}{2g}, \qquad x^{\pm}_j = x\left(u_j \pm \frac{i}{2}\right).  
\end{align}  
Expressed in terms of these variables, the momentum and $S$-matrix read
\begin{align}  
    e^{i p(u_j)} = \frac{x^{+}_j}{x^{-}_j}, \quad S(u_j, u_k) = \frac{u_j - u_k + i}{u_j - u_k - i} \left[ \frac{1 - \frac{1}{x_j^- x_k^+}}{1 - \frac{1}{x_j^+ x_k^-}} \right]^2 \sigma^2(u_j, u_k)\,,
\end{align}  
where \( \sigma^2(u, v) \) is the BES dressing phase, which only start to contribute at three-loop order and can be neglected in this work. 
To perform perturbative checks, we need to solve the asymptotic BAE \eqref{asyBAE} perturbatively, we first expand the Zhukovsky variable in powers of $g$:
\begin{align}
    x(u)=\frac{u}{g}-\frac{g}{u}-\frac{g^3}{u^3}+\mathcal{O}(g^5)
\end{align}
and write the Bethe roots in a perturbative ansatz
\begin{align}
    u_j=u_j^{(0)}+g^2u_j^{(1)}+g^4 u_j^{(2)}+\mathcal{O}(g^6)\,.
\end{align}
Substituting these expansions into the asymptotic BAE \eqref{asyBAE}, we obtain the following structure
\begin{itemize}
    \item $u_j^{(0)}$ satisfy the leading-order BAE;
    \item $u_j^{(1)}$ are determined by one-loop BAE, which depends on $u_j^{(0)}$;
    \item $u_j^{(2)}$ are determined by two-loop BAE involving both $u_j^{(0)}$ and $u_j^{(1)}$\,.
\end{itemize}
This structure allows an iterative determination of the perturbative solutions $u_j$. We substitute these perturbative solution into the physical quantities, expand in $g$ and $\mu$, and then read off the corresponding coefficients to obtain the sum rule $\mathcal{P}_{L,S}^{(a,b)}$. When doing so, there are two  points that one has to take into account.
\begin{enumerate}
\item \textbf{Selection rule}: The structure constant $\mathsf{d}_{L,S}(\mathbf{u})$ is nonzero only when the Bethe roots are  parity symmetric, \emph{i.e.} $\mathbf{u}=\{u_1,-{u}_1,\dots,u_{S/2},-{u}_{S/2}\}$. 
\item \textbf{Sign ambiguity}: The asymptotic formulas for $\mathsf{d}_{L,S}$ and $\mathsf{c}_{L,S}$ contain square roots which may cause sign ambiguity in the calculation. It is therefore better to use an alternative expression for $\mathsf{d}_{L,S}\mathsf{c}_{L,S}$ (See \ref{dc} in Appendix \ref{Asymptotic formula}) which is unambiguous.
\end{enumerate}
For example, when $L = 4$ and spin $S = 4$, the asymptotic BAE \eqref{asyBAE} admit five solutions, three of which respect parity symmetry. One such solution is given by
\begin{align}
   u_1&= 1.3614692731543783275+2.49683037536767001 g^2-5.025233398579724 g^4,\nonumber\\
   u_2&=0.57684315550370651842+2.37408583427923996 g^2-5.342414619692888 g^4,\nonumber\\
   u_3&=-0.57684315550370651842-2.37408583427923996g^2+ 5.342414619692888 g^4,\nonumber\\
   u_4&=-1.3614692731543783275-2.49683037536767001 g^2+ 5.025233398579724 g^4.\nonumber
\end{align}
Plugging the three parity symmetric Bethe roots into the asymptotic formula and summing the contributions,
we obtain
\begin{align}
  \sum_{\mathbf{u}}\mathsf{d}_{L,S}(\mathbf{u})\mathsf{c}_{L,S}(\mathbf{u})e^{\delta\Delta_{L,S}(\mathbf{u})\,\mu}= -\frac{1}{63}+\frac{1549}{1458}g^4-\frac{176}{81}g^4\mu+ \frac{10 }{9}g^4\mu^2+\mathcal{O}(g^6)\,,
\end{align}
which agrees with the result of conformal block expansion. The remaining $SL(2)$ asymptotic sum rules, presented in Appendix \ref{sumruledata} also agrees perfectly with the field theory results.

\paragraph{Wrapping corrections} In some cases, we need to consider the wrapping correction already at two-loop order. As illustrated in Figure \ref{sumrulep}, we refer to the OPE channel between $\mathcal{D}$ and $\mathcal{O}_p$ as the adjacent channel, with twist $L=2l$, while the channel connecting the two $\mathcal{D}$ and two $\mathcal{O}_p$ are called the opposite channels. For the giant graviton OPE coefficient $\mathcal{D}$, the wrapping correction only start to contribute from three-loop order, therefore here it is sufficient to use the asymptotic formula \eqref{sumrule} in this work. For the OPE coefficeint of the single-trace operators, when $p-l$ is small, opposite wrapping corrections between the two $\mathcal{O}_p$ operators need to be taken into account, whereas for small $l$ we must consider also adjacent wrapping corrections. Adjacent wrapping corrections first appear at three loops, so up to two-loop we only need to consider the opposing wrapping corrections.\par 

The opposite wrapping correction $\delta(\mathsf{c}_{L,S})^2$ to the structure constant $(\mathsf{c}_{L,S})^2$ must be considered when $\ell = p-l = 1$ at two-loop order. Let us consider $p=2,~l=1$,  the opposite wrapping corrections have been computed in  \cite{Basso:2015zoa,Basso:2015eqa} and we quote here in Table~\ref{tab:botWrap}.
\begin{table}[htbp]
\centering
\renewcommand{\arraystretch}{2}  
    \setlength{\tabcolsep}{6pt}
\begin{tabular}{c|c|c|c|c p{10cm}<{}}
\hline
Spin $S$ & 2 & 4 & 6 & 8 \\
\hline
$\delta(\mathsf{c}_{L,S})^2$   & $12\zeta(3)$  & $\frac{1}{6}+\frac{10\zeta(3)}{7}$   &  $\frac{199}{7920}+\frac{7\zeta(3)}{55}$  & $\frac{1721}{655200}+\frac{761\zeta(3)}{75075}$   \\
\hline
\end{tabular}
\caption{The opposite wrapping corrections for $p=2,~l=1$}
\label{tab:botWrap}
\end{table}
Including these corrections, we obtain the complete results of $SL(2)$ sum rules for $\langle\mathcal{D}\mathcal{D}\mathcal{O}_2\mathcal{O}_2\rangle$ (see Table \ref{P002wrap} in Appendix~\ref{sumruledata}), which match the results obtained from the conformal block expansion.

\section{Conclusions and discussions}
\label{sec:conclude}

In this work, we computed the four-point functions involving two maximal giant gravitons and two arbitrary-length single-trace half-BPS operators up to two-loop order. This is achieved using the harmonic PCGG method, a technique combining harmonic superspace with partially contracted giant graviton approach. This approach drastically reduces the number of Feynman diagrams at each order, enabling an efficient evaluation of the correlation functions.\par

Our one- and two-loop results reveal an intriguing structure at the integrand level, which we interpret as a defect analogue of higher-dimensional conformal symmetry. This structure was first identified in four-point functions of single-trace operators and has recently been observed at strong coupling for giant graviton correlators. The conformal data extracted from our four-point function calculation are in perfect agreement with integrability predictions.\par

This study opens several avenues for future research. The harmonic PCGG method can be applied to the four-point functions of maximal giant gravitons. While the one-loop result was previously obtained using standard PCGG methods \cite{Jiang:2019xdz}, our harmonic formalism simplifies the calculation and paves the way for a tractable two-loop computation, which we will present in a forthcoming publication.\par

A natural extension is to correlators involving non-maximal giant gravitons (with a smaller size in the internal space) and dual giant gravitons (extended in AdS). We anticipate that a tree-level generalized PCGG, combined with the Lagrangian insertion method and the $\mathcal{N}=2$ harmonic superspace formalism, can efficiently handle these computations. This would elucidate the size dependence of giant correlators and provide valuable insights for strong-coupling analyses.\par

The demonstrated efficiency of the harmonic PCGG method at one and two loops invites its application to higher orders, beginning with three loops. A direct three-loop calculation of the correlator with two length-2 single-trace operators would provide a crucial cross-check of an existing bootstrap result \cite{Jiang:2023uut} and offer a stringent test of the proposed higher-dimensional conformal symmetry at higher perturbative orders.

\acknowledgments
We are indebted to Junding Chen and Xinan Zhou for very helpful discussions, in particular on hidden higher dimensional symmetry. The work of Y.J. is supported by Startup Fund No.4007022326 of Southeast University and National Natural Science Foundation of China through Grant No.12575073 and 12247103.  YZ is supported by National Natural Science Foundation of China through Grant No. 12575078 and
12247103.

\appendix

\section{Explicit examples for $\mathcal{N}=2$ reduction}
\label{app:reduce}
In this appendix, we provide more details on $\mathcal{N}=2$ reduction along with some explicit examples.\par

\subsection{Pure projection}
The projection we choose depends on the channel to be recovered. The simplest cases are the channels of $\mathcal{X}^m \mathcal{Y}^n$ and $\mathcal{X}^m \mathcal{Z}^n$. In such cases, pure projection which corresponds to $\langle\det(\tilde{q}_1)\det(q_2) \text{tr}(\tilde{q}_3^p)\text{tr}(q_4^p)\rangle$, is enough to recover the complete information of the $\mathcal{N}=4$ SYM correlator. Therefore, for $p=2\,,3$, it is sufficient to obtain the full correlator through pure projection. Below we present examples which are mentioned in the main text. 
\paragraph{$p=2$}
For $p=2$, the loop correction of the correlator is formulated as
\begin{equation}
    G_{\{2,2\}}=\mathcal{R}_{\mathcal{N}=4} d_{12}^{N-2}F_{[0,0,0]}.
\end{equation}

Replacing those $y_{ij}^2$ in $\mathcal{R}_{\mathcal{N}=4}$ and the factor $d_{12}^{N-2}$ by $[i\bar{j}]+[\bar{i}j]$, we further pick up the channels without $[1\bar{j}],\,[3\bar{j}],\,[\bar{2}j],\,[\bar{4}j]$ for arbitrary $j$. Following this, the $\mathcal{N}=2$ reduced correlator is 
\begin{equation}
   G_{\{2,2\}}|_{\text{pure projection}}=\mathcal{R}_{\mathcal{N}=2}\left(\frac{[12]}{x_{12}^2}\right)^{N-2}F_{[0,0,0]},
\end{equation}
while
\begin{equation}
        \langle\det(\tilde{q}_1)\det(q_2)\text{tr}(\tilde{q}_3^2)\text{tr}(q_4^2)\rangle=\mathcal{R}_{\mathcal{N}=2}\left(\frac{[12]}{x_{12}^2}\right)^{N-2}f_{[0,0,0]}.
\end{equation}
Thus we can determine that
\begin{equation}
    F_{[0,0,0]}=f_{[0,0,0]}.
\end{equation}

\paragraph{$p=3$}
For $p=3$, the $\mathcal{N}=4$ SYM correlator reads
\begin{equation}
    G_{\{3,3\}}=\mathcal{R}_{\mathcal{N}=4}d_{12}^{N-3}(\mathcal{X} F_{[1,0,0]}+\mathcal{Y}F_{[0,1,0]}+\mathcal{Z}F_{[0,0,1]}),
\end{equation}
After pure projection, the reduced correlator gives
\begin{equation}
    G_{\{3,3\}}|_{\text{pure projection}}=\mathcal{R}_{\mathcal{N}=2}(X F_{[1,0,0]}+Z F_{[0,0,1]}),
\end{equation}
while
\begin{equation}
         \langle\det(\tilde{q}_1)\det(q_2)\text{tr}(\tilde{q}_3^3) \text{tr}(q_4^3)\rangle=\mathcal{R}_{\mathcal{N}=2}(X f_{[1,0,0]}+Z f_{[0,0,1]}).
\end{equation}
Thus we have
\begin{equation}
    F_{[0,0,1]}=f_{[0,0,1]},\quad\,F_{[1,0,0]}=f_{[1,0,0]}.
\end{equation}

\subsection{$r=1$ projection}
For larger $p$, it requires other projections different from the pure projection to recover the full information of $\mathcal{N}=4$ SYM correlators. We take $p=4$ as an example to showcase the procedure.
In $\mathcal{N}=4 $ SYM, the correlator is expressed as
\begin{equation}
    G_{\{4,4\}}=\mathcal{R}_{\mathcal{N}=4}d_{12}^{N-4}(\mathcal{X}^2F_{[2,0,0]}+\mathcal{Y}^2F_{[0,2,0]}+\mathcal{Z}^2F_{[0,0,2]}+\mathcal{X}\mathcal{Y}F_{[1,1,0]}+\mathcal{Y}\mathcal{Z}F_{[0,1,1]}+\mathcal{X}\mathcal{Z}F_{[1,0,1]}).
\end{equation}
In this case, the channels of $\mathcal{Y}\mathcal{Z}$ is abscent under the pure projection and thus we need to choose the projection with $r=1$. Under this projection, the $\mathcal{N}=2$ reduced correlator is
\begin{equation}
    G_{\{4,4\}}|_{\text{$r=1$ projection}}=YZ^3(F_{[0,1,1]}-F_{[0,0,2]})+...,
\end{equation}
and at the same time, the $\mathcal{N}=2$ correlator gives

\begin{equation}
     \langle\det(\tilde{q}_1)\det(q_2)\text{tr}(\tilde{q}_3^3q_3) \text{tr}(q_4^3\tilde{q}_4)\rangle=YZ^3 f_{[0,1,1]}+...\,.
\end{equation}
Matching the coefficients of the channel $YZ^3$, we have 
\begin{equation}
    f_{[0,1,1]}=F_{[0,1,1]}-F_{[0,0,2]}.
\end{equation}

\section{More Sum rules}
\label{sumruledata}

In this appendix, we list the sum rules obtain from field theory calculations and integrability predictions. Here we used the generating function in \cite{Basso:2017khq,Jiang:2019xdz} to package the sum rules up to two loops
\begin{align}    \mathbb{P}_{L,S}^{[n,m]}=\sum_{a=0}^{2}\sum_{b=0}^{a}\mathcal{P}_{nm,L,S}^{(a,b)}~g^{2a}~\mu^{b}.
\end{align}
$\mathbf{Table}$~$\ref{P002}$-$\ref{p226}$ are asymptotic data which are obtained from $\langle\mathcal{D}\mathcal{D}\mathcal{O}_5\mathcal{O}_5\rangle$. $\mathbf{Table}$~$\ref{P002wrap}$ shows the data with the wrapping correction, obtained from $\langle\mathcal{D}\mathcal{D}\mathcal{O}_2\mathcal{O}_2\rangle$. The blue colored data corresponds to the cases that only requires asymptotic contributions in the  $SL(2)$ sector, while the red colored  data corresponds to the cases that also requires including the wrapping correction at two-loop order in the $SL(2)$ sector. 
\begin{table}[h!]
\centering
\setlength{\tabcolsep}{6pt}
\renewcommand{\arraystretch}{2.5}
\begin{tabular}{|c|c|}
\hline
Spin $S$ & $l=1,\quad \ell>1$ \\
\hline
2 & \color{blue}\makecell[c]{
$\frac{1}{3}-4g^2+4g^2\mu+(56-12\zeta(3))g^4-64 g^4 \mu+24g^4\mu^2$
} \\
\hline
4 & \color{blue}\makecell[c]{
$\frac{1}{35}-\frac{205}{441}g^2+\frac{10}{21}g^2+\Big(\frac{143525}{18522}-\frac{10\zeta(3)}{7}\Big)g^4-\frac{4280}{441}g^4 \mu+\frac{250}{63} g^4 \mu^2$
} \\
\hline
6 & \color{blue}\makecell[c]{
$\frac{1}{462}-\frac{1106}{27225}g^2+\frac{7}{165}g^2\mu
+\Big(\frac{199888687}{269527500}-\frac{7\zeta(3)}{55}\Big)g^4$\\[3mm]$-\frac{396851}{408375} g^4 \mu+\frac{343}{825} g^4 \mu^2$
} \\
\hline
8 & \color{blue}\makecell[c]{
$\frac{1}{6435}-\frac{14380057}{4509004500}g^2+\frac{761}{225225}g^2\mu+\Big(\frac{10545232129372049}{170610810470100000}-\frac{761\zeta(3)}{75075}\Big)g^4$\\[3mm]
$-\frac{29591140129}{355084104375}g^4 \mu+\frac{579121}{15765750}g^4 \mu^2$
} \\
\hline
\end{tabular}
\caption{${\mathbb{P}^{[0,0]}_{2,S}}$. These sum rules can be recovered by asymptotic Bethe ansatz in the $SL(2)$ sector using integrability.}
\label{P002}
\end{table}

\begin{table}[h!]
    \centering
    \renewcommand{\arraystretch}{2.5}  
    \setlength{\tabcolsep}{6pt}  
    \begin{tabular}{|c|c|}
        \hline
       Spin $S$ & $l=2,\quad \ell>1$ \\
        \hline
        0 &$-\frac{1}{30}+\frac{7}{25}g^2-\frac{2}{5}g^2\mu+\Big(\frac{6 \zeta
   (3)}{5}-\frac{3269}{750}\Big)g^4+\frac{472}{75}g^4\mu-\frac{46}{15}g^4\mu^2$ \\
        \hline
       2 &  $-\frac{1}{378}+\frac{127}{7938}g^2-\frac{2}{63}g^2\mu +\Big(\frac{2 \zeta (3)}{21}+\frac{345862}{750141}\Big)g^4-\frac{11686
   }{11907}g^4\mu+\frac{83}{189}g^4\mu^2$\\
        \hline
        4 &\makecell[c]{$-\frac{1}{5148}+\frac{1865}{2208492}g^2-\frac{1}{429}g^2\mu+ g^4 \Big(\frac{\zeta
   (3)}{143}+\frac{93310002367}{568465840800}\Big)g^4$\\[3mm]$-\frac{2737324}{8281845}g^4 \mu+\frac{1237}{7722}g^4\mu^2$}\\
        \hline
       6 &\makecell[c]{$-\frac{1}{72930}+\frac{24471}{590976100}g^2-\frac{2}{12155}g^2\mu+ \Big(\frac{6 \zeta
   (3)}{12155}$\\[3mm]$+\frac{61708430516499713}{2280846018611160000}\Big)g^4-\frac{15332580503}{279236207250}g^4\mu+\frac{6657}{243100}g^4\mu^2$} 
        \\
        \hline
    \end{tabular}
    \caption {$\mathbb{P}^{[0,0]}_{4,S}$.}
    \label{P004}
\end{table}

\begin{table}[h!]
    \centering
    \renewcommand{\arraystretch}{2.5}  
    \setlength{\tabcolsep}{6pt}  
    \begin{tabular}{|c|c|}
        \hline
       Spin $S$ & $l=3,\quad  \ell> 1$ \\
        \hline
        0 &$\frac{1}{105}-\frac{218}{2205}g^2+\frac{4}{35}g^2\mu+\Big(\frac{1204447}{555660}-\frac{12 \zeta
   (3)}{35}\Big)g^4-\frac{21746}{6615}g^4\mu+\frac{54}{35}g^4\mu^2$ \\
        \hline
       2 & \makecell[c]{ $\frac{1}{660}-\frac{30607}{1633500}g^2+\frac{101}{4950}g^2\mu+ \Big(\frac{21480952129}{32343300000}-\frac{101 \zeta
   (3)}{1650}\Big)g^4$\\[3mm]$-\frac{1320709}{1225125}g^4\mu+\frac{4093}{8250}g^4\mu^2$}\\
        \hline
        4 & \makecell[c]{$\frac{2}{10725}-\frac{176412889}{67635067500}g^2+\frac{9419}{3378375}g^2\mu+
   \Big(\frac{309198608850475343}{2559162157051500000}$\\[3mm]$-\frac{9419 \zeta
   (3)}{1126125}\Big)g^4-\frac{1079572353157}{5326261565625}g^4\mu+\frac{3681994}{39414375}g^4\mu^2$}\\
        \hline
       6 & \makecell[c]{$\frac{1}{50388}-\frac{1701615497}{5598385949520}g^2+\frac{8549}{26453700}g^2\mu+
   \Big(\frac{1299750824747981890597}{77751461756229042600000}$\\[3mm]$-\frac{8549 \zeta
   (3)}{8817900}
   \Big)g^4-\frac{105165881428613
   }{3673940779372500}g^4\mu+\frac{110477629}{8332915500}g^4\mu^2$} 
        \\
        \hline
    \end{tabular}
    \caption {$\mathbb{P}^{[0,0]}_{6,S}$.}
    \label{P006}
\end{table}

\begin{table}[h!]
    \centering
    \renewcommand{\arraystretch}{2.5}  
    \setlength{\tabcolsep}{6pt}  
    \begin{tabular}{|c|c|}
        \hline
       Spin $S$ & $l=2,\quad \ell> 1$ \\
        \hline
        2 &\color{blue}$ -\frac{1}{5}+0\times g^2+0\times g^2\mu+8 g^4-16 g^4 \mu+8 g^4 \mu^2$ \\
        \hline
       4 &  \color{blue}$-\frac{1}{63}+0\times g^2+0\times g^2\mu+\frac{7381}{1458} g^4-\frac{824}{81}g^4 \mu+\frac{46}{9} g^4 \mu^2$\\
        \hline
        6 &\color{blue}$-\frac{1}{858}+0\times g^2+0\times g^2\mu+\frac{26683960663}{23686076700}g^4-\frac{31574008}{13803075}g^4\mu+\frac{7472}{6435}g^4\mu^2$\\
        \hline
       8 &\color{blue}\makecell[c]{$-\frac{1}{12155}+0\times g^2+0\times g^2\mu+\frac{9752108060381}{57121112412000}g^4$\\[3mm]$-\frac{376954397}{1076944050}g^4\mu+\frac{555}{3094}g^4\mu^2$} 
        \\
        \hline
    \end{tabular}
    \caption {${\mathbb{P}^{[1,1]}_{4,S}}$. These sum rules can be recovered by asymptotic Bethe ansatz in the $SL(2)$ sector using integrability.}
    \label{P114}
\end{table}

\begin{table}[h!]
    \centering
    \renewcommand{\arraystretch}{2.5}  
    \setlength{\tabcolsep}{6pt}  
    \begin{tabular}{|c|c|}
        \hline
       Spin $S$ & $l=3,\quad \ell > 1$ \\
        \hline
        0 &$\frac{2}{35}-\frac{2}{5}g^2+\frac{2}{5}g^2\mu+\Big(\frac{33623}{3430}-\frac{6 \zeta
   (3)}{5}\Big)g^4-\frac{3508}{245}g^4\mu+\frac{214}{35}g^4\mu^2$ \\
        \hline
       2 &  $\frac{1}{110}-\frac{521}{5445}g^2+\frac{16}{165}g^2\mu+\Big(\frac{1998358}{539055}-\frac{16\zeta(3)}{55}\Big)g^4-\frac{97742
   }{16335}g^4\mu+\frac{147}{55}g^4\mu^2$\\
        \hline
        4 &\makecell[c]{$\frac{4}{3575}-\frac{10909}{760500}g^2+\frac{43}{2925}g^2\mu+\Big(\frac{49938672636803}{71058230100000}-\frac{43 \zeta
   (3)}{975}
   \Big)g^4$\\[3mm]$-\frac{2436555227}{2070461250}g^4\mu+\frac{19113}{35750}g^4\mu^2$}\\
        \hline
       6 &\makecell[c]{$\frac{1}{8398}-\frac{4415079}{2556060500}g^2+\frac{101}{56525}g^2\mu+\Big(\frac{94670690730795592489}{959894589583074600000}-\frac{303 \zeta
   (3)}{56525}\Big)g^4$\\[3mm]$-\frac{5740278579554}{34017970179375}g^4\mu+\frac{23918623}{308626500}g^4\mu^2$ }
        \\
        \hline
    \end{tabular}
    \caption {$\mathbb{P}^{[1,1]}_{6,S}$.}
    \label{P116}
\end{table}

\begin{table}[h!]
    \centering
   \renewcommand{\arraystretch}{2.5}  
    \setlength{\tabcolsep}{6pt}  
    \begin{tabular}{|c|c|}
        \hline
       Spin $S$ & $l=3,\quad \ell > 1 $\\
        \hline
        0 &$-\frac{1}{15}+\frac{2}{3}g^2-\frac{2}{3}g^2 \mu+\Big(2 \zeta
   (3)-\frac{28}{3}\Big)g^4+\frac{32}{3}g^4\mu-4 g^4\mu^2$ \\
        \hline
       2 &  $-\frac{1}{77}+\frac{521}{3267}g^2-\frac{16}{99}g^2\mu+\Big(\frac{16 \zeta (3)}{33}-\frac{1051549}{431244}\Big)g^4+\frac{9574}{3267}g^4\mu-\frac{343}{297}g^4\mu^2$\\
        \hline
        4 &\makecell[c]{$-\frac{1}{585}+\frac{10909}{456300}g^2-\frac{43}{1755}g^2 \mu+ \Big(\frac{43\zeta
   (3)}{585}-\frac{3114671447}{8008065000}\Big)g^4$\\[3mm]$+\frac{2480956}{5133375}g^4\mu-\frac{5183}{26325}g^4\mu^2$}\\
        \hline
       6 &\makecell[c]{$-\frac{2}{10659}+\frac{1471693}{511212100}g^2-\frac{101}{33915}g^2 \mu+
   \Big(\frac{101\zeta(3)}{11305}$\\[3mm]$-\frac{38610116719604041}{786440719731240000}\Big)g^4+\frac{22647560078}{362321575875}g^4\mu-\frac{1115083}{42732900}g^4\mu^2$} 
        \\
        \hline
    \end{tabular}
    \caption {$\mathbb{P}^{[2,0]}_{6,S}$.}
    \label{p206}
\end{table}

\begin{table}[h!]
    \centering
\renewcommand{\arraystretch}{2.5}  
    \setlength{\tabcolsep}{6pt}  
    \begin{tabular}{|c|c|}
        \hline
       Spin $S$ & $l=3,\quad \ell>1$ \\
        \hline
        2 &$\color{blue}\frac{4}{7}-4g^2+4g^2\mu+(56-12\zeta(3)) g^4-64g^4\mu+24 g^4 \mu^2$ \\
        \hline
       4 &  $\color{blue}\frac{1}{11}-\frac{1042}{1089}g^2+\frac{32}{33}g^2\mu+\Big(\frac{1051549}{71874}-\frac{32\zeta(3)}{11}\Big)g^4-\frac{19148}{1089}g^4\mu+\frac{686}{99} g^4 \mu^2$\\
        \hline
        6 &\color{blue}\makecell[c]{$\frac{8}{715}-\frac{10909}{76050}g^2+\frac{86}{585}g^2\mu+\Big(\frac{3114671447
   }{1334677500}-\frac{86\zeta(3)}{195} \Big)g^4$\\[3mm]$-\frac{4961912}{1711125}g^4\mu+\frac{10366}{8775}g^4\mu^2$}\\
        \hline
       8 & \color{blue}\makecell[c]{$\frac{5}{4199}-\frac{4415079}{25560605}g^2+\frac{202}{11305}g^2\mu+\Big(\frac{38610116719604041
   g^4}{131073453288540000}$\\[3mm]$-\frac{606\zeta (3)}{11305}\Big)g^4-\frac{45295120156}{120773858625}g^4\mu+\frac{1115083 }{7122150}g^4\mu^2$} 
        \\
        \hline
    \end{tabular}
    \caption {${\mathbb{P}^{[2,2]}_{6,S}}$. These sum rules can be recovered by asymptotic Bethe ansatz in the $SL(2)$ sector using integrability.}
    \label{p226}
\end{table}

\begin{table}[h!]
    \centering
    \renewcommand{\arraystretch}{2.5}  
    \setlength{\tabcolsep}{6pt}  
    \begin{tabular}{|c|c|}
        \hline      
       Spin $S$ & $l=1,\quad \ell=1$ \\
        \hline
        2 &\color{red}$\frac{1}{3}-4g^2+4g^2\mu+56g^4-64 g^4 \mu+24g^4\mu^2$ \\
        \hline
       4 &\color{red}  $\frac{1}{35}-\frac{205}{441}g^2+\frac{10}{21}g^2\mu+\frac{73306} {9261}g^4-\frac{4280}{441}g^4\mu+\frac{250}{63} g^4\mu^2$\\
        \hline
        6 &\color{red}$\frac{1}{462}-\frac{1106}{27225}g^2+\frac{7}{165}g^2\mu+\frac{826643623}{1078110000}g^4-\frac{396851}{408375}g^4\mu+\frac{343}{825} g^4\mu^2$\\
        \hline
       8 &\color{red}\makecell[c]{$\frac{1}{6435}-\frac{14380057}{4509004500}g^2+\frac{761}{225225}g^2\mu+\frac{2748342985341731}{42652702617525000}g^4$\\[3mm]$- \frac{29591140129 }{355084104375}g^4\mu+\frac{579121
 }{15765750} g^4\mu^2$} \\
 \hline
    \end{tabular}
    \caption {${\mathbb{P}^{[0,0]}_{2,S}}$. These sum rules can be recovered by asymptotic Bethe ansatz and the bottom wrapping corrections in the $SL(2)$ sector using integrability.}
    \label{P002wrap}
\end{table}

\clearpage


\section{Asymptotic OPE formula from integrability}
\label{Asymptotic formula}
The formula for $\mathsf{d}_{L,S}$ and $\mathsf{c}_{L,S}$ involve square root expressions, which might cause an inherent sign ambiguity. In this appendix, we give the product formula without such square roots.
\paragraph{The anomalous dimension} $\delta\Delta_{L,S}$ in the $SL(2)$ sector is given by
\begin{align}
    \delta\Delta_{L,S}=2i g\sum_{j=1}^{S}\Big(\frac{1}{x_j^{+}}-\frac{1}{x_j^{-}}\Big)\label{delatk}
\end{align}

\paragraph{The OPE coefficients} The coefficients $\mathsf{d}_{L,S}$ in the $SL(2)$ sector can be expressed as
\begin{align}
   \mathsf{d}_{L,S}=-\frac{i^{L}+(-i)^{L}}{2^S\sqrt{L}}\sqrt{\Big(\prod_{j=1}^{S/2}\frac{u_{j}^{2}+1/4}{u_{j}^{2}}\sigma_{B}^{2}(u_{j})\Big)\frac{\det G_{+}}{\det G_{-}}}\label{dk}
\end{align}
where $G_{\pm}$ are $\frac{S}{2}\times\frac{S}{2}$ Gaudin-like matrices whose elements are \cite{Jiang:2019zig}
\begin{align}
    \left(G_{\pm}\right)_{ij}=\left[L\partial_{u}p(u_{i})+\sum_{k=1}^{\frac{S}{2}}K_{+}(u_{i},u_{k})\right]\delta_{ij}-K_{\pm}(u_{i},u_{j}),
\end{align}
with $K_{\pm} $ given by
\begin{align}
    K_{\pm}(u,v)=\frac{1}{i}\partial_u\log S(u,v)\pm\partial_u\log S(u,-v).
\end{align}
$\sigma_B(u)$ is the boundary dressing phase given by \cite{Jiang:2019xdz}
\begin{align}
     \sigma_B(u)=\frac{1}{4}-g^4\frac{12\zeta(3)}{4u^2+1}+g^6\left(\frac{64(1-12u^2)\zeta(3)}{(1+4u^2)^3}+\frac{30\zeta(5)}{1+4u^2}\right)+O(g^8).
\end{align}
The structure constant  $\mathsf{c}_{L,S}$ in the $SL(2)$ sector is given by \cite{Vieira:2013wya,Basso:2015zoa,Basso:2017khq}
\begin{align}
    \mathsf{c}_{L,S} = 
    \sqrt{\frac{\prod_{k=1}^S\mu(u_k)\prod_{i<j}h(u_i,u_j)h(u_j,u_i)}{\det\partial_{u_i}\phi_j\prod_{i<j}S(u_i,u_j)}}\mathcal{A}_{l},\label{ck}
\end{align}
where $\mu(u)$ is defined as
\begin{align}
     \mu(u)=\frac{\left(1-\frac{1}{x^{+}(u)x^{-}(u)}\right)^2}{\left(1-\frac{1}{(x^{+}(u))^2}\right)\left(1-\frac{1}{(x^{-}(u))^2}\right)},
\end{align}
and the fundamental building block $h(u,v)$ is defined as
\begin{align}
   h(u,v) = \frac{x^{-}(u) - x^{-}(v)}{x^{-}(u) - x^{+}(v)} 
    \frac{1 - \frac{1}{x^{-}(u)x^{+}(v)}}{1 - \frac{1}{x^{+}(u)x^{+}(v)}}\frac{1}{\sigma_B(u)}.
\end{align}
$\phi(u_j)$ is the scattering phase defined by
\begin{align}
    e^{i\phi(u_j)}\equiv e^{ip_jL}\prod_{k\neq j}S(u_j,u_k).
\end{align}

The most nontrivial component is $\mathcal{A}_l$, which depends on the three-point function configuration through the integer $l$. It involves a sum over all bipartitions $\alpha \cup \bar{\alpha} = \{u_j\}$ of the Bethe roots:
\begin{align}   \mathcal{A}_{l}=\sum_{\alpha\cup\bar{\alpha}=\{u_j\}}(-1)^{|\bar{\alpha}|}\prod_{j\in\bar{\alpha}}e^{ip(u_j)l}\prod_{i\in\alpha,j\in\bar{\alpha}}\frac{1}{h(u_i,u_j)}.
\end{align}

By multiplying the two expressions and using the parity symmetry of the Bethe roots, we can simply the product $\mathsf{d}_{L,S}\mathsf{c}_{L,S}$ as follow:
\begin{align}
\mathsf{d}_{L,S}\mathsf{c}_{L,S}=-(i^L+(-i)^L)\frac{1}{\det G_-}\left(\prod_{k=1}^{\frac{S}{2}}\tilde{\mu}(u_k)\right)\left(\prod_{1\leq i<j\leq\frac{S}{2}}H(u_i,u_j)\right)\mathcal{A}_{l},\label{dc}
\end{align}
with
\begin{align}
\tilde{\mu}(u)
&=\sigma_B(u)\frac{\left(1+\frac{1}{(x^+(u))^2}\right)\left(1+\frac{1}{(x^-(u))^2}\right)}{\left(1-\frac{1}{(x^+(u))^2}\right)\left(1-\frac{1}{(x^-(u))^2}\right)}\left(\frac{1-\frac{1}{x^+(u)x^-(u)}}{1+\frac{1}{x^+(u)x^-(u)}}\right)^2,
\end{align}
and
\begin{align}
H(u,v)&=\frac{(u^2-v^2)^2}{((u-v)^2+1)\left((u+v)^2+1\right)}\\\nonumber
&\quad\times\left[\frac{\left(1+\frac{1}{x^+(u)x^+(v)}\right)\left(1+\frac{1}{x^-(u)x^-(v)}\right)}{\left(1-\frac{1}{x^+(u)x^+(v)}\right)\left(1-\frac{1}{x^-(u)x^-(v)}\right)}\frac{\left(1-\frac{1}{x^+(u)x^-(v)}\right)\left(1-\frac{1}{x^-(u)x^+(v)}\right)}{\left(1+\frac{1}{x^+(u)x^-(v)}\right)\left(1+\frac{1}{x^-(u)x^+(v)}\right)}\right]^2.
\end{align}

\bibliographystyle{JHEP}
\bibliography{yunfeng.bib}

\end{document}